\documentclass[Chicago,csreport]{fsuthesis}
\usepackage{epsfig,contigrefs}
\include{Laithdefs}
\fourlevels
\title{PHOTOPRODUCTION OF PSEUDOSCALAR MESONS FROM NUCLEI}
\author{Laith J. Abu-Raddad}
\uauthor{LAITH J. ABU-RADDAD}
\thesistype{dissertation}
\department{Department of Physics}

\headofdept{Kirby W.\ Kemper}
\deptheadtitle{Chairman}
\deanofschool{Donald J. Foss}
\degree{Doctor of Philosophy}
\gradyear{2000}
\dept{Physics}
\majorprof{Jorge Piekarewicz}
\majorprofdegree{Ph.D.}
\outcommmember{Gregory Riccardi}
\commmembera{Simon Capstick}
\commmemberc{Adam Sarty}
\commmemberb{Adriana Moreo}

\semester{Summer}
\college{Arts and Sciences}
\defensedate{APRIL 27, 2000}

\begin{document}
\dedication{For my parents Huda and Jamal, my wife Mercedes, and my son Ommar.}

\def\acknowledgementtext{
\hskip\parindent
My dissertation could not have been achieved without the support, patience,
and guidance of Professor Jorge Piekarewicz in every detail of this
research work. I am deeply indebted for his close personal interaction
and concern. My sincere thanks to every member of the Nuclear Theory
Group at Florida State University for making this doctoral study
such an educational and enjoyable experience. I would like also to
extend my sincere appreciation to Professor Simon Capstick,  Professor
Adriana Moreo, Professor Gregory Riccardi, and Professor Adam Sarty
for reviewing this manuscript. 

This work was supported in part by the United
States Department of Energy under Contracts Nos. DE-FC05-85ER250000
and DE-FG05-92ER40750.
}

\def\abstracttext{%
The subject of this doctoral study is the pseudoscalar meson
photoproduction from nuclei. For simplicity, we study this process only
from spherical nuclei. Two types of processes are
investigated in this regard: coherent and quasifree processes. 
In the case of the coherent process, we study it for the
photoproduction of $\pi$ and $\eta$ mesons. We place special emphasis on
the various sources that put into question earlier
nonrelativistic-impulse-approximation calculations. These include:
final-state interactions, relativistic effects, off-shell ambiguities,
and violations of the impulse approximation. By far the largest uncertainty emerges from the ambiguity in
extending the many on-shell-equivalent representations of the
elementary amplitude off the mass shell. In the case of
the quasifree process, we study it for the photoproduction of $K^+$ meson.
We compute the recoil polarization of the lambda-hyperon and 
the photon asymmetry as well as the differential cross section.
By introducing the notion 
of a ``bound-nucleon propagator'' we exploit Feynman's trace 
techniques to develop closed-form, analytic expressions for all 
photoproduction observables. Our results indicate that 
polarization observables are sensitive only to
the fundamental physics, making them ideal tools for the study 
of modifications to the elementary process in the nuclear medium.

}

\dedicationtrue
\titlepage
\maketitle
\dedicationpage
\prefacesection{Acknowledgements}\acknowledgementtext
\tableofcontents
\listoftables
\listoffigures
\newpage
\singlespace
\abstractsection{ABSTRACT}
\singlespace
\maintext
\singlespace
\chapter{Introduction}
\label{Introduction}
\pagenumbering{arabic}

Before describing my doctoral research I would like to point out that this
manuscript has been written with the following philosophy in my mind:
I aspire to provide the reader with a comprehensive overview of my
research that stresses the fundamental physics and avoids unnecessary
details. In many occasions, insignificant intricacies were sacrificed
for a logical flow of ideas.

This research is concerned with the pseudoscalar meson
photoproduction from nuclei. A meson is a particular kind of
fundamental particle (as the pion, eta, and kaon) made up of a quark
and an antiquark~\cite{webster}. Quarks are the elementary particles
that constitute, as we believe today, the fundamental building blocks
of matter. Pseudoscalar mesons form a subgroup of mesons that have
zero spin (and thus called scalars) and behave in a certain
well-defined fashion under the action of symmetry operations. More
specifically, the pseudoscalar-meson wavefunction $\phi$ transforms to
$-\phi$ under the symmetry operation of spatial inversion. We study in
this manuscript photoproduction processes of three pseudoscalar
mesons: the kaon, pion, and eta. Table \ref{tab:quarkcontent} illustrates
the quark content of the different states of these mesons. In this
table $u$, $d$, and $s$ stand for up, down, and strange quarks
respectively, while $\bar{u}$, $\bar{d}$, $\bar{s}$ stand for the
corresponding antiparticles (antiquarks) of these quarks.
\begin{table}
\caption{Quark content of the kaon, pion, and eta pseudoscalar mesons.}
\vspace{.5cm}
\begin{center}
\thicklines
\begin {tabular} { l c c }
\hline
\hline
Pseudoscalar Meson &  Quark & Content \\
\hline
\thicklines
Kaon & $K^+ \sim \bar{s}u$ & $K^- \sim \bar{u}s$ \\
     & $K^\circ \sim \bar{s}d$ & $\bar{K}^\circ \sim \bar{d}s$ \\
Pion & $\pi^+ \sim \bar{d}u$ & $\pi^- \sim \bar{u}d$ \\
     & $\pi^\circ \sim \frac{1}{\sqrt{2}} (\bar{u}u - \bar{d}d) $ & \\
Eta  & $\eta \sim \frac{1}{\sqrt{6}} (\bar{u}u + \bar{d}d - 2
     \bar{s}s) $\\
\hline
\hline
\end{tabular}
\end{center}
\label{tab:quarkcontent}
\end{table}

Photoproduction describes a process where elementary particles (such
as mesons) are
produced as a result of the action of photons
(electromagnetic waves) on atomic nuclei~\cite{webster}. The basic
interaction in this work is as following: a photon is incident on a
target nucleus and interacts with its constituents. As a result, a
pseudoscalar meson is produced along with other particles. For
simplicity, we investigate here photoproduction processes only from spherical nuclei. We study
here two kinds of processes depending on the nature of the other
particles produced in this interaction: coherent and quasifree
processes.

In the coherent processes, the meson is produced with the
target nucleus maintaining its initial character. Thus we start the
interaction with a photon and some nucleus, and end up with a meson
and the same nucleus we started with. The process is labeled
``coherent'' because all protons and neutrons (referred to
collectively as
``nucleons'') in the nucleus participate in the process, leading to a
coherent sum of these individual nucleon contributions.

In the quasifree processes, the nucleus ruptures and thus fails to
maintain its initial identity. The meson is produced in association
with a nucleon (or an excited state of the nucleon like the lambda
hyperon) and some new recoil ``daughter'' nucleus. Thus, we start the interaction
with a photon and some nucleus, and end up with a meson, a free
nucleon (or an excited state of it), and a new nucleus. The process is
labeled as ``quasifree'' because it occurs in kinematic and physical
circumstances similar to those of the process that produces a meson from
a free unbound nucleon.

It is appropriate here to try to place these interactions to the bigger
picture of general physics research. Studying these processes is one facet
of the physicists' quest to understand the fundamental strong force
which plays the prominent role in interactions between
elementary particles at very small distance scales. In our current
understanding of physics, there are four forces that drive all
interactions in nature: gravitational, electromagnetic, weak, and
strong forces. Of these, we understand to a great extent the
nature of the electromagnetic and the weak forces, while the gravitational and
the strong still elude satisfactory and complete description. We do have a theory for the strong
interactions --- Quantum Chromodynamics (QCD) --- but this theory is
formidable to solve. As a result, a large chunk of the scientific
research in physics today, whether in experiment or theory, is devoted
to understanding this strong force. This effort is so extensive that
it encompasses thousands of scientists in the fields of elementary
particle and nuclear physics. This work is one minute step in
this grand path, in the subfield called medium-energy nuclear
physics. Our study attempts to provide a theoretical understanding of
experiments that have been conducted or planned to be conducted in
several laboratories: in the USA [such as the Thomas Jefferson
National Accelerator Facility (TJNAF)], in Europe [such as the Mainz Microtron
Laboratory (MAMI)], or in Japan [at the Research Center for Nuclear
Physics (RCNP)].

The study I presented here assists in understanding
several issues regarding this grand path of comprehending QCD. One of
these is the structure and nature of the QCD bound
states. There are two kinds of bound states in QCD: mesons (like the
pion or the kaon), and hadrons, which includes nucleons (protons or
neutrons) and nucleon resonances (excited states of nucleons) such as the lambda
or delta particles. The processes of meson photoproduction are
excellent tools in studying these states since these reactions
proceed through the exchange of QCD bound states. For example, the
pion photoproduction in a certain energy regime occurs as a result of
the exchange of a delta resonance. By studying this process, we can
have insights into the nature of this resonance and the mechanisms by
which it interacts and decays.

Many research projects have been devoted to studying
these kinds of meson photoproduction processes. Most studies have
concentrated on studying the photoproduction from free nucleons. Such
a process is labeled as ``free'' or ``elementary'' to distinguish it
from the same process from a nucleus. An enormous amount of knowledge has
been accumulated as a result, but this is still
insufficient.

In this work, we go a step further by
studying these processes from nuclei, because the nucleus in the coherent process acts as a ``filter'' that
allows certain physical mechanisms that occur in the elementary
process to go through, while blocking others. An example of this is the
$S_{11}$ resonance that dominates the elementary
process of eta photoproduction from a nucleon, but is almost perfectly
suppressed in the process from a spherical nucleus due to this filtering. Thus,
other mechanisms (such as the $D_{13}$ resonance) that are overshadowed
by the $S_{11}$ and cannot be disentangled in the elementary process,
in fact
dominate the process from a nucleus. Another manifestation of
this filtering is that the process from a spherical nucleus depends only on one
of the four amplitudes that drive the elementary process. Indeed, the
nucleus here acts as a laboratory to probe what we cannot study
otherwise.

As the name conveys, the quasifree process from nuclei is the
closest physically to the elementary or free process. The process can
be viewed as the elementary one but now in a nuclear
medium rather than in a free space. We can use this
reaction to investigate the changes of the elementary process in the
nuclear medium. One example is the pion quasifree process. As
pointed out above, the pion elementary process is driven by delta
resonance propagation in free space. In the quasifree process, however,
this resonance propagates in a nuclear medium and so interacts through
the strong force with the constituents of the nucleus, resulting in
modifications to its basic properties. Understanding these
modifications can elucidate some aspects of QCD.

So far I may have given an inaccurate impression that this work illuminates parts
of our knowledge concerning only the ``very small'' scales of time and
space. The processes that
drive the ``very small'' also propel the ``very large''. Indeed, our
impetus to study the quasifree process is because it is a basic
building block toward the bigger goal of assessing the possibility of
kaon condensation in neutron stars. Neutron stars are dense
celestial objects that consist primarily of closely packed neutrons
and result from the collapse of a supernova~\cite{webster}. These stars are among the most dense systems that
we can find in nature; their densities are about ten times that of the
nucleus, which is the most dense system in our solar system. Inquiries regarding the
nature, structure, and stability of these objects are among the most
intriguing questions in astrophysics today. One of the scenarios that
may be able to explain their existence is that these stars consist of a
new state of matter: strange matter. Strange matter refers to a form
of matter where there is a significant presence of strange quarks. Although strange matter has been observed in
laboratories --- as in the production of hypernuclei --- this matter has not
yet been observed as a stable state in nature. Kaon condensation in neutron stars
describes a hypothetical mechanism where, due to the very high density, it becomes
energetically favorable to produce strange particles like the kaon
(strange meson) or
the lambda (strange nucleon resonance). Thus, we have a stable
matter that is a ``condensate'' of ``strangeness''. Much work has been
devoted to this possibility and this scenario has yet to be confirmed
or refuted conclusively.

The bulk of this dissertation is essentially a reproduction of several
publications by the author and the
collaboration~\cite{pisabe97,apsm98,raddad99,radcar,abpi2000}\footnote{Copyright The American Physical Society 1997, 1998, 1999, and
2000. All rights reserved. Except as provided under U.S. copyright
law, this work may not be reproduced, resold, distributed or modified
without the express permission of The American Physical Society. The
archival versions of these works were published
in~\cite{pisabe97,apsm98,raddad99,abpi2000}}. Since
it is tedious and pointless to keep referring to these publications
throughout the manuscript, I only referred to them when I determine it
to be appropriate to do so. The reader should bear in mind however
that much of this work has its origin in these publications.

I would like to ask the reader for forgiveness for any repetitions in
this manuscript. In several occasions, I had to repeat certain aspects
because of appropriateness or significance in context.

Throughout this work (unless otherwise stated) we adopt the natural
system of units where $\hbar = c = 1$. This system is the appropriate
and standard one in all studies involving quantum field theory.

\section{Outline of Thesis}

The dissertation is divided into three parts: preliminaries, coherent
process, and quasifree process. The preliminaries part includes
Chapters \ref{ch:ElementaryProcess} and \ref{ch:NuclearStructure}. 
Chapter
\ref{ch:ElementaryProcess} describes the basic ideas behind
what is referred to as the elementary process: a pseudoscalar meson is
photo-produced from a free nucleon. Understanding this process is the foundation for understanding the same process from nuclei. Since
I will study processes from nuclei, I have to build the nuclear
structure for several nuclei. This is done in Chapter
\ref{ch:NuclearStructure}, where a relativistic nuclear structure
formalism is developed.

In the second part of the dissertation that encompasses Chapters
\ref{ch:CoherentTheory}, \ref{ch:Distortions},
\ref{ch:CoherentResults}, and \ref{ch:ConclusionsCoherent}, I
concentrate on the coherent process. I study this process for two
kinds of mesons: the pion ($\pi$) and the eta ($\eta$). In Chapter
\ref{ch:CoherentTheory}, I develop the basic theory where no
final-state interactions are assumed between the emitted meson and the
recoil nucleus. Then, I incorporate these interactions in Chapter
\ref{ch:Distortions}. In Chapter \ref{ch:CoherentResults}, I present
our results for this kind of process and discuss them. Finally in
Chapter \ref{ch:ConclusionsCoherent}, I draw conclusions.

The third part of the manuscript follows in a similar fashion to the
second one, but here I investigate the quasifree process. This is done
in Chapters
\ref{ch:QuasifreeTheory}, \ref{ch:ResultQuasifree}, and
\ref{ch:ConclusionQuasifree}.  I study this interaction only for one
kind of meson: the kaon ($K^+$). In Chapter \ref{ch:QuasifreeTheory},
I sketch the theory behind this process, while I present and discuss
the results in Chapter \ref{ch:ResultQuasifree}, and finally conclude
in Chapter \ref{ch:ConclusionQuasifree}.

\section{Technical Introduction and Background for the Coherent Process}

The coherent photoproduction of pseudoscalar mesons has been
advertised as one of the cleanest probes for studying how
nucleon-resonance formation, propagation, and decay get modified in
the many-body environment of nuclear matter; for current experimental
efforts see Ref.~\cite{Sambeek98}. The reason behind such optimism is
the perceived insensitivity of the reaction to nuclear-structure
effects. Indeed, many of the earlier nonrelativistic calculations
suggest that the full nuclear contribution to the coherent process
appears in the form of its matter
density~\cite{bofmir86,cek87,bentan90,ndu91,tryfik94,fix97}---itself
believed to be well constrained from electron-scattering experiments
and isospin considerations.

Recently, however, this simple picture has been put into question.
Among the many issues currently addressed---and to a large extent
ignored in all earlier analyses---are: background (non-resonant)
processes, relativity, off-shell ambiguities, non-localities, and
violations of the impulse approximation. We discuss each one of them
in this manuscript. For example, background contributions to the
resonance-dominated process can contaminate the analysis due to
interference effects. This has been shown recently for the
$\eta$-photoproduction process, where the background contribution
(generated by $\omega$-meson exchange) is in fact larger than the
corresponding contribution from the $D_{13}(1520)$
resonance~\cite{pisabe97}. We suggest in our study that---by using a
relativistic and model-independent parameterization of the elementary
amplitude---the nuclear-structure information becomes sensitive to
off-shell ambiguities. Further, the local assumption implicit in most
impulse-approximation calculations, and used to establish that all
nuclear-structure effects appear exclusively via the matter density,
has been lifted by Peters, Lenske, and
Mosel~\cite{Peters98a,Peters98b}. An interesting result that emerges
from their work on coherent $\eta$-photoproduction is that the
$S_{11}(1535)$ resonance---known to be dominant in the elementary
process but predicted to be absent from the coherent
reaction~\cite{bentan90}---appears to make a non-negligible
contribution to the coherent process in the case of non-spin-saturated
but spherical nuclei such as ${}^{12}$C. Spin-saturated nuclei represent one
type of nuclei where all states corresponding to one orbital angular
momentum are
completely filled. Finally, to our knowledge, a comprehensive study of possible
violations to the impulse-approximation, such as the modification to
the production, propagation, and decay of nucleon resonances in the
nuclear medium, has yet to be done.

In this work we concentrate---in part because of the expected
abundance of new, high-quality experimental data---on the coherent
photoproduction of neutral pions. The central issue to be addressed
here is the off-shell ambiguity that emerges in relativistic
descriptions and its impact on extracting reliable resonance
parameters; no attempt has been made here to conduct a quantitative
and detailed study of possible violations
of the impulse approximation or to the local assumption. These
violations have been studied only qualitatively. Indeed, we
carry out our calculations within the framework of a relativistic
impulse approximation model. However, rather than resorting to a
nonrelativistic reduction of the elementary amplitude, we keep intact
its full relativistic structure~\cite{cgln57}. As a result, the lower
components of the in-medium Dirac spinors are evaluated dynamically in
the Walecka model~\cite{serwal86}.

Another important ingredient of the calculation are the final-state
interactions of the outgoing meson with the nucleus. We address the
mesonic distortions via an optical-potential model of the
meson-nucleus interaction. For example, we use earlier models of the
pion-nucleus interaction plus isospin symmetry---since these models
are constrained mostly from charged-pion data---to construct the
neutral-pion optical potential.  However, since we are unaware of a
realistic optical-potential model that covers the $\Delta$-resonance
region, we have extended the low-energy work of Carr, Stricker-Bauer,
and McManus~\cite{SMC} to higher energies. In this way we have
attempted to keep to a minimum the uncertainties arising from the
optical potential, allowing concentration on the impact of the
off-shell ambiguities to the coherent process.

\section{Technical Introduction and Background for the Quasifree Process}

Impelled by recent experimental advances, there is an increasing
interest in the study of strangeness-production reactions from
nuclei. These reactions form our gate to the relatively unexplored
territory of hypernuclear physics. Moreover, these reactions
constitute the basis for studying novel physical phenomena, such as
the existence of a kaon condensate in the interior of neutron
stars\cite{kn86}. Indeed, the possible formation of the condensate
could be examined indirectly by one of the approved
experiments\cite{chw91} at the Thomas Jefferson National Accelerator
Facility (TJNAF). This experimental approach is reminiscent of the
program carried out at the Los Alamos Meson Physics Facility (LAMPF)
where pion-like modes were studied extensively through the quasifree
$(\vec{p},\vec{n})$ reaction~\cite{McCl92,Chen93}. These measurements
placed strong constraints on the (pion-like) spin-longitudinal
response and showed conclusively that the long-sought pion-condensed
state does not exist.

The work presented here is a small initial step towards a more
ambitious program that concentrates on relativistic studies of
strangeness in nuclei. Our aim in this manuscript is the study of the
photoproduction of kaons from nuclei in the quasifree regime. This
investigation helps us in two fronts. First, it sheds light on the
elementary process, $\gamma p \rightarrow K^{+} \Lambda$, by providing
a different physical setting (away from the on-shell point) for
studying the elementary amplitude. Second, it will enable us, in a
future study, to explore modifications to the kaon propagator in the
nuclear medium and to search for those observables most sensitive to
the formation of the condensate. To achieve these goals we focus on
the study of polarization observables. Polarization observables have
been instrumental in the understanding of elusive details about
subatomic interactions, as they are much more effective
discriminators of subtle physical effects than the traditional
unpolarized cross section. Moreover, quasifree polarization
observables might be one of the cleanest tools for probing nuclear
dynamics. For example, the reactive content of the process is simple,
being dominated by the quasifree production and knockout of a
$\Lambda$-hyperon. Further, free polarization observables provide a
baseline, against which possible medium effects may be inferred.
Deviations of polarization observables from their free values are
likely to arise from a modification of the interaction inside the
nuclear medium or from a change in the response of the target.
Indeed, relativistic models of nuclear structure predict medium
modifications to the free observables stemming from an enhanced lower
component of the Dirac spinors in the nuclear medium~\cite{serwal86}.
Finally, nonrelativistic calculations of the photoproduction of
pseudoscalar mesons suggest that, while distortion effects provide an
overall reduction of the cross section, they do so without
substantially affecting the shape of the
distribution\cite{lwb93,lwbt96,blmw98}. Indeed, these nonrelativistic
calculations show that two important polarization observables --- the
recoil polarization of the ejected baryon and the photon asymmetry ---
are largely insensitive to distortion effects. Moreover, they seem to
be also independent of the mass of the target nucleus.

An insensitivity of polarization observables to distortion effects is
clearly of enormous significance, as one can unravel distortion
effects from those effects arising from relativity or from the
large-momentum components in the wavefunction of the bound nucleon.
Indeed, relativistic plane-wave impulse approximation (RPWIA)
calculations have been successful in identifying physics not present
at the nonrelativistic level~\cite{gp94,cdmu98}. Finally, neglecting
distortions allows the computation of all polarization observables in
closed form~\cite{gp94} by using the full power of Feynman's trace
techniques.

\chapter{Photoproduction of pseudoscalar mesons from free nucleons}
\label{ch:ElementaryProcess}

Any investigation of the processes of meson
photoproduction from nuclei must start with a study of the
photoproduction from a single free nucleon. This
process from a free nucleon is usually labeled as elementary to distinguish it from other
processes from an interacting or bound nucleon. It is
appropriate here to stress that it is not the purpose of this work to
investigate the photoproduction interactions from free nucleons;
this topic has been extensively studied by many scientific groups and
is an ``industry'' of its own. It is imperative, however, to examine
these processes to incorporate them in our investigation of the
photoproduction reactions from nuclei. We start this chapter by
describing the basic formalism of any elementary process of meson photoproduction from a free nucleon.

\section{Elementary Process: Model Independent Formalism}

In the elementary process a photon is absorbed by a free nucleon (a
proton or a neutron) to yield a pseudoscalar meson in addition to a
nucleon (or a hyperon).  Figure \ref{fig:ElementaryProcess} illustrates
this process.
\BFIG
\centerline{\psfig{figure=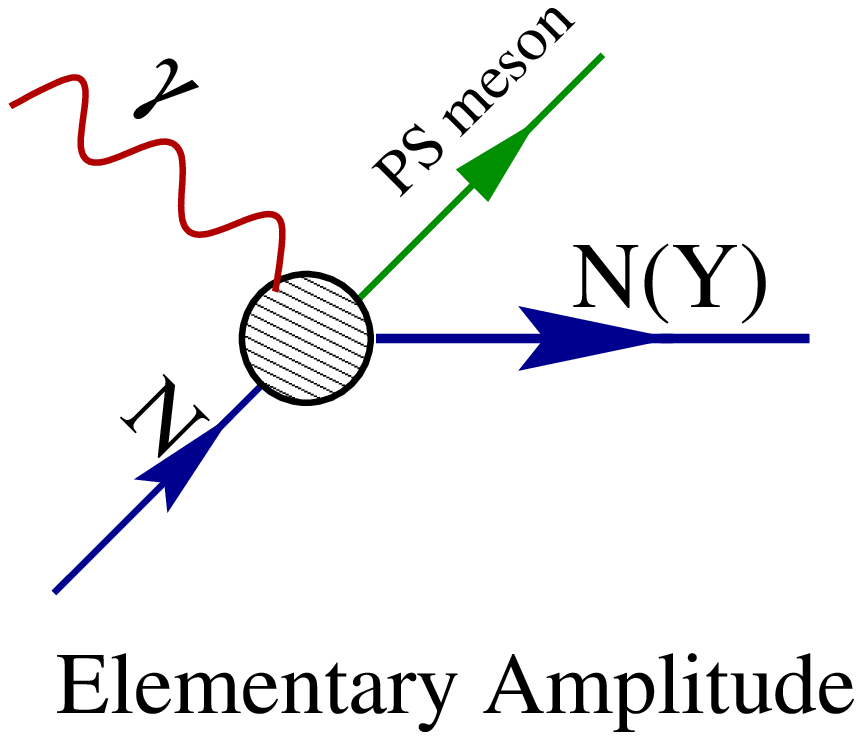,height=9.0cm,width=10.5cm}}
\caption[Schematic diagram of the process $N( \gamma , PS\,\,meson )N (Y)$]
{The elementary process of pseudoscalar meson photoproduction from a
single free nucleon. A photon is absorbed by the nucleon to yield a
pseudoscalar meson in addition to a nucleon (or a hyperon).}
\label{fig:ElementaryProcess}
\EFIG
The most general expression for the scattering matrix element using
perturbation theory can be
written as a multiple integral in the following form:
\begin{equation}
\int d^4x_1 \ldots d^4x_N \; \overline{{\psi}} A^\mu
J_\mu (x_1,\ldots,x_N){\psi} \phi\;,
\label{scatmatele}
\end{equation}
where $\psi$ is the Dirac spinor
for a free nucleon, $A^\mu$ 
is the photon wavefunction (field), and $\phi$ is the pseudoscalar
meson wavefunction (field). The expression clearly includes the
electromagnetic contraction between the photon field and the conserved
electromagnetic current $J_\mu (x_1,...,x_N)$. The number $N$ of
independent variables to be integrated over, depends on the nature of
the effective field theory employed. In other words, it depends on the
number of vertices in each Feynman diagram derived from this effective
field theory. From this most general form, it can be shown that the model independent
parameterization for this interaction is given in terms of four Lorentz-
and gauge-invariant amplitudes (matrices) in the space of Dirac
spinors as\cite{bentan90,cgln57,wcc90,dfls96}
\begin{equation}
T[\gamma N \rightarrow PS\, meson \, N (Y)] = \sum_{i=1}^{4}
A_{i}(s,t) M_i\,,
\label{scatmat}
\end{equation}                             
where the invariant matrices have the form
\begin{eqnarray}
M_1 & =& - \gamma^5 \rlap/{\varepsilon} \rlap/k\,, \nonumber \\ 
M_2 &=&
2 \gamma^5 [(\varepsilon \cdot p_1) (k\cdot p_2 ) - (\varepsilon \cdot
p_2) (k\cdot p_1)]\,,\nonumber \\ 
M_3 &=& \gamma^5 [\rlap/{\varepsilon}
((k\cdot p_1) - \rlap/k (\varepsilon \cdot p_1)]\,,
\nonumber \\ 
M_4 &=& \gamma^5 [\rlap/{\varepsilon} ((k\cdot p_2 ) - \rlap/k
(\varepsilon \cdot p_2)]\,,
\end{eqnarray}
and where $\varepsilon$ and $k$ are the polarization and four-momenta
of the photon, and ${p_1}$ and ${p_2}$ are the four momenta of the
struck nucleon and recoil nucleon (hyperon) respectively. The terms
$\rlap/{\varepsilon}$ and $\rlap/k$ stand for $\gamma^\mu
\varepsilon_mu$ and $\gamma^\mu k_\mu$ respectively. Here, the
kinematic quantities $s$ and $t$ are the Mandelstam variables $s
\equiv (p_1 + k)^2$ and $t \equiv (k - k')^2$. 

This is the standard, but not unique, parameterization of the elementary
process. There are many other possible parameterizations which are
equivalent provided the struck nucleon is free
(on-shell). Unfortunately, it is not clear how we can apply this
parameterization to a bound nucleon (off-shell nucleon) without a
detailed microscopic model for this process. We will come back to this
point in Chapter \ref{ch:CoherentTheory}.

We choose to transform this standard form into a more suitable
one\cite{pisabe97} by using the identity\cite{yndurain}
\begin{equation}
\gamma^5 \gamma^\mu \gamma^\nu = \gamma^5 g^{\mu \nu} + \frac {1}{2 i}
\varepsilon^{\mu \nu \alpha \beta} \gamma_\alpha \gamma_\beta\,,
\end{equation}
to rewrite the term $M_1 = - \gamma^5 \rlap/{\varepsilon} \rlap/k$ as
\begin{eqnarray}
M_1 & =& - \gamma^5 \rlap/{\varepsilon} \rlap/k \\ & = & \frac {1}{2}
\varepsilon^{\mu \nu \alpha \beta} \varepsilon_\mu k_\nu
\sigma_{\alpha \beta}\,,
\end{eqnarray}
where we have used the convention of $\varepsilon^{0 1 2 3} = -1$ for
the Levi-Civita tensor. Consequently, the parameterization of the
elementary process is rewritten as
\begin{equation}
T[\gamma N \rightarrow PS\, meson \, N (Y)] = F^{\alpha \beta}_{T}
\sigma_{\alpha \beta} + F_P i \gamma_5 + F^{\alpha}_{A} \gamma _\alpha
\gamma_5\,,
\label{lorentzrep}
\end{equation}
where tensor, pseudoscalar, and axial-vector coefficients have been
  introduced as following
\begin{eqnarray}
F^{\alpha \beta}_{T} &=& \frac {1}{2} \varepsilon^{\mu \nu \alpha
\beta}
\varepsilon_\mu k_\nu A_1(s,t)\,, \nonumber \\
F_P &=& - i\mbox{ } 2 \mbox{ }[(\varepsilon \cdot p_1) (k\cdot p_2) -
(\varepsilon \cdot p_2) (k\cdot p_1)] A_2(s,t)\,, \nonumber \\
F^{\alpha}_{A} &=& [(\varepsilon \cdot p_1) k^\alpha - (k\cdot p_1)
\varepsilon ^\alpha] A_3(s,t) + [(\varepsilon \cdot p_2) k^\alpha -
(k\cdot p_2) \varepsilon ^\alpha] A_4(s,t)\,.
\label{lorentzrepcoef}
\end{eqnarray}
This form manifests nicely the Lorentz and parity transformation
properties of the different bilinear covariants.

\section{Elementary Process: Model Dependent Formalism}

The parameterization developed above is model
independent and applies to any process of pseudoscalar meson
photoproduction from a single free nucleon. This parameterization,
however, is in terms of four unknown amplitudes: $\{ A_1, A_2, A_3,
A_4\}$. These amplitudes can be determined using different methods. In
one method, we can simply extract them from experimental data for the
observables of these processes. Another method, which is a fundamental one, is
to study the physical processes behind each photoproduction process,
and thus to construct a microscopic model for this process in terms of
the fundamental degrees of freedom in this interaction. Since these degrees of freedom involve quarks and
gluons, this approach is simply intractable at this point. An
alternative approach is to build a microscopic model that accommodates
all the symmetries of the problem while describing the interaction in
terms of effective degrees of freedom. This is in fact what is done by
many researchers in this field using effective Lagrangian field
theories.

In an effective Lagrangian field theory, one postulates a Lagrangian
that encompasses physically reasonable degrees of freedom. Then from
this Lagrangian one finds the field equations of the system. Since
solving these equations is still obstinate, one resorts to perturbation
theory to determine the dynamics of the system. This involves the
generation of Feynman diagrams describing the process. By calculating
these diagrams we can determine the observables. From
experimental data for these observables, one fits the unknown
parameters in the effective theory. In the following three
subsections, I will present briefly three effective Lagrangian
theories for the photoproduction of $\eta$, $\pi$, and $K$ mesons
respectively from a free nucleon.

\subsection{Eta Photoproduction from a Free Nucleon: $N( \gamma , \eta )N$}

This process is assumed to proceed in the s- and u-channels through
the exchange of nucleons (Born terms) and nucleon excited states
(resonances) like the $S_{11}(1535)$ and $D_{13}(1520)$ resonances~\cite{pisabe97,Peters98b,benm92,bmz95}. In
the t-channel we have vector-meson exchanges like the $\omega$ and
$\rho$ mesons. Figure \ref{fig:EtaElementaryProcess} lists the Feynman
diagrams for this process.
\BFIG
\centerline{\psfig{figure=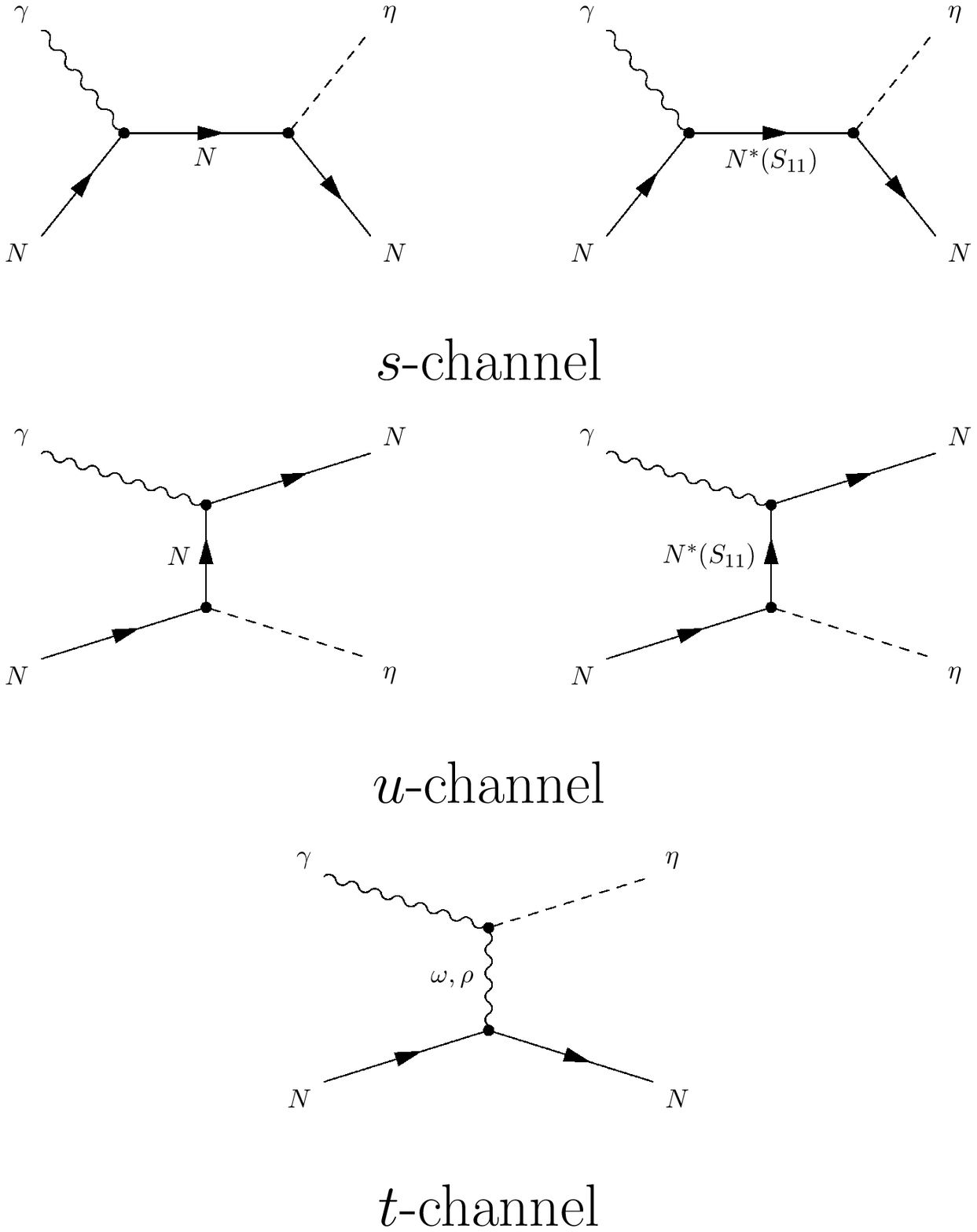,height=6.5in,width=5.5in}}
\caption[Feynman diagrams for the process $N( \gamma , \eta )N$]
{The $\eta$ photoproduction process from a free nucleon: $N( \gamma ,
\eta )N$. The process 
proceeds through the exchange of nucleons and resonances in the s- and
u-channels, as well as vector-meson exchanges in the t-channel.}
\label{fig:EtaElementaryProcess}
\EFIG
Of all of these diagrams, it turns out that the process is
strongly dominated by only one of them: the s-channel resonance
diagram in terms of the $S_{11}(1535)$ resonance. This contribution
overshadows all other Feynman diagrams. Figure
\ref{fig:EtaElementaryProcessData} illustrates this dominance where
measurements are shown of the differential cross section as a function
of incident-photon energy and at different scattering angles for this process from a proton or a
neutron~\cite{pisabe97}.
\BFIG
\centerline{\psfig{figure=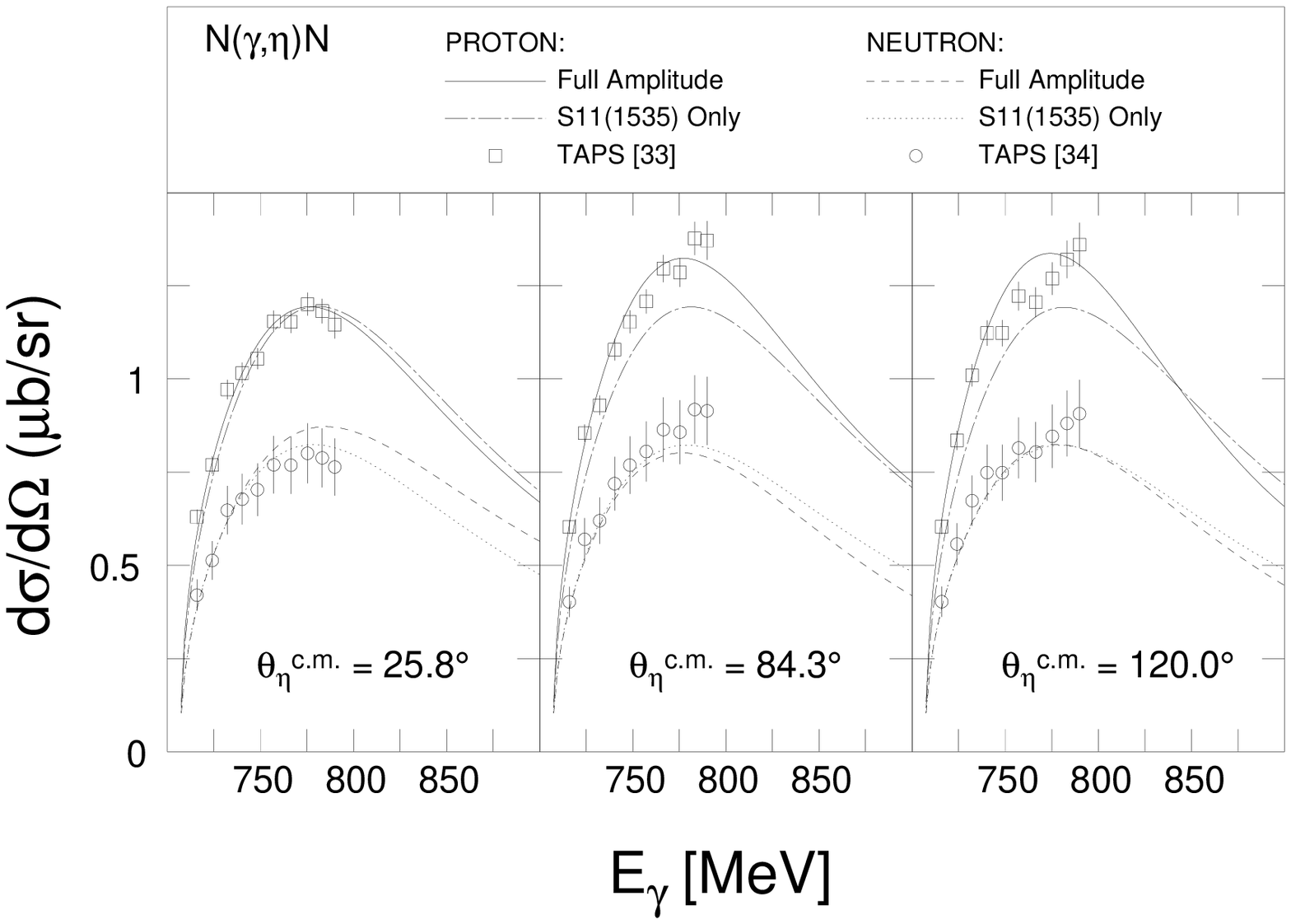,height=5.0in,width=8.0in}}
\caption[Differential cross section and the $S_{11}(1535)$ dominance for
the processes $p( \gamma , \eta )p$ and $n( \gamma , \eta )n$]
{The differential
cross section as a function of incident-photon energy at different
scattering angles for the processes $p( \gamma ,
\eta )p$ and $n( \gamma , \eta )n$~\cite{pisabe97}. The figure includes the theoretical
calculations for this process with all Feynman diagram contributions
included (Full Amplitude), and with only the $S_{11}(1535)$ resonance
contribution ($S_{11}(1535)$ only). It also includes experimental
measurements for these processes from Ref.'s ~\cite{krus95a} (proton)
and ~\cite{krus95b} (extracted neutron).}
\label{fig:EtaElementaryProcessData}
\end{figure}
The figure also includes the theoretical calculations for the
differential cross section
with all Feynman diagram contributions included (Full Amplitude), and
with only the $S_{11}(1535)$ resonance contribution. It is clear that
the $S_{11}(1535)$ alone can almost explain the total magnitude
of the cross section.

\subsection{Pion Photoproduction from a Free Nucleon: $N( \gamma , \pi )N$}

In a similar fashion to the $\eta$ elementary process, one can develop
an effective field theory for the pion photoproduction from a free
nucleon. Then, we find that the $\eta$ and $\pi$ processes have
similar Feynman diagrams, but in the case of the pion it is the
$\Delta$ resonance that dominates this interaction~\cite{ndu91,Peters98a}. Figure
\ref{fig:PionElementaryProcess} lists the different Feynman diagrams for this process.
\BFIG
\centerline{\psfig{figure=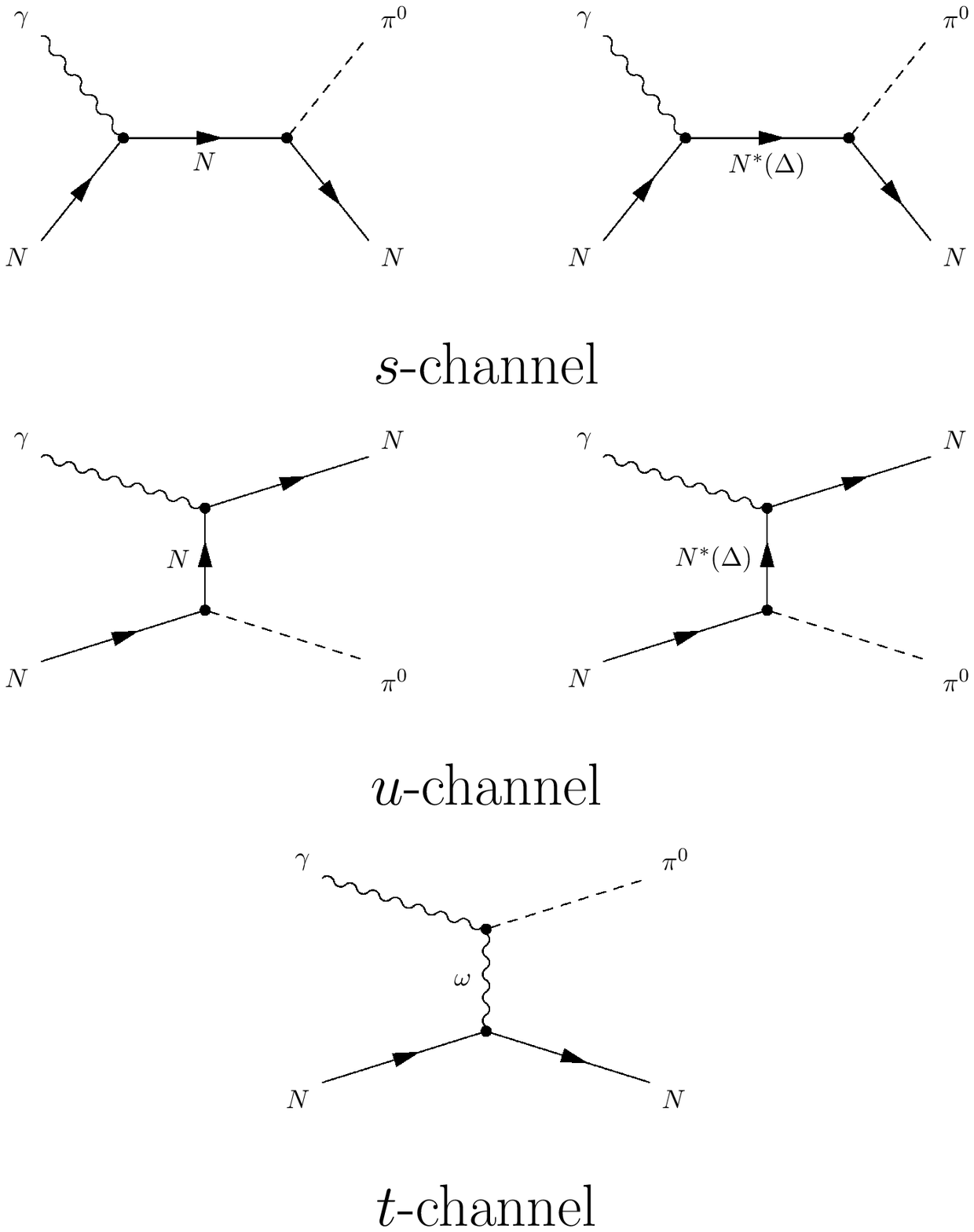,height=6.5in,width=5.5in}}
\caption[Feynman diagrams for the process $N( \gamma , \pi )N$]
{The $\pi$ photoproduction process from a free nucleon: $N( \gamma ,
\pi )N$. The process 
proceeds through the exchange of nucleons and resonances in the s- and
u-channels, as well as vector-meson exchanges in the t-channel.}
\label{fig:PionElementaryProcess}
\EFIG

Finally, we chose
to extract the the amplitudes $A_1, A_2, A_3,$ and $A_4$ from experimental data using the most recent
phase-shift analysis of Arndt, Strakovsky, and Workman~\cite{Arnphn}.

\subsection{Kaon Photoproduction from a Free Nucleon: $p( \gamma ,
K^+ )\Lambda$}
						
The microscopic model for the $K^+$ elementary process is somewhat
different from the one for the $\eta$ (or $\pi$) meson. The reason is
that we have here a strangeness production in the final state: a
$\Lambda$ hyperon
(strange nucleon) and a $K^+$ (strange meson) have been formed. These two particles have a net strange-quark content and
thus labeled as strange particles. As a result of this strangeness
production, the u and t-channels have to proceed now through strange
particles. Thus we have a u-channel proceeding through the exchange of
hyperons like the $\Lambda$ or $\Sigma$, as well as through resonances
of these hyperons ($Y^*$), while the t-channel proceeds through
the exchange of strange scalar mesons~\cite{wcc90,dfls96,mfls98}. Figure
\ref{fig:KaonElementaryProcess} lists the different Feynman diagrams for this process.
\BFIG
\centerline{\psfig{figure=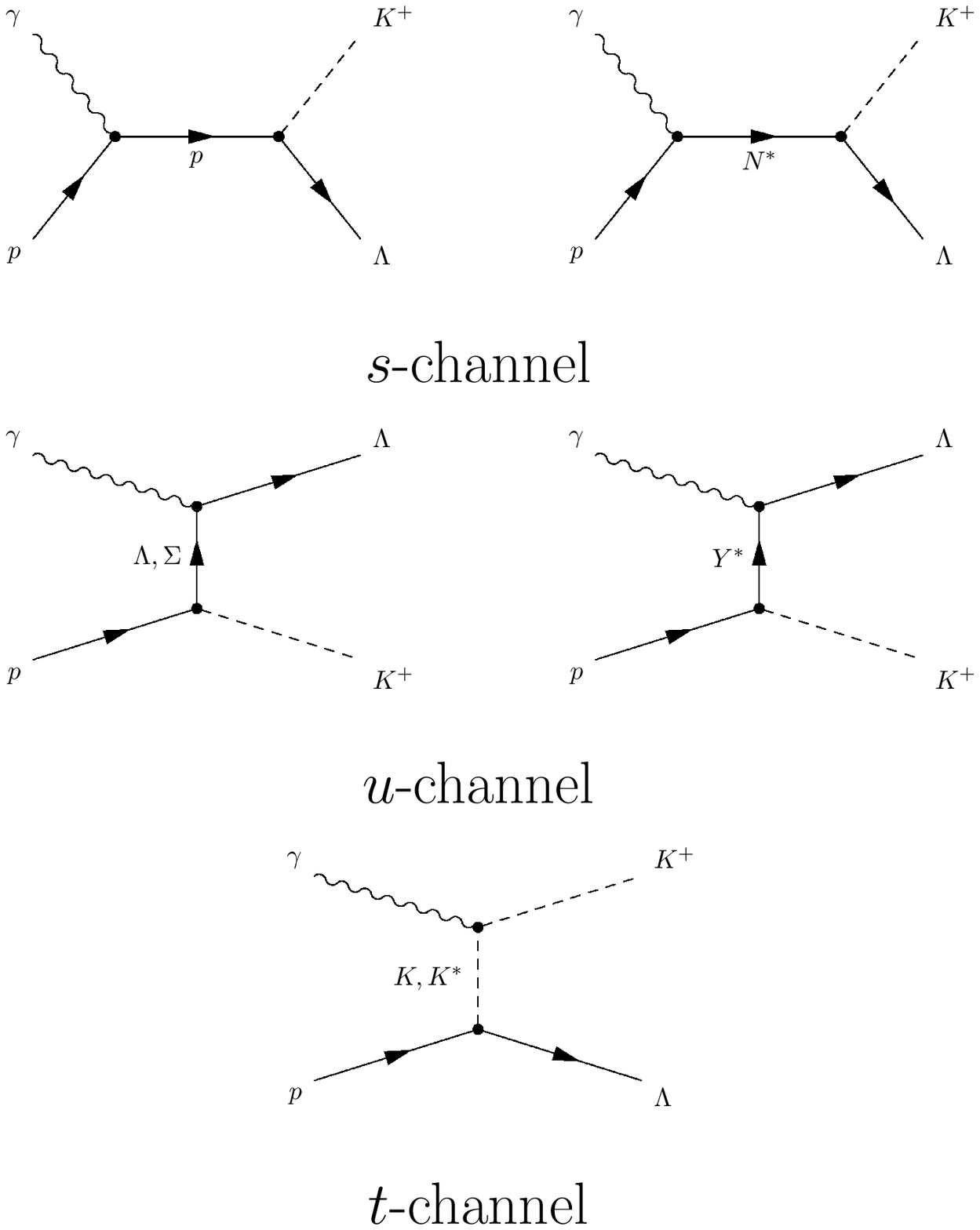,height=6.5in,width=5.5in}}
\caption[Feynman diagrams for the process $p( \gamma , K^+ )\Lambda$]
{The $K^+$ photoproduction process from a free nucleon: $p( \gamma ,
K^+ )\Lambda$. The process proceeds through the exchange of nucleons
and resonances in the s-channel, through the exchange of hyperons like
$\Lambda$ and $\Sigma$ and their resonances $Y^*$ in the u-channel, as
well as strange-meson exchanges in the t-channel.}
\label{fig:KaonElementaryProcess}
\EFIG

\chapter{Relativistic Nuclear Structure}
\label{ch:NuclearStructure}

Relativistic nuclear structure formalisms represent a growing field of
study where the nuclear structure is determined using fully
relativistic models. It has been argued for a long time that due to
the relatively small binding of the nucleons in nuclei, nonrelativistic
formalisms should be adequate to describe the nuclear structure. This
assertion is impressively challenged in the relativistic treatments,
where it has been suggested that the small binding energy is a result
of a cancellation between two large potentials with different Lorentz
transformation properties, with one of the potentials being attractive
and the other repulsive.

Not only do the relativistic formalisms point to the importance of
relativistic effects, but they also provide us
with a more credible and aesthetic theory. This is because the
relativistic formalism is an effective field theory as opposed to the
``ad hoc'' potential-based nonrelativistic formalisms. Thus the theory is
physical and consistent with quantum-field-theory
principles. Furthermore, aspects of the nuclear force that have always
been put in the nonrelativistic formalisms by hand and with no
basis, appear naturally in the relativistic
formalisms. Examples of these include spin-orbit coupling and
three-body forces.

Relativistic treatments have enjoyed a great success in recent years in
their description of the nuclear structure. They do have a number of
pitfalls that are systematically being surmounted and resolved. The
bottom line, however, lies in the experimental verification of these
formalisms. To this end, there are various experimental approaches 
that may decisively prove the validity and applicability of these
formalisms.

\section{Quantum Hadrodynamics}

Quantum hadrodynamics (QHD) is a model for the study of the
relativistic nuclear many-body problem through an effective Lagrangian
field theory. The model was introduced by J. D. Walecka in
1974~\cite{serwal86}. It describes nuclear matter as resulting from
interactions between nucleons (baryons) in the nucleus through the
exchange of neutral scalar $\sigma$ and vector $\omega$
mesons. The couplings of these mesons to the baryon fields is achieved
by the minimal substitution as can be seen in Table \ref{QHD-I
parameters}. In this table, $g_s$ is the scalar coupling constant and $
g _v $ is the vector coupling constant. The model suggests a
nucleon-nucleon force which is attractive at large separations and
repulsive at short ones. Other mesons can be included in this
formalism but their contributions are rather small --- at least in the
mean-field picture which we adopt here. For example, the contribution of the pion
vanishes in the mean-field approximation as a result of its negative
parity. 
\begin{table}
\caption{The fields in quantum hadrodynamics.}
\vspace{.5cm}
\begin{center}
\thicklines
\begin {tabular} { clccc }
\hline
\hline
\thinlines
Field & Description & Particle & Mass & Coupling \\
\hline
\thicklines
$ \psi $ & Baryon & p, n,... & $M$ & \\ $ \phi $ & Neutral scalar
meson & $ \sigma $ & $ m _s $ & $ g _s \bar \psi \phi \psi $\\ $ V
_\mu $ & Neutral vector meson & $ \omega $ & $ m _v $ & $ g _v \bar
\psi \gamma ^ \mu V_\mu \psi $ \\
\hline
\hline
\end{tabular}
\end{center}
\label{QHD-I parameters}
\end{table}
The Lagrangian for this system is as following:
\begin{equation}
{\cal L} = {\bar \psi} [ \gamma _ \mu ( i \partial ^ \mu - g_v V^ \mu)
- (M - g _s \phi ) ] \psi + \frac {1} {2} (\partial _ \mu \phi
\partial ^ \mu \phi - m _s ^2 \phi ^2) - \frac {1}{4} F_{\mu \nu}
F^{\mu \nu} + \frac {1} {2} m_v ^2 V_ \mu V^ \mu\,,
\end{equation}
where
\begin{equation}
F_{\mu \nu} \equiv \partial_\mu V_\nu - \partial _\nu V_\mu\,.
\end{equation}

The field equations can be derived then from the Lagrangian and one
obtains
\begin{equation}
(\partial_\mu \partial^\mu + {m_s}^2)\phi = g_s {\bar\psi} \psi\,,
\label{scalarequation}
\end{equation}
\begin{equation}
(\partial_\mu \partial^\mu + {m_v}^2) V^\nu= g_v {\bar\psi} \gamma^\nu
\psi\,,
\label{vectorequation}
\end{equation}
\begin{equation}
[\gamma _\mu(i\partial ^\mu - g_v V^\mu) - ( M - g _s \phi)] \psi =
0\,.
\label{nucleonequation}
\end{equation}
Hence we have a system of three coupled nonlinear differential
equations. Since solving these equations exactly is a formidable task,
one resorts to approximations like the mean-field picture known also
as the Hartree approximation. In this picture, the scalar and vector
fields are treated as classical fields, and one solves this system by
finding the configuration of these fields that solves all three
equations simultaneously. That is one finds a self-consistent solution
for this system. As a result, the nucleon equation
\ref{nucleonequation} becomes a one-body Dirac equation with a scalar
$g_s \phi$ and a vector $ g_v V^\mu$ potentials. One also finds that
the spatial components of the vector field have a vanishing
contribution in the static limit as a result of current
conservation. This is because we are restricting our discussion to
spherically symmetric nuclei with a total angular momentum of
zero. The mean-field equation then reads as
\begin{equation}
[i \gamma ^ \mu \partial _ \mu - g_v \gamma ^ 0 V({\mathbf x} ) - ( M - g
_s \phi ( {\mathbf x}) ) ] \psi = 0\,.
\end{equation}
The theory has three free parameters to be determined: $\{g_s,
g_v,m_s\}$; the $m_v$ is chosen as the physical mass of $\omega$ meson
since this neutral vector meson is the natural degree of freedom in
this effective field theory. These are resolved using basic properties of
finite nuclei and infinite matter like the saturation density and the rms
charge radius of $^{40}$Ca.

In using the QHD model (known also as Walecka model) one finds that it
can successfully explain and predict many physical features of nuclei
with impressively a minimal number of phenomenological parameters that
are determined from only bulk properties of nuclei. The relativistic
structure is a keystone of this model. There are many consequences of
this relativistic treatment~\cite{serwal86}. One of them is the
existence of a nuclear shell model with the experimentally observed
level orderings, spacings, and major shell closures in nuclei.

Another consequence is the saturation of nuclear matter. This
saturation explains the stability of only a limited number of nuclei
which is what is observed in nature. The relatively small nuclear
binding energy of saturation is the result of a very delicate and fine
cancellation between a large scalar attraction and a large vector
repulsion.

A third consequence of the relativistic structure is the spin-orbit
splitting. The splitting here appears naturally and within the
structure of the theory unlike the nonrelativistic treatments. In
fact, we find a large spin-orbit splitting in this model as is
experimentally verified.

A final consequence of this model is the density dependence of the
interaction as the vector and scalar potentials have different density
dependences. This difference is the reason for the nuclear saturation
in this model. In the nonrelativistic treatments the density
dependence must be included phenomenologically.

The natural remarkable consequences of the QHD Hartree model testify
to its physical validity and to the ``smartness'' of the Dirac equation
which, within its simple but illusive structure, can produce many
physical effects that are never dreamed of in the nonrelativistic
treatments.

\section{Extensions to Quantum Hadrodynamics}

Since Walecka introduced the QHD model many extensions have been
achieved to improve its predictions. As a result, the original QHD
model presented above is now labeled as QHD-I. One extension is the
QHD-II introduced by Serot~\cite{serwal86} that incorporates charged
vector ${\rho}$ and charged pseudoscalar ${\pi}$ mesons in
addition to the $\sigma$ and $\omega$ mesons. The model also
incorporates the electromagnetic interaction through the photon field
$A^\mu$ to account for the Coulomb repulsion between protons in
nuclei. Table \ref{QHD-II parameters} lists the ingredients of this
model.
\begin{table}
\caption{The fields in QHD-II.}
\vspace{.5cm}
\begin{center}
\thicklines
\begin {tabular} { clccc }
\hline
\hline
\thinlines
Field & Description & Particle & Mass & Coupling \\
\hline
\thicklines
$ \psi $ & Baryon & p, n,... & $M$ & \\

$ \phi $ & Neutral scalar meson & $ \sigma $ & $ m _s $ & $ g _s \bar
\psi \phi \psi $\\ 

$ V_\mu $ & Neutral vector meson & $ \omega $ & $ m _v $ & $ g _v \bar
\psi \gamma ^ \mu V_\mu \psi $ \\

\boldmath${\pi}$ \unboldmath& Charged pseudoscalar meson & $ \pi $ & $ m _\pi
$ & $ ig _\pi \bar \psi \gamma ^5
\mbox{\boldmath$\tau\cdot{\pi}$\unboldmath} \psi $ \\

$ {\mathbf b_\mu} $ & Charged vector meson & $ \rho $ & $ m _\rho $ &
$ \frac {1}{2} g_\rho \bar \psi \gamma^\mu$
\boldmath$\tau\cdot{b_\mu}$\unboldmath $\psi$ \\

$ {A_\mu} $ & Photon & $ \gamma $ & $ m_\gamma = 0$ & $ e
\bar \psi \gamma^\mu \frac {1}{2}(1 + \tau_3) A_\mu \psi $ \\
\hline
\hline
\end{tabular}
\end{center}
\label{QHD-II parameters}
\end{table}
Other extensions that incorporate nonlinear terms for the meson fields
have also been introduced. QHD theory with these extensions provide us
today with a very successful model for describing nuclear matter in an
impressively transparent formalism. For the sake of brevity I will not
elaborate on these extensions but it is appropriate to mention that I
use only two versions of the QHD theory throughout this work namely
QHD-I and QHD-II.

Finally, we have used a standard set
of parameters for the Walecka model: $g_{ s}^{2}=109.63$,
$g_{ v}^{2}=190.43$, $g_{\rho}^{2}=65.23$, $m_{ s}=520$~MeV,
$m_{ v}=783$~MeV, and $m_{\rho}=770$~MeV.

\section{Mean Field Approximation to ${}^4$He}

In our study of meson photoproduction processes we have used ${}^4$He
as a nuclear target. In doing so, we needed to have a reasonable
description of the nuclear structure of ${}^4$He. We determined this
structure using a mean-field approximation to the Walecka model.  Even
though the use of this approximation to describe a nucleus as small as
${}^{4}$He should be suspect, we feel justified in adopting this
choice. The reason is that the photoproduction processes we studied
are sensitive only to the bulk properties of ${}^4$He --- which can be
constrained by experiment. Consequently, in order to reproduce the
experimental charge density of ${}^{4}$He, we have modified the mass
of the $\sigma$ meson to $m_{ s}=564$~MeV --- while keeping
constant the ratio of $g_{ s}^{2}/m_{ s}^{2}$. Figure
\ref{fig:HeFormFact} shows the ${}^4$He form factor (the Fourier
transform of the proton density normalized to one)
\BFIG
\centerline{\psfig{figure=HeFormFact.pstex,height=6.0in,width=6.0in}}
\caption[${}^4$He form factor using mean-field approximation to
Walecka model] {The ${}^4$He form factor as
calculated using the original parameters of Walecka model (QHD-II),
and then using the modified ones to fit the experimental form factor.}
\label{fig:HeFormFact}
\end{figure}
as a function of momentum transfer ($q \equiv |{\bf q}|$) as calculated using the original parameters of Walecka model (QHD-II),
and then using the modified ones, to fit the experimental form factor
(included also in the figure). It is remarkable that by a small change in
only one of the parameters, we can fit the experimental form factor
almost perfectly. To be noted here that the calculation using QHD-I
gives also identical results to the QHD-II ones.

The experimental form factor (in the rest frame of the nucleus) in Figure \ref{fig:HeFormFact} is produced
using a phenomenological fit to data over a
wide range of momentum transfers and is parameterized according to the
following equation~\cite{cek87,bentan90}:
\begin{equation}
F(q) = - \frac{3 \pi b \left[ cos(qc) - \pi b\,sin(qc) \, coth(\pi b q)/c
\right]} {q c^2 sinh(\pi b q) [ 1 + \pi^2 b^2 / c^2] }\;.
\end{equation}
Here the parameters $b$ and $c$ are given in Table
\ref{Tab:formfactorpar} for the three nuclei: ${}^4$He, ${}^{12}$C and
${}^{40}$Ca.
\begin{table}
\caption{Parameters of the nuclear form factor for ${}^4$He, ${}^{12}$C, and
${}^{40}$Ca.}
\vspace{.5cm}
\begin{center}
\thicklines
\begin {tabular} { cccc }
\hline
\hline
\thinlines
 & $b$ (fm) & $c$ (fm) \\
\hline
\thicklines
${}^4$He & 0.406 & 1.231 \\ 
${}^{12}$C & 0.478 & 2.220 \\ 
${}^{40}$Ca & 0.537 & 3.573 \\
\hline
\hline
\end{tabular}
\end{center}
\label{Tab:formfactorpar}
\end{table}

\section{Bound Nucleon Wavefunction}

As evident in the previous sections, the QHD theory reduces to finding
a solution to the nucleon Dirac equation
\ref{nucleonequation} with scalar and vector potentials in such away
that this solution is also self-consistent with the field equations
\ref{scalarequation} and
\ref{vectorequation}. It is proper here to give a brief idea of
the solutions to the Dirac equation with scalar and vector fields.

Concentrating on spherically symmetric nuclei one finds that the
fields $\phi$ and $V^0$ must be spherically symmetric too. Hence, we
can rewrite equation \ref{nucleonequation} for a certain energy
eigenvalue $E$ with the new definitions of $S(r) = g_s \phi (r)$ and
$V(r) = g_v V^0 (r)$ as
\begin{equation}
H \psi = E \psi\,,
\end{equation}
where
\begin{equation}
H = \mbox{\boldmath$\alpha \cdot p$\unboldmath} + \beta \left[\,M - S(r)\,\right] +
V(r)\,.
\end{equation}
In this equation
\begin{equation}
\alpha _i \equiv \gamma _0 \gamma _ i\,,
\end{equation}
and
\begin{equation}
\beta \equiv \gamma _0\,.
\end{equation} 

We can find a set of commuting operators that also commute with the
Hamiltonian ($H$) of this equation. Consequently, these operators
provide us with constants of the motion that can be used to
characterize the energy eigenfunctions. Examples of these operators
include ${\mathbf J^2}$ (total angular momentum squared), $J_z$ (z-axis
projection of the total angular momentum), and $ K$ ~\cite{sak73} an
operator that is defined as
\begin{equation}
K \equiv \beta (2 {{\mathbf S} \cdot {\mathbf J} } - \frac {1} {2} ) =
\beta ({2 {\mathbf S} \cdot {\mathbf L}}+ 1)\,.
\end{equation}
The operator $K$ determines in the nonrelativistic limit whether the
projection of the spin is parallel or anti-parallel to the total
angular momentum. The eigenvalues for these operators are $ j(j+1)$
for $ {\mathbf J^2} $, $m$ for $J_z$, and $ -\kappa$ for $ K$.

It can be shown that there is a relationship between $ \kappa $ and
$j$ which is
\begin{equation}
\kappa = \pm (j + \frac {1} {2} )\,.
\end{equation}
Hence $ \kappa $ is a nonzero integer which can be positive or
negative. The sign of $ \kappa $ determines whether the spin is
parallel (positive) or anti-parallel (negative) in the nonrelativistic
limit.

We can write the four-component eigenfunction $ \psi $ as a vector of
two-component spinors
\begin{equation}
\psi = {\psi _A \choose \psi _B}\,.
\end{equation}
By this decomposition one can show that even though the four component
eigenfunction is not an eigenfunction of $ {\mathbf L^2} $, the
spinors $
\psi _A$ and $ \psi _B$ are separately eigenfunctions with eigenvalues
$ l_A(l_A + 1)$, and $ l_B(l_B + 1)$ respectively. It can be shown
also that these eigenvalues are related to $ \kappa $ and $j$ through
the equations
\begin{equation}
- \kappa = j(j+1) - l_A(l_A+1) + \frac {1} {4}\,,
\end{equation}
and
\begin{equation}
\kappa = j(j+1) - l_B(l_B+1) + \frac {1} {4}\,.
\end{equation}
As a result any energy eigenfunction can be uniquely characterized by
only $E$, $ \kappa $, and $m$.

	The above analysis enables us to write $\psi$ as
\begin{equation}
\psi = {\psi _A \choose \psi _B} =  {g_{E\kappa}(r) {y } {^{m} _{j l_A}}  \choose i f_{E\kappa}(r) {y } {^{m} _{j l_B}}}\,,
\end{equation}
where ${y } {^{m} _{j l}} $ are the normalized spin-angular functions
constructed by the addition of Pauli spinors to the spherical
harmonics of order $l$. The inclusion of $i$ with $f_{E\kappa}(r)$
is in order to make $f_{E\kappa}(r)$ and $g_{E\kappa}(r)$ real for
bound-state solutions. Substituting this result back into the Dirac
equation and performing some algebra, we arrive at the coupled
equations:
\begin{equation}
-\frac {df} {dr} - \frac {(1-\kappa) } {r} f(r) = \left[ E - V(r) - M +
S(r)\,\right] g(r)\,,
\end{equation}
\begin{equation}
\frac {dg} {dr} + \frac {(1+\kappa) } {r} g(r) = \left[ E - V(r) + M -
S(r)\,\right] f(r)\,.
\end{equation}
Now writing these equations in terms of
\begin{equation}
F(r) = r f(r)\,,
\end{equation}
\begin{equation}
G(r) = r g(r)\,,
\end{equation}
we have
\begin{equation}
 \frac {dF} {dr} - \frac {\kappa} {r} F(r) = - \left[ E - V(r) - M + S(r)\,\right] G(r)\,,
\end{equation}
\begin{equation}
 \frac {dG} {dr} + \frac {\kappa} {r} G(r) = \left[ E - V(r) + M -
 S(r)\, \right] F(r)\,.
\end{equation}
These are two coupled differential equations which can be solved
numerically by the Runge-Kutta method. It is worth noting that there
is an implicit symmetry in these equations as $ F
\rightarrow G$ and $ G \rightarrow F$ if $ E \rightarrow -E$, $ V(r)
\rightarrow -V(r)$, and $ \kappa \rightarrow -\kappa $.

\section{Nuclear Densities in the Relativistic Formalism}

Nuclear densities in the relativistic formalisms are a vivid example
of the richness of relativity. While we have essentially only one
ground state density in the nonrelativistic formalisms: the vector
(matter) density, the relativistic treatments provide us with the
possibility of having up to five different densities: vector (matter),
tensor, scalar, axial-vector, and pseudoscalar. This richness is a result of the fact that in the space of Dirac spinors
we can have up to 16 linearly-independent matrices. These
form the set: $\{1, \gamma ^\mu, \gamma ^\mu \gamma ^5, i \gamma ^5,
\sigma ^{\mu \nu} \}$ of bilinear covariants. The covariants transform
as 
scalar, vector, axial-vector, pseudoscalar, and tensor respectively
under Lorentz transformations (Poincar\'e group). It is important to
note that the densities are truly independent and constitute
fundamental nuclear-structure quantities. The fact that in the
nonrelativistic framework only one density survives is due to the
limitation of the approach. Indeed, in the nonrelativistic framework
one employs the free space relation to relate the lower to the upper
component of the Dirac spinor instead of determining the lower component
dynamically through the Dirac equation. Hence, any evidence of possible
medium modifications to the ratio of lower-to-upper components of the
Dirac spinors is lost.

Using the QHD theory developed above one finds that there are three
non-vanishing ground state densities for spherical and spin-saturated
nuclei. These are the conventional matter (vector) density defined by
\begin{equation}
  \rho_{\lower 3pt \hbox{$\scriptstyle V$}}(r) = \sum_{\alpha}^{\rm
  occ} \overline{{\cal U}}_{\alpha}({\bf x})\,
  \gamma^{{\scriptscriptstyle 0}} \, {\cal U}_{\alpha}({\bf x}) \,,
  \quad
\end{equation}
which leads to the vector density given by
\begin{equation}
   \rho_{\lower 3pt \hbox{$\scriptstyle V$}}(r) = \sum_{a}^{\rm occ}
   \left({2j_{a}+1 \over 4\pi r^{2}}\right) \Big(g_{a}^2(r) +
   f_{a}^2(r)\Big) \,,
\label{rhov}
\end{equation}
where ${\cal U}_{\alpha}({\bf x})$ is a single-particle Dirac spinor
(solution to Dirac equation) for the bound nucleon, $g_{a}(r)$ and
$f_{a}(r)$ are the radial parts of the upper and lower components of
the Dirac spinor, respectively, and the above sums run over all the
occupied single-particle states in the nucleus. Analogously, the
scalar density is defined by
\begin{equation}
  \rho_{\lower 3pt \hbox{$\scriptstyle S$}}(r) = \sum_{\alpha}^{\rm
  occ} \overline{{\cal U}}_{\alpha}({\bf x})\, {\cal U}_{\alpha}({\bf
  x}) \,, \quad
\end{equation}
leading to it given as
\begin{equation}
   \rho_{\lower 3pt \hbox{$\scriptstyle S$}}(r) = \sum_{a}^{\rm occ}
   \left({2j_{a}+1 \over 4\pi r^{2}}\right) \Big(g_{a}^2(r) -
   f_{a}^2(r)\Big) \,.
\label{rhos}
\end{equation}
Finally, we have the tensor density defined by
\begin{equation}
  \Big[ \rho_{\lower 3pt \hbox{$\scriptstyle T$}}(r)\,\hat{r}
    \Big]^{i} = \sum_{\alpha}^{\rm occ} \overline{{\cal
    U}}_{\alpha}({\bf x})\, \sigma^{{\scriptscriptstyle 0}i} \, {\cal
    U}_{\alpha}({\bf x}) \,, \quad
\end{equation}
resulting in the following expression for $\rho_{\lower 3pt
\hbox{$\scriptstyle T$}}(r)$
\begin{equation}
   \rho_{\lower 3pt \hbox{$\scriptstyle T$}}(r) = \sum_{a}^{\rm occ}
   \left({2j_{a}+1 \over 4\pi r^{2}}\right) 2g_{a}(r)f_{a}(r) \,.
   \label{rhotr}
\end{equation}
The axial-vector density as well as the pseudoscalar density can be
defined in analogous fashion to these three non-vanishing densities. In Chapter \ref{ch:CoherentTheory} the consequences of having
three fundamentally different and non-vanishing densities and their role in the
photoproduction process will be clarified.

\section{An Example of a Relativistic Nuclear Structure Calculation: ${}^{40}$Ca}

In this section, I will discuss a specific example of a nuclear
structure calculation in order to present a manifestation of using
this formalism. Figure \ref{fig:protonElevelsCa40} illustrates a
comparison between our calculations and the experimental data for the
proton level diagram of ${}^{40}$Ca.
\BFIG
\centerline{\psfig{figure=Ca40EneLevel.pstex,height=6.0in,width=4.50in}}
\caption[Energy level diagram for ${}^{40}$Ca]
{A comparison between our calculations (QHD-II) and the experimental data
for the proton level diagram of ${}^{40}$Ca. The experimental measurements are obtained from ($p, 2p$)~\cite{ray79} and ($e,
e^{\prime} p$)~\cite{naket76,mouet76} experiments.}
\label{fig:protonElevelsCa40}
\EFIG
The experimental measurements are obtained from ($p, 2p$)~\cite{ray79} and ($e,
e^{\prime} p$)~\cite{naket76,mouet76} experiments. The theoretical calculations for this figure are done using the QHD-II
model. As can be seen, this model predicts properly the shell
structure of ${}^{40}$Ca with accurate level ordering and spacing
as well as the proper magnitude of the spin-orbit splitting.

Figure \ref{fig:protonQHD-IvsQHD-II} shows the proton spectrum as
calculated using QHD-I and QHD-II models.
\BFIG
\centerline{\psfig{figure=pQHD-IvsQHD-II.pstex,height=6.0in,width=4.5in}}
\caption[Comparison between QHD-I and QHD-II results for the
proton spectrum in ${}^{40}$Ca] {A comparison between QHD-I and QHD-II
results for the proton spectrum in ${}^{40}$Ca. The QHD-II results are
shifted positively in energy due to the Coulomb repulsion.}
\label{fig:protonQHD-IvsQHD-II}
\EFIG
It is evident that apart from an overall positive shift of the
energies in the QHD-II model calculations, the two level diagrams are
essentially identical. This shift is a realization of including the
Coulomb repulsion in the QHD-II model.  Figure
\ref{fig:neutronQHD-IvsQHD-II} displays the same comparison but this
time for the neutron spectrum.
\BFIG
\centerline{\psfig{figure=nQHD-IvsQHD-II.pstex,height=6.0in,width=4.50in}}
\caption[Comparison between QHD-I and QHD-II results for the
neutron spectrum in ${}^{40}$Ca] {A comparison between QHD-I and
QHD-II results for the neutron spectrum in ${}^{40}$Ca. Since neutrons
are not affected by the Coulomb repulsion, the two results are
essentially identical.}
\label{fig:neutronQHD-IvsQHD-II}
\EFIG
Since neutrons do not feel the Coulomb repulsion, the spectrum using
QHD-II is identical to that using QHD-I apart from minute differences.
The differences arise from the inclusion of the $\rho$ meson which
couples differently to the protons and neutrons as well as from
indirect nonlinear effects originating from the Coulomb repulsion in
the proton sector.

Figure \ref{fig:Ca40desities} exhibits various nuclear densities for
${}^{40}$Ca determined using the QHD-II model.
\BFIG
\centerline{\psfig{figure=Ca40densities.pstex,height=5.5in,width=6.0in}}
\caption[Various nuclear densities in ${}^{40}$Ca]
{Various nuclear densities in ${}^{40}$Ca: the vector (matter) density
$\rho_v$, the scalar density $\rho_s$, the iso-vector density
$\rho_{iso}$ (the difference between the proton and the neutron vector
densities in the nucleus), and the charge density $\rho_{ch}$.}
\label{fig:Ca40desities}
\EFIG
It includes the vector (matter) density $\rho_v$, the scalar density
$\rho_s$, the iso-vector density $\rho_{iso}$ (the difference between
the proton and the neutron vector densities in the nucleus), and the
charge density $\rho_{ch}$. These densities generate the four
different potentials in the nucleus: the $\sigma$ scalar potential
$g_s \phi(r)$, the $\omega$ vector potential $g_v V(r)$, the $\rho$ vector
potential $\frac{1}{2} g_\rho {b(r)}$, and the photon (electromagnetic) vector
potential $e A(r)$ respectively. These potentials are shown in Figure
\ref{fig:Ca40potentials}.
\BFIG
\centerline{\psfig{figure=Ca40potential.pstex,height=5.5in,width=6.0in}}
\caption[The different potentials in ${}^{40}$Ca]
{The different potentials in ${}^{40}$Ca: the $\sigma$ scalar
potential $g_s\phi(r)$, the $\omega$ vector potential $g_v V(r)$, the
$\rho$ vector potential $\frac{1}{2} g_\rho b(r)$, and the photon (electromagnetic)
vector potential $e A(r)$. The $b(r)$ and $A(r)$ potentials have been
magnified by a factor of ten for better display.}
\label{fig:Ca40potentials}
\EFIG
In this figure, the $b(r)$ and $A(r)$ potentials have been magnified
by a factor of ten for a better display.

\chapter{Theory of the Coherent Pseudoscalar Meson Photoproduction
from Nuclei}
\label{ch:CoherentTheory}

In the four forthcoming chapters of this manuscript including this one, I will develop and discuss the first part of this doctoral study:
the coherent pseudoscalar-meson photoproduction from
nuclei. This process consists of a photon ($\gamma$-ray) incident on a
nucleus. The photon interacts with the nucleus and as a result a
pseudoscalar meson is produced (like $\pi$ or $\eta$ mesons) in
addition to the recoil nucleus. In this way, we start the interaction
with a photon and some nucleus, and end up with a meson and the same nucleus we started with. The
process is labeled as ``coherent'' because all nucleons participate in the
process leading to a coherent sum of these individual nucleon
contributions.

\section{Ingredients}

The basic tenet of this theoretical study is the relativistic impulse
approximation. It consists of the assumption that the
process proceeds through the interaction of the incident photon with individual nucleons in the nucleus as opposed to interacting with
the nucleus as a whole. Furthermore, the approximation assumes that
the nature of the interaction between the photon and the
bound nucleon is identical to the nature of the interaction between a
photon and a free nucleon, apart from including the binding aspect of
the nucleon. Figure \ref{fig:CoherentImpulse} sketches
this process within this approximation.
\BFIG
\centerline{\psfig{figure=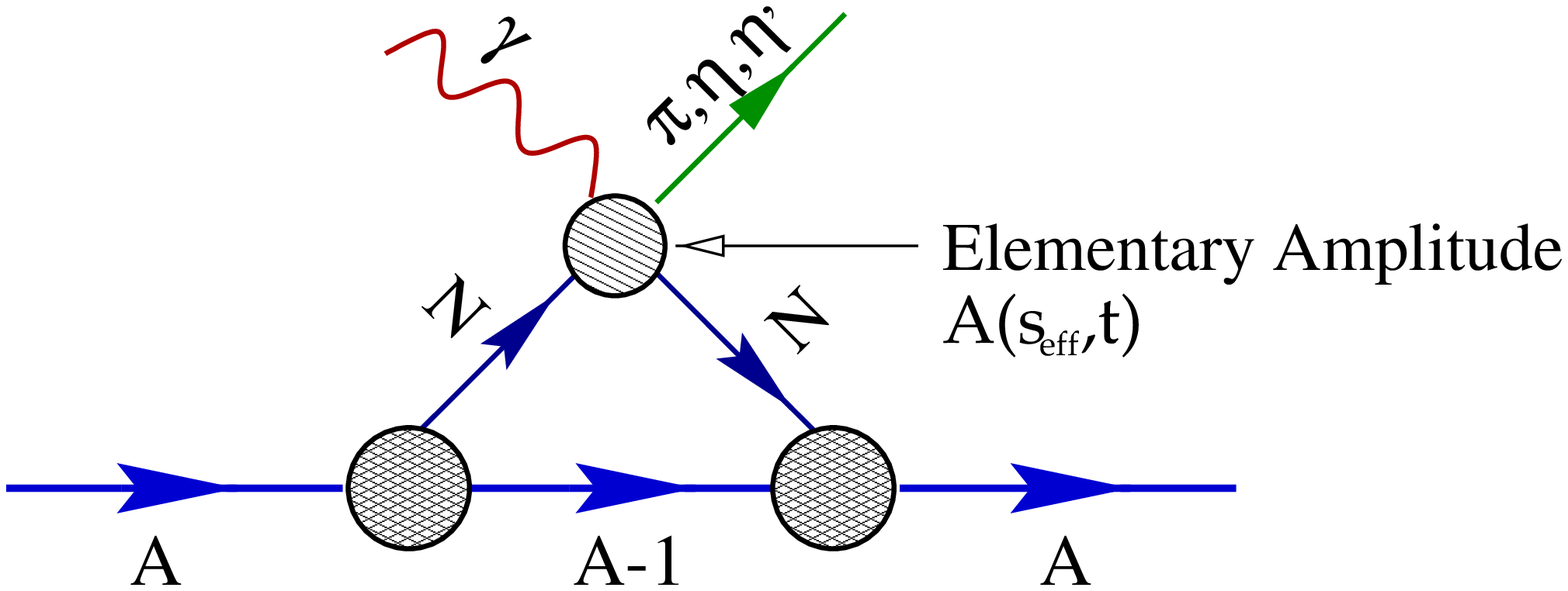,height=9.0cm,width=14.5cm}}
\caption[Schematic diagram for the coherent process $A( \gamma ,
PS\,\,meson )A$]{A schematic diagram
for the coherent process within the framework of the impulse
approximation. The incident photon is assumed to interact with 
individual nucleons in a nucleus ($A$) leading to the production of a
pseudoscalar meson and the same, but now recoil, nucleus.}
\label{fig:CoherentImpulse}
\EFIG

In our formalism we maintain the full relativistic
structure whether in the elementary photoproduction
process or in the nuclear structure. This approach forms a major
departure from the traditional
studies~\cite{bofmir86,cek87,bentan90,ndu91,tryfik94,fix97} of this subject where one resorts to non-relativistic reduction of the
elementary photoproduction amplitude and uses non-relativistic
models for the nuclear structure to simplify the formalism. In this
regard we use the Walecka model for the nuclear structure that we developed in Chapter
\ref{ch:NuclearStructure}.

Since mesons do in principle interact strongly with nucleons and
nuclei, we have to account for the final-state interaction between the
emitted meson and the recoil nucleus. This kind of final-state
interaction is usually labeled as ``distortion'', because instead of
having a plane wave ($e^{-i k'x}$) describing the meson wavefunction, we have a wave
``distorted'' from its plane-wave limit due to the presence of these
interactions. We incorporate distortions
through an optical potential formalism that will be
the subject of the next chapter. To distinguish between two types of
impulse approximation and to follow the conventional terminology in
the literature, we label the relativistic impulse
approximation with no final-state interactions as the relativistic plane-wave
impulse approximation (RPWIA), while we refer to the approximation in the
presence of distortions as the relativistic distorted-wave impulse
approximation (RDWIA).

\section{Differential Cross Section for the Coherent Process}

The expression for the differential cross section has been derived
using well established procedures for the case of two incoming
particles and two outgoing ones~\cite{ManSha}. Thus, we have the
following form for the cross section in the center-of-momentum frame
(c.m.)
\begin{equation}
   \left({d\sigma \over d\Omega}\right)_{\rm c.m.} = \left({M_{\lower
   1pt \hbox{$\scriptstyle T$}} \over 4\pi W}\right)^{2}
   \left({{k}'_{\rm c.m.} \over {k}_{\rm c.m.}}\right) {1 \over
   2}\sum_{\lambda} |{\mathcal T}_{\lower 2pt
   \hbox{$\scriptstyle\lambda$}}|^2 \;, \label{dsigmaa}
\end{equation}
where $M_{\lower 1pt \hbox{$\scriptstyle T$}}$ is the mass of the
target nucleus, $W$ is the total energy in the c.m. frame, while
${k}_{\rm c.m.} \equiv |{\bf k}_{\rm c.m.}|$ and ${k}'_{\rm
c.m.} \equiv |{\bf k}'_{\rm c.m.}|$ are the three-momenta of the
photon and $\eta$-meson in the c.m. frame, respectively. This
expression is independent of the mass of the produced meson and so it
is applicable to the coherent photoproduction of any pseudoscalar
meson. Restricting our formalism to coherent processes from nuclei
with zero angular momentum and zero isospin
($J^{\pi}\!=\!0^{+};T\!=\!0$), the scattering matrix element
$T_{\lower 2pt \hbox{$\scriptstyle\lambda$}}$ is given by
\begin{equation}
 {\mathcal T}_{\lower 2pt \hbox{$\scriptstyle\lambda$}} =
 \epsilon_{\mu}(\hat{\bf k},\lambda) \langle A(p'); \eta(k') | J^{\mu} |
 A(p) \rangle \;, \label{tampl}
\end{equation}
This expression is nothing but the standard contraction in
electrodynamics between the photon polarization
$\epsilon_{\mu}(\hat{\bf k},\lambda)$ and the conserved
electromagnetic current $\langle A(p'); \eta(k') | J^{\mu} | A(p)
\rangle$. Now using basic symmetry considerations that include parity
and Lorentz covariance, we can write a model-independent form for the
current matrix element as
\begin{equation}
 \langle A(p'); \eta(k') | J^{\mu} | A(p) \rangle =
 \varepsilon^{\mu\nu\alpha\beta} k_{\nu}k'_{\alpha}p_{\beta} {1 \over
 W} F_{0}(s,t) \;.  \label{emcurrent}
\end{equation}
Here $p(p'\!=\!p\!+\!k\!-\!k')$ is the four momentum of the
initial(final) nucleus and $\varepsilon^{\mu\nu\alpha\beta}$ is the
relativistic Levi-Civita symbol ($\varepsilon^{0123}\equiv-1$). It is
evident in this expression, and in fact a remarkable result, that the
cross section cannot depend in this process on more than one
Lorentz-invariant form factor $F_{0}(s,t)$, which is a function of the
Mandelstam variables $s=(k+p)^{2}$ and $t=(k-k')^{2}$. All
dynamical information in this process must be contained in this form
factor. Now substituting Equation \ref{emcurrent} in Equation
\ref{dsigmaa} and doing some algebraic manipulations we arrive at the
following expression for the cross section
\begin{equation}
   \left({d\sigma \over d\Omega}\right)_{\rm c.m.} = \left({M_{\lower
   1pt \hbox{$\scriptstyle T$}} \over 4\pi W}\right)^{2}
   \left({k'_{\rm c.m.} \over k_{\rm c.m.}}\right) \left({1 \over
   2}{k_{\rm c.m.}}^{2}{k'_{\rm c.m.}}^{2} \sin^{2}\theta_{\rm
   c.m.}\right)|F_{0}(s,t)|^{2} \;, \label{dsigmab}
\end{equation}
where $\theta_{\rm c.m.}$ is the scattering angle (between ${\bf k}$ and
${\bf k}'$) in the c.m. frame.

\section{Determination of $F_{0}(s,t)$ in a Relativistic Impulse
Approximation Approach}

The most general expression for the scattering matrix element in the
framework of the relativistic plane-wave impulse approximation can be
written as a multiple integral in the following form
\begin{equation}
\sum_\alpha \int d^4x_1 \ldots d^4x_N \; \overline{{\cal U}}_{\alpha} A^\mu
J_\mu (x_1,\ldots,x_N){\cal U}_{\alpha} \phi\;,
\label{impulematele1}
\end{equation}
where ${\cal U}_{\alpha}$ is the single-particle Dirac spinor for the
bound nucleon with a set of quantum numbers $\alpha$, $A^\mu$ is the
photon wavefunction (field), and $\phi$ is the pseudoscalar meson
wavefunction (field). The number $N$ of the
independent variables to be integrated over depends on the nature of
the effective field theory employed. In other words, it depends on the
number of vertices in each Feynman diagram derived from this effective
field theory. The sum $\sum_\alpha$ runs over all occupied states in
the nucleus. It can be shown then that this expression can be reduced
to the following form:
\begin{equation}
\sum_\alpha \int \; \frac{d^3p_1}{(2\pi)^3} \frac{d^3p_2}{(2\pi)^3}
(2\pi)^4 \delta^4 (k + p_1 - k' - p_2)
\overline{{\cal U}}_{\alpha}{(\bf p_1)}  T (p_1,p_2,k,k') {\cal U}_{\alpha}{(\bf p_2)}\;,
\label{impmatele2}
\end{equation}
where ${\cal U}_{\alpha}({\bf p})$ is the Dirac spinor in momentum
space and where $T (p_1,p_2,k,k) \equiv \epsilon_{\mu} J^{\mu}
(p_1,p_2,k,k')$ is the scattering matrix introduced in Equation
\ref{scatmat} of Chapter \ref{ch:ElementaryProcess}. The momenta $p_1,
p_2, k,$ and $k'$ are the four momenta of the struck nucleon, outgoing
nucleon, incident photon, and emitted meson respectively. Note that
here the struck and outgoing nucleons are bound and so they are not in
a specific momentum state but have a momentum distribution.

The evaluation of this integral is involved. A great
simplification ensues if one uses the factorization approximation
(also called optimal approximation) of Gurvitz, Dedonder, and
Amado~\cite{GDA79}. The approximation is standard in this kind of
study and consists of evaluating $T (p_1,p_2,k,k')$ at certain
optimal (effective) value of $p_1$ to enable us to ``factorize'' $T
(p_1,p_2,k,k')$ from the integral in such a way that minimizes any
correction from the Fermi motion of the bound nucleon. More details on
this optimal prescription will be presented in the next section. Thus,
the approximation works best if $T (p_1,p_2,k,k')$ is a slowly varying
function of $p_1$. Using this approximation and replacing the
$\delta$-function by its integral representation $ \frac{1}{(2\pi)^4}
\int d^4x \; e^{i(k + p_1 - k' - p_2)\cdot x}$ one can arrive at the
following form for the scattering matrix element
\begin{equation}
\sum_\alpha \delta (p_1^0 + k^0 - p_2^0 -k'^0) \; \int d^3 x \; e^{i {\bf q \cdot x}} \overline{{\cal U}}_{\alpha}{(\bf x)} T (p_1,p_2,k,k') {\cal U}_{\alpha}{(\bf x)}\;,
\label{impmatele3}
\end{equation}
where ${\bf q} = {\bf k - k'}$ is the momentum transfer. It is evident in
this expression that the combination of the impulse and factorization
approximations is effectively achieved by simply sandwiching the
scattering matrix for the on-shell nucleons between bound-nucleon
spinors instead of free spinors as is the case in the elementary
process.

Now replacing $T$ by its expression in terms of the bilinear
covariants (Chapter \ref{ch:ElementaryProcess}):
\begin{equation}
T[\gamma N \rightarrow PS\, meson \, N (Y)] = F^{\alpha \beta}_{T}
\sigma_{\alpha \beta} + F_P i \gamma_5 + F^{\alpha}_{A} \gamma _\alpha
\gamma_5\;,
\end{equation}
and taking advantage of basic definitions for the nuclear densities in the
relativistic formalism (see Chapter \ref{ch:NuclearStructure}) we
arrive at
\begin{equation}
\delta (p_1^0 + k^0 - p_2^0 -k'^0) \; \int d^3 x \; e^{i {\bf q \cdot
x}} \{F^{\alpha \beta}_{T} {\rho_{\alpha \beta}}_{\lower 3pt
\hbox{$\scriptstyle T$}}({\bf x}) + F_P i {\rho}_{\lower 3pt
\hbox{$\scriptstyle PS$}}({\bf x}) + F^{\alpha}_{A} {\rho_{\alpha}}_{\lower 3pt
\hbox{$\scriptstyle AV$}}({\bf x)}\}\;.
\label{impmatele4}
\end{equation}
Thus the coherent process probes three nuclear densities in the
nucleus: the tensor (T), pseudoscalar (PS), and axial-vector (AV)
densities. However, as has been indicated in Chapter
\ref{ch:NuclearStructure}, the pseudoscalar and axial-vector densities
vanish for ($J^{\pi}\!=\!0^{+};T\!=\!0$) nuclei. In fact even all
components of the tensor density vanish except the three ${\rho_{0 i}}_{\lower 3pt \hbox{$\scriptstyle T$}}$ where $i = 1, 2,
3$. Substituting the expressions for $F^{\alpha \beta}_{T}$ and
${\rho_{\alpha \beta}}_{\lower 3pt
\hbox{$\scriptstyle T$}}({\bf x})$ and carrying out some algebraic
manipulations we arrive at a remarkably simple expression for the the
coherent-process scattering matrix element in the c.m. frame:
\begin{equation}
   \delta (p_1^0 + k^0 - p_2^0 -k'^0) \, i A_{1}(s,t) \frac {\rho_{\lower
   3pt \hbox{$\scriptstyle T$}}(q)}{q} \Big\{ k_{} k'_{} {\bf
   \hat{k'}}\cdot \left[ {\bf \hat{k}} \times \mbox{\boldmath
   $\epsilon$ \unboldmath} (\hat{\bf k},\lambda)\right] \Big\}\;, 
\label{impmatele5}
\end{equation} 
where
\begin{equation}
    \rho_{\lower 3pt \hbox{$\scriptstyle T$}}(q) = 4\pi
     \int_{0}^{\infty} dr \, r^2 j_{\lower 2pt \hbox{$\scriptstyle
     1$}}(qr) \rho_{\lower 3pt \hbox{$\scriptstyle T$}}(r)\;.
     \label{rhotq}
\end{equation}
Taking this form, we can then
find the expression for the differential cross section in the
relativistic plane-wave-impulse-approximation approach. By comparing
this expression to the model-independent one given by Equation
\ref{dsigmab}, we can extract the value of the Lorentz-invariant form factor
$F_{0}(s,t)$ to be
\begin{equation}
  F_{0}^{\scriptscriptstyle PW}(s,t) = i A_{1}(s,t)\frac {\rho_{\lower 3pt
   \hbox{$\scriptstyle T$}}(q)} {q}\;. 
\label{fpwia}
\end{equation} 
It is important to stress here that the analysis I sketched here is
valid only in the plane-wave limit where no distortions for the
emitted meson have been incorporated. Including these distortions
spoils this simple and elegant result. This will be the subject
of the next chapter.

\section{More on the Factorization Approximation: the Optimal
Prescription}

In the previous section I have hinted at the basic idea of the
factorization approximation. What remains is to find the optimal value
for ${\bf p}_1$ which is determined using what is called the ``optimal
prescription''~\cite{GDA79}. Since in Equation \ref{impmatele5} all kinematic
quantities are fixed except for $s$ ($t$ and $q$ are determined from
the measured ${\bf k}$ and ${\bf k'}$), we are trying effectively to
find the optimal value for $s$ (call it $\tilde s$) for the coherent process.

The optimal value of ${\bf p}_1$ is determined by the principle of
``democratic'' sharing of momentum expressed as:
\begin{equation}
\frac {{\bf p_1} + {\bf p_2}}{2} = {\bf P}_{avg}\;,
  \label{demsha}
\end{equation} 
where ${\bf P}_{avg}$ is the average momentum carried by a spectator
nucleon during the collision. Since only one nucleon participates in
the interaction in the impulse-approximation picture, ${\bf P}_{avg}$ is the
average momentum of the other nucleons in the nucleus. Using this condition and the conservation of momentum
during the collision
\begin{equation}
{\bf P_A} + {\bf k} = {\bf P'_A} + {\bf k'}\;,
 \label{conmom}
\end{equation} 
where ${\bf P_A}$ (${\bf P'_A}$) is the momentum of the nucleus before
(after) the collision, as well as the conservation of momentum at the
interaction vertex (see Figure \ref{fig:CoherentImpulse})
\begin{equation}
{\bf p_1} + {\bf k} = {\bf p_2} + {\bf k'}\;, 
\label{vertexconmom}
\end{equation} 
one can show that the effective momentum of the struck nucleon is
given by (in the c.m frame)
\begin{equation}
{\bf p_1} = - \frac{A + 1}{2 A} {\bf k}_{cm} + \frac{A - 1}{2 A} {\bf
  k'}_{cm}\;, 
\label{p1}
\end{equation} 
while the effective momentum of the outgoing nucleon is expressed as
\begin{equation}
{\bf p_2} = \frac{A - 1}{2 A} {\bf k}_{cm} - \frac{A + 1}{2 A} {\bf
  k'}_{cm}\;.  
\label{p2}
\end{equation} 
As a result, it is straightforward to find the
optimal value $\tilde s$ as
\begin{equation}
{\tilde s} = {M^2}_N + 2 k_{cm}( E_1 - p_1 cos\theta_1 )\;,
  \label{stilda}
\end{equation} 
where
\begin{eqnarray}
E_1 & = & ( {M^2}_N + \alpha^2 {k}_{cm}^2 + \alpha'^2 {k'}_{cm}^2 - 2
\alpha \alpha' k_{cm} {k'}_{cm} cos \theta_{cm} )^\frac{1}{2}\;, \\ \nonumber
p_1cos\theta_1 & = & - \alpha {k}_{cm} + \alpha' {k'}_{cm}
cos\theta_{cm}\;, \\ \nonumber
\alpha &\equiv& \frac{A+1}{2A}\;, \\ \nonumber
\alpha' &\equiv& \frac{A-1}{2A}\;,
\end{eqnarray}
where $M_N$ is the mass of the nucleon, and ${p_1}$, ${k}_{cm}$, and
${k}^\prime_{cm}$ are the three-momenta of the bound nucleon, incident photon, and
emitted meson respectively. Moreover, $\theta_1$ is the angle between
${\bf k}$ and ${\bf p_1}$, and $\theta_{cm}$ is the scattering angle
between ${\bf k}$ and ${\bf k'}$.

\section{Off-Shell Ambiguity}
\label{sec:offshellambiguity}

The study of the coherent reaction represents a challenging
theoretical task due to the lack of a detailed microscopic model of
the process. Indeed, most of the models used to date rely on the
impulse approximation assumption that the elementary amplitude remains
unchanged as the process is embedded in the nuclear medium. Yet, even
a detailed knowledge of the elementary amplitude does not guarantee a
good understanding of the coherent process.  The main difficulty stems
from the fact that there are, literally, an infinite number of
equivalent on-shell representations of the elementary amplitude.
These different representations---although
equivalent on-shell---can give very different results when evaluated
off-shell. We will present in this section an example of this
ambiguity and later on in our discussion of the results (Chapter
\ref{ch:CoherentResults}) we will show how two equivalent
parameterizations on-shell can give results that are an order of
magnitude apart for off-shell spinors. Of course, this uncertainty is present in many other kinds
of nuclear reactions, not just in the coherent photoproduction
process.  Yet, this off-shell ambiguity comprises one of the biggest,
if not the biggest, hurdle in understanding the coherent
photoproduction of pseudoscalar mesons.

In Chapter \ref{ch:ElementaryProcess}, I have included the standard
form for the amplitude of the elementary process $\gamma N \rightarrow
PS\, meson \, N (Y)$ as
\begin{equation}
T[\gamma N \rightarrow PS\, meson \, N (Y)] = \sum_{i=1}^{4}
A_{i}(s,t) M_i\,,
\label{scatmatcopy}
\end{equation}                             
where the invariant matrices have the form
\begin{eqnarray}
M_1 & =& - \gamma^5 \rlap/{\varepsilon} \rlap/k\,, \nonumber \\ M_2
&=& 2 \gamma^5 [(\varepsilon \cdot p_1) (k\cdot p_2 ) - (\varepsilon
\cdot p_2) (k\cdot p_1)]\,,\nonumber \\ M_3 &=& \gamma^5
[\rlap/{\varepsilon} ((k\cdot p_1) - \rlap/k (\varepsilon \cdot
p_1)]\,,
\nonumber \\ 
M_4 &=& \gamma^5 [\rlap/{\varepsilon} ((k\cdot p_2 ) - \rlap/k
(\varepsilon \cdot p_2)]\;.
\end{eqnarray}
I indicated then that this form although complete and standard, is not
unique. Many other choices---all of them equivalent on shell---are
possible. Indeed, we could have used the relation---valid only on the
mass shell,
 \begin{eqnarray} {M}_{1} &=& -\gamma^{5}\rlap/{\epsilon}\;\rlap/{k} =
  {1\over 2} \varepsilon^{\mu\nu\alpha\beta}\, \epsilon_{\mu}\,k_{\nu}
  \sigma_{\alpha \beta} = {i\over 2} \varepsilon^{\mu\nu\alpha\beta}\,
  \epsilon_{\mu}\,k_{\nu}\,{q_{\alpha}\over M_N}\gamma_{\beta}
  \nonumber \\ &-& {1 \over 2M_N}\gamma^{5} \Big[\rlap/{\epsilon}\,(k
  \cdot p) - \rlap/{k}(\epsilon \cdot p) \Big] - {1 \over
  2M_N}\gamma^{5} \Big[\rlap/{\epsilon}\,(k \cdot p') -
  \rlap/{k}(\epsilon \cdot p') \Big]\;, 
\end{eqnarray}
to obtain the following representation of the elementary amplitude:
\begin{equation}
   T[\gamma N \rightarrow PS\, meson \, N (Y)] = \sum_{i=1}^{4} B_{i}(s,t) {N}_{i}
   \;,  
\label{telemavec}
\end{equation}
where the new invariant amplitudes and Lorentz structures are now
defined as:
\begin{eqnarray} B_{1} &=& A_{1} \;; \qquad \phantom{-,A_{1}/2M} \;\;
    N_{1} = {i\over 2} \varepsilon^{\mu\nu\alpha\beta}\,
    \epsilon_{\mu}\,k_{\nu}\,{q_{\alpha}\over M_N}\gamma_{\beta} \;,
    \\ B_{2} &=& A_{2} \;; \qquad \phantom{-,A_{1}/2M_N} N_{2} =
    M_{2}=2\gamma^{5} \Big[(\epsilon \cdot p)(k \cdot p') - (\epsilon
    \cdot p')(k \cdot p) \Big] \;, \\ B_{3} &=& A_{3}-A_{1}/2M_N \;;
    \qquad N_{3} = M_{3}=\gamma^{5} \Big[\rlap/{\epsilon}\,(k \cdot p)
    - \rlap/{k}(\epsilon \cdot p) \Big] \;, \\ B_{4} &=&
    A_{4}-A_{1}/2M_N \;; \qquad N_{4} = M_{4}=\gamma^{5}
    \Big[\rlap/{\epsilon}\,(k \cdot p') - \rlap/{k}(\epsilon \cdot p')
    \Big] \;.  
\end{eqnarray}
Although clearly different, Equations \ref{scatmatcopy} and \ref{telemavec}
are totally equivalent on-shell: no observable measured in the
elementary process could distinguish between these two forms. We could
go on. In fact, it is well known that a pseudoscalar and a
pseudovector representation are equivalent on shell. That is, we could
substitute the pseudoscalar vertex in $N_{2}$ and $M_{2}$ by a
pseudovector one:
\begin{equation}
  \gamma^{5}={\rlap/{\!q}\over 2M_N}\gamma^{5}\;.
\end{equation}
The possibilities seem endless.

Given the fact that there are many---indeed infinite---equivalent
parameterizations of the elementary amplitude on-shell, it becomes
ambiguous on how to take the amplitude off the mass shell. The
question that arises here: are these equivalent representations
on-shell, still equivalent when we consider bound nucleons; nucleons
that are off their mass shell? The answer is negative. In this work we
have examined this off-shell ambiguity by studying the coherent
process using the ``tensor'' parameterization, as in
Equation \ref{scatmatcopy}, and the ``vector'' parameterization, as in
Equation \ref{telemavec}. Denoting these parameterizations as tensor and
vector originates from the fact that for the coherent process from
spherical nuclei (such as the ones considered here) the respective
cross sections become sensitive to only the tensor and vector (matter)
densities, respectively. Indeed, we have seen in the previous section
that the standard form for the amplitude resulted in the process
probing the tensor density of the nucleus.

It is important to note here that the vector and tensor densities are
fundamentally different quantities and that this off-shell
ambiguity is a direct consequence of using the fully relativistic
formalism. Had we elected to use non-relativistic
formalisms~\cite{bofmir86,cek87,bentan90,ndu91,tryfik94,fix97}, we would
have found that the process is probing the vector (matter) density and
that there is no off-shell ambiguity. This is, however, due to the
limitations of the non-relativistic nuclear structure formalism which
cannot produce more than one nuclear density due to the arbitrary
neglect of any medium modifications to the ratio of lower-to-upper
components of the Dirac spinors as a result of using the free-space
relation to relate these components to each other.

Since the substance of the difference between the tensor and vector
parameterizations lies in the use of the tensor as opposed to the vector
density of the nucleus, it is instructive to find the relationship
between these two quantities. This can be most easily seen by assuming
the free-space relation between the upper and lower components of the
Dirac spinors. In this case the tensor density can be written in terms
of the vector density as
\begin{equation}
  \rho_{\lower 3pt \hbox{$\scriptstyle T$}}(q) = -{q \over 2M_{N}}
  \rho_{\lower 3pt \hbox{$\scriptstyle V$}}(q) +
  \sum_{\alpha}^{\rm{occ}} {{\kappa +1} \over M_{N}} \int_{0}^{\infty}
  dr {{g^{2}_{\alpha}(r)} \over r^2} j_1(qr)\;, 
\label{trhotv}
\end{equation}
where $\kappa$ is the generalized relativistic angular momentum (see
Chapter \ref{ch:NuclearStructure}), ${g_{\alpha}(r)}$ is the upper
component of the Dirac spinor, and $j_1(qr)$ is the Bessel function of
order one. The second term in the above expression is negligible for
closed-shell (spin-saturated) nuclei; this term is proportional to the
difference between the square of the wavefunctions of spin-orbit
partners (such as $p^{3/2}$ and $p^{1/2}$ orbitals) which is very
small even in the Walecka model. Hence, for closed shell nuclei---and
adopting a free-space relation---the tensor density becomes
proportional to the vector density. Thus we have produced the
non-relativistic limit of the tensor density. However, for open-shell
nuclei such as ${}^{12}$C, the second term in Equation \ref{trhotv} is
no longer negligible and leads to an additional enhancement of the
tensor density---above and beyond the one obtained from the the
dynamic enhancement of the lower component of the Dirac spinor. We
label this additional enhancement of the cross section as ``open-shell
effect'' to distinguish it from the dynamic enhancement.

To a provide a feeling for the nature of the nuclear tensor density
($\rho_T$) and its dependence on the nucleus radius,
Figure \ref{fig:TensorDenisty} displays the proton and neutron tensor densities in
\BFIG
\centerline{\psfig{figure=TensorDenisty.pstex,height=6.0in,width=6.0in}}
\caption[Nuclear tensor density ($\rho_T$) in ${}^{40}$Ca]
{The proton and neutron tensor densities [$\rho_T (r)$] in ${}^{40}$Ca as
calculated using the QHD-II model for the nuclear structure.} 
\label{fig:TensorDenisty}
\end{figure}
${}^{40}$Ca. As evident in this figure, the tensor density has a
different behavior compared to the vector and scalar densities;
it is appreciable only at the surface of the nucleus and
vanishing elsewhere
(compare to Figure \ref{fig:Ca40desities}).  The densities in the figure are calculated using the QHD-II
model for the nuclear structure (Chapter \ref{ch:NuclearStructure}). QHD-I evaluation gives
identical results.

\section{Inclusion of Isospin}
\label{sec:isospin}

Recall that the elementary process parameterization contains four
amplitudes: $\{ A_1, A_2, A_3, A_4\}$. These have different values
depending on the kind of nucleon target: a proton (p) or a neutron
(n). Since nuclei include both of these nucleons, we have to modify
our formalism to incorporate the isospin aspect of the problem. Thus,
the $T$ matrix is modified as
\begin{equation}
T \Longrightarrow T_p \frac{1}{2} (1 + \tau_z) + T_n \frac{1}{2} (1 -
\tau_z)\;.
\end{equation}
Now substituting this form in our formalism for the coherent process
results in the scattering matrix element depending on
two combinations of the $A_1$ amplitude for the proton (${A_1}_p$) and the
neutron (${A_1}_n$) : $A_s$ and $A_v$ as
\begin{equation}
T_{\mbox{coherent}} \sim A_s + A_v \tau_z\;,
\end{equation}
where $A_s = \frac{1}{2} ( {A_1}_p + {A_1}_n )$ and $A_v = \frac{1}{2}
( {A_1}_p - {A_1}_n )$. Hence, it is clear that the $A_s$ part carries the
isoscalar component of the matrix element while the $A_v$ part
includes the isovector component. We arrive then at an expression for
the matrix element (in the tensor parameterization) of the form
\begin{equation}
T_{\mbox{coherent}} \sim A_s {\rho_{T}}_s + A_v {\rho_{T}}_v\;,
\end{equation}
where $ {\rho_{T}}_s = {\rho_T}_p + {\rho_T}_n$ and ${\rho_{T}}_v = {\rho_T}_p -
{\rho_T}_n$. That is the matrix element depends on two combinations of
the proton and neutron tensor densities. Analogous expressions hold
if we would have used the vector parameterization.

For the nuclei that we studied in this work, the proton and neutron
numbers are equal. Therefore, ${\rho_T}_p
\approx {\rho_T}_n$ and so $\rho_v \rightarrow 0$. That is the isovector
component vanishes. Note that although $N_p = N_n$,
for these nuclei, the cancellation between ${\rho_T}_p$ and
${\rho_T}_n$ is not perfect because of isospin symmetry violation in
the Hamiltonian of the nucleus due mainly to the Coulomb repulsion of the
protons. It is found however that
this cancellation is almost exact and affect minimally the coherent
process (see Chapter \ref{ch:CoherentResults}).

\chapter{Distortions and the Coherent Process}
\label{ch:Distortions}

In the previous chapter, I have depicted the basic formalism for the
coherent process in the limit of no final-state interactions between
the emitted meson and the recoil nucleus. The expressions that we
reached are elegant and transparent, but this beauty cannot survive the
hammering of distortions. In this chapter, I will outline the
modifications to this formalism in the presence of distortions. In the
first section I will describe the basic mechanisms behind the
meson-nucleus interaction while in the second one I will lay out
the changes to the coherent-process formalism.

\section{Optical Potential Formalism}

Mesonic distortions play a critical role in all studies involving
meson-nucleus interactions. These distortions are strong, and thus
modify significantly any process relative to its naive plane-wave
limit. Indeed, it has been shown in earlier studies of the coherent processes---and verified
experimentally~\cite{ndu91}---that there is a large modification of
the plane-wave cross section once distortions are
included. Fortunately, the meson-nucleus interaction is short range
and present only in the close vicinity of the collision. The
long-range Coulomb
distortions do not play a role here since the
emitted meson must be electrically neutral due to charge
conservation. Because of the importance of mesonic distortions,
any realistic study of the coherent reaction must invoke them from the
outset. However, since a detailed microscopic model for distortions has yet to be developed, I have resorted to a
semi-phenomenological method: optical-potential formalism.

\subsection{Equation of Motion for the Meson Field}

In this section, the equation of motion for the meson will be
discussed. Since the mass of the emitted meson is comparable to the
momentum carried by this particle, the meson must be treated
relativistically. On the other hand, the nucleus has a much larger
mass compared to its momentum and thus can be treated
non-relativistically at least in the low-energy scattering processes.

There are at least three approaches to write the effective equation of
motion for the meson-nucleus interaction. The simplest one is to
consider the nucleus as a static source of potential in which
the meson travels through. Consequently, we have a one-body
Klein-Gordon equation of the form:
\begin{equation}
\left(D_\mu D^\mu + m^2 \right) \phi = 0\;,
\end{equation}
where $D_\mu = i \partial_\mu - V_\mu(x)$, $V_\mu$ is the interaction
potential, and $m$ is the mass of the meson~\cite{ndu91,GreRel95}.

Another approach is to write a ``relativized'' Schrodinger-like
equation for the system as
\begin{equation}
\left[ (-{\nabla^2}_x + m^2 )^{\frac{1}{2}} + M_A + \frac{-{\nabla^2}_{x'}}{2
M_A} + V({\bf x,x'}) \right]
\psi = i \partial_t \psi\;,
\label{schrodingerlike}
\end{equation}
where in this equation $\nabla^2 = \partial_i \partial^i$, $x$ denotes the meson coordinates while $x'$ denotes
the nucleus coordinates, $M_A$ is the nucleus mass, $V({\bf x,x'})$ is the
interaction potential, and $\psi$ is the meson-nucleus system wavefunction~\cite{SMC,GolWat}. It is clear in the equation that the
kinetic energy term for the meson has been relativized $(-{\nabla^2}_x
+ m^2 )^\frac{1}{2}$, while the one for the nucleus has its
non-relativistic form ($\frac{-{\nabla^2}_x'}{2 M_A}$).

A third approach is to write a Klein-Gordon-like equation for the
system as~\cite{SMC}
\begin{equation}
 \left( -{\nabla^2}_x + m^2 \right)\psi = \left(i \partial_t - V({\bf x,x'}) - M_A -
 \frac{-{\nabla^2}_{x'}}{2 M_A}\right)^2 \psi\;.
\label{kleingordonSMC}
\end{equation}
Goldberger and Watson~\cite{GolWat} have shown that the second and
third of these approaches (Equations \ref{schrodingerlike} and
\ref{kleingordonSMC}) are equivalent for certain class of potentials.

Starting with Equation \ref{kleingordonSMC}, one can arrive at an
effective one-body equation for the meson field by absorbing the
nucleus degrees of freedom and using several assumptions about the
nature of the interaction, to yield the eigenvalue equation:
\begin{equation}
 \left[ {\nabla^2} + {\bf k}^2 - 2\omega U(r)\right] \phi = 0\;,
\label{effkleingordonSMC}
\end{equation}
where ${\bf k}$ is the meson asymptotic momentum in the center-of-momentum
frame (c.m.), and $2\omega U(r)$ is the effective potential for this
one-body equation. This potential is an involved nonlocal function of the
potential $V$ and several kinematic variables in the
problem~\cite{SMC}. We adopt this third approach for our studies.

The potential $2\omega U(r)$ is independent of the angular coordinates
$(\Phi,\theta)$ and thus we can separate the angular parts from the
radial part in the equation. The angular parts reduce to the orbital-angular-momentum equations which have the spherical harmonics as their
solutions, while the radial part reduces to the following equation:
\begin{equation}
\left[ \frac{d^2}{dr^2} - \frac{l(l+1)}{r^2} + {\bf k}^2 \right] u_{nl}(r) = r 2
\omega U(r) \, \left[ \frac{u_{nl}(r)}{r}\right] \;,
\label{radialKG}
\end{equation}
where $l$ is the orbital angular momentum quantum number, $n$ is the
energy quantum number, and $u_{nl}(r)$ is the radial part for a
specific $l$-partial wave (angular-momentum channel) of the meson wavefunction. Consequently, the meson wavefunction is given by the
expansion
\begin{equation}
\phi_{\mathbf k'} =  \sum_{l,m} 4 \pi (i)^l \; \frac{u_{nl} (r) }{r}\;
 Y_{lm}(\hat{\mathbf{r}}) \; Y_{lm}^{*} (\hat{\mathbf{k'}})\;,
\label{phiexpansion}
\end{equation}
where $m$ is the quantum number for the z-axis projection of the angular
momentum and $Y_{lm}$ is the spherical harmonic with $l$ and $m$ orders.

\subsection{Meson-Nucleus Optical Potential Form}

The potential $2\omega U(r)$ for the kind of applications that we are
considering here has the formal form:
\begin{equation}
2 \omega U (r) = f_1(r) + \nabla^2 f_2(r) + \mbox{\boldmath ${\nabla}$
\unboldmath} \cdot f_3(r)
\mbox{\boldmath ${\nabla}$ \unboldmath}\;,
\end{equation}
where $f_i(r)$ are some functions. Such form can also be rewritten as
\begin{equation}
2 \omega U (r) = f_1(r) + \frac{2}{r} \frac{d f_2(r)} {dr} +
\frac{d^2 f_2(r)}{dr^2} + f_3(r) \nabla^2 + \frac{d f_3(r)}{dr} \frac
{d}{dr}\;. 
\end{equation}

We have studied in this work the coherent process for the production of
$\eta$ and $\pi$ mesons. Thus, I will outline here two kinds of
optical potentials: the $\eta$-nucleus and the $\pi$-nucleus optical
potentials.

\subsubsection{Eta-Nucleus Optical Potential}

We have been very fortunate to find a simple and local form for the
$\eta$-nucleus optical potential in the literature; a fortune that we
lacked in the case of the $\pi$-nucleus
potential. This simplicity is due the fact that the $\pi$-nucleus
interaction is far stronger and sophisticated than the
$\eta$-nucleus interaction. For the $\eta$-nucleus interaction in the low energy
regime of our interest, $s$-wave components dominates, and $p$-wave and
$d$-wave contributions are very small. This in turn is a result of
the fact that the $\eta$ ($0$-isospin) can couple only to
$\frac{1}{2}$-isospin nucleon resonances like the $S_{11}$, and cannot
couple to the $\Delta$-resonance which has an isospin of
$\frac{3}{2}$. Consequently, there are only few resonances that the
$\eta$ can couple to leading to a simple form for the $\eta$-nucleus
interaction. This situation is in sharp contrast to the $\pi$-nucleus
interaction presented in the next section where the pion can couple strongly to several nucleon resonances.

The optical potential expression is constructed from the scattering
amplitude of the process $\eta N \rightarrow \eta N$ to fit a simple
$t\rho$ form~\cite{bentan90} as following :
\begin{equation}
  2 \omega U(r) = - b \rho_{\lower 3pt \hbox{$\scriptstyle V$}}(r) \;.
  \label{trho}
\end{equation}
Here $\rho_{\lower 3pt \hbox{$\scriptstyle V$}}(r)$ is the vector
density of the nucleus and and $b$ is a complex two-body ($\eta N$) parameter
that is given by the following:
\begin{eqnarray}
  b(p_{\rm lab}) & \equiv & (\alpha + \beta p_{\rm lab} + \gamma
  p_{\rm lab}^{2})^{-1} \;, \\ \alpha &=& (+0.136,-0.052)\,{\rm
  fm}^{-1} \;, \\ \beta &=& (+0.035,-0.072) \;, \\ \gamma &=&
  (-0.061,+0.009)\,{\rm fm} \;.  \label{bfit}
\end{eqnarray}

Having this simple form for the potential, one can solve numerically
the radial part of the Klein-Gordon equation \ref{radialKG} for each
partial wave to obtain the full outgoing $\eta$-meson wavefunction
(Equation \ref{phiexpansion}).

\subsubsection{Pion-Nucleus Optical Potential}
 
Constructing the $\pi$-nucleus optical potential proved to be a
difficult task. Admittedly, there is a lot of work in the literature
that covers this issue. Nevertheless, most studies concentrated on the
low energy optical potential and I was not able to find any work that
derives the optical potential in the $\Delta$-resonance
region. Consequently, J. Carr and I 
extended earlier studies~\cite{SMC} on the low-energy $\pi$-nucleus
optical potential to higher energies so that they cover the $\Delta$
resonance region~\cite{radcar}. A pleasant by-product emerged from
our study: we were able to update earlier studies with our newly
extracted optical potential parameters from recent state-of-the-art
experimental measurements~\cite{Arnpin}. In this regard, this project
can serve as a current comparative view of earlier attempts to extract
these parameters. Furthermore, we make no recourse to nonrelativistic
approximations (as opposed to the earlier low-energy treatments), and
include the full relativistic nucleus recoil.

The derivation of the optical potential form is a challenging endeavor
for the following reasons: first, the $\pi$-nucleus interaction is
very strong which renders the fine details in the potential
significant. Second, the first-order impulse-approximation form of the
potential is not adequate as one has to incorporate many corrections
stemming from the many-body nature of the interaction like multiple
scattering and pion absorption. Indeed, pion absorption is crucial in
the $\Delta$ resonance. Finally, the nature of the potential is
complicated as it involves local and nonlocal terms. These
complications arise in fact from the essence of the fundamental
process that drives the interaction in this energy regime: the
$\Delta$ resonance formation. Since the procedures for this derivation
are very involved, for the purpose of this manuscript, I will give
only an overview of the derivation as well as the final form of the
optical potential.

The $\pi$-nucleus optical potential is derived using a
semi-phenomenological formalism that originates in the $\pi-N$
interaction scattering amplitude. This amplitude is given
by~\cite{SMC}
\begin{equation}
f(\pi N \rightarrow \pi N) = b_0 + b_1 \; {\bf t} \cdot
\mbox{\boldmath $\tau$ \unboldmath} + (c_0 + c_1 \;
{\bf t} \cdot \mbox{\boldmath$\tau$\unboldmath} ) \; {\bf k}.{\bf k}^\prime\;,
\label{pinamp}
\end{equation}
where ${\bf t}$ and \boldmath $\tau$ \unboldmath are the pion and nucleon isospin
operators, ${\bf k}$ and ${\bf k}^\prime$ are the incoming and
outgoing pion momenta, $b_0$ and $b_1$ are the s-wave parameters and
$c_0$ and $c_1$ are the p-wave parameters. In this form the small
spin-dependent term has been neglected. The s- and p-wave parameters
are determined from the phase shifts. In earlier
treatments~\cite{SMC}, these parameters were determined initially from
a phase shift analysis performed by Rowe, Salomon, and Landau~\cite{RSL}. The parameters then were slightly modified to
obtain the best fit for the $\pi$-nucleus scattering and pionic atom
data. Our treatment differs from the previous studies in two aspects:
first, we determine them from the state-of-the-art experimental
measurements and phase shift analysis (SP98) of Arndt,
Strakovsky, Workman, and Pavan from the Virginia Tech
SAID program~\cite{Arnpin}.  Second, we keep these parameters intact
by not attempting to change them to fit any specific data. In doing so
we have kept the theoretical basis for the optical potential
unblemished. Nonetheless, the parameters determined by the two methods
match nicely in the low-energy limit.

After adopting the $\pi-N$ scattering amplitude of Equation
\ref{pinamp} in the center-of-momentum frame (c.m.), the first step in
the derivation is to transform the scattering amplitude to the
$\pi$-nucleus c.m. frame. This is done using the
relativistic potential theory of Kerman, McManus, and
Thaler~\cite{KeMcTh59}. The kinematic arguments of the scattering
amplitude are then expressed in terms of the appropriate kinematic
quantities in the $\pi$-nucleus c.m. frame using what is referred to as
the angle transformation. In this manner, we would have achieved most
of the
first class of modifications to the scattering amplitude: kinematic
corrections.

By invoking the impulse approximation, the resultant form for the
amplitude is then sandwiched between bound-nucleon states and the
expression is summed over all occupied states of the nucleus. Hence,
one obtains the $\pi$-nucleus interaction amplitude in momentum
space. Now taking the Fourier transform, we obtain an expression for
the optical potential form.

This impulse-approximation form still lacks the second class of
modifications: physical corrections resulting from many-body
processes. These corrections modify the scattering amplitude
parameters like $b_0$ and $c_0$ and add new terms to the optical
potential. The first of these corrections are the multiple scattering
ones. It has been found that the second-order corrections for the
$s$-wave terms as well as higher order corrections for $p$-wave terms,
are necessary. Therefore, the multiple scattering series for the
p-wave is summed
partially to all orders. This introduces the Ericson-Ericson
effect~\cite{erer66} which is analogous to the Lorentz-Lorenz effect
in electrodynamics~\cite{jackson62}. This effect adds a nonlocal term
to the potential of the form $\mbox{\boldmath ${\nabla}$ \unboldmath}
\cdot f(r) \mbox{\boldmath ${\nabla}$ \unboldmath}$. The
Ericson-Ericson term is further modified to account for short-range
correlations between nucleons.

A second physical correction is the absorption correction. This
one gives the potential its name as an ``optical'' potential
since it implies the existence of an imaginary part in the
potential. There are two types of absorption. The first
one arises from the fact that there are many open inelastic channels
in the $\pi$-nucleus interaction like nucleon knock-out. Accordingly,
part of the incoming flux is absorbed by these processes leading to an
imaginary part in the potential. This kind of absorption is naturally
included in the impulse-approximation form for the potential. The
second type of absorption arises from many-body mechanisms like the
two-nucleon absorption where the pion is scattered from one nucleon
but then absorbed by another. This is in fact the dominant many-body
absorption mechanism. Another less important absorption mechanism, is
the quasi-elastic charge exchange process. All of these many-body absorption
mechanisms are referred to as true absorption to distinguish them from
the inelastic (type one) absorptions. Ironically, the
$\Delta$-resonance formation that drives strongly the elementary
process $\pi N \rightarrow
\pi N$, dampens it in the nucleus through the absorption
mechanisms.

Another alteration to the potential is the Pauli correction. Due to
the Pauli principle, the number of available final states for the struck
nucleon in the nucleus is reduced by Pauli blocking leading to this
kind of correction. Another correction is the Coulomb one originating
from the fact that the incoming charged pion (in $\pi$-nucleus
scattering) is accelerated or
decelerated depending on its charge, by the Coulomb field of the
nucleus before interacting with the nucleus through the strong
interaction. This correction is of no impact in our study as we are
considering coherent processes where the emitted pion is always
neutral. Finally, it is noteworthy to mention that there are also
other kinematic corrections stemming from transformation properties of
the many-body subsystems like the $\pi-2N$ subsystem in the $\pi-2N$
interaction mechanisms.

After implementing these corrections to the impulse-approximation
expression, we arrive at a pion-nucleus optical potential---applicable
from threshold up to the delta-resonance region---of the form:
\begin{eqnarray}
        2 \omega U &\!=\!&\!-4\pi \Big[ p_1 b(r) \!+\! p_2 B(r)\!-\!
        \mbox{\boldmath ${\nabla}$
        \unboldmath} Q(r) \cdot \mbox{\boldmath ${\nabla}$
        \unboldmath}  \\ \nonumber
&&  \!-\! \; \frac {1}{4} p_1 u_1
        {\nabla}^2 c(r)\!-\!  \frac {1}{4} p_2 u_2 {\nabla}^2
        C(r)\!+\!  p_1 y_1 \widetilde{K}(r) \Big]\;,
\end{eqnarray} 
where
\begin{eqnarray}
   b(r) &=& \bar{b}_0\rho(r)-\epsilon_\pi {b_1} {\delta}\rho(r) \;, \\
   B(r) &=& {B_0}\rho^2(r)-\epsilon_\pi {B_1}\rho(r)\delta\rho(r) \;,
   \\ c(r) &=& c_0\rho(r)-\epsilon_\pi {c_1}{\delta}\rho(r)\;, \\ C(r)
   &=& C_0\rho^2(r)-\epsilon_\pi{C_1}\rho(r){\delta}\rho(r)\;, \\ Q(r)
   &=& \frac{ L(r) }{1 + \frac {4 \pi}{3} \lambda L(r)} + p_1 x_1
   \acute{c} \rho(r)\;, \\ L(r) &=& p_1 x_1 c(r) + p_2 x_2 C(r) \;, \\
   \widetilde{K}(r) &=& \frac {3}{5} \left({\frac{3
   \pi^2}{2}}\right)^{2/3} c_0\rho^{5/3}(r)\;,
\end{eqnarray}
and with
\begin{eqnarray}
   \bar{b}_0 &=& {b_0} - p_1 \frac {A-1}{A} (b^2_0 + 2 b^2_1) I \;, \\
   \acute{c} &=& p_1 x_1 \frac {1}{3} k^2_o (c^2_0 + 2 c^2_1) I \;.
\end{eqnarray}
In the above expressions, the set \{$p_1, u_1, x_1,$ and $y_1$\}
represents various kinematic factors in the effective $\pi-N$ system
(pion-nucleon mechanisms), and the set \{$p_2, u_2,$ and $x_2$\}
represents the corresponding kinematic factors in the $\pi\!-\!2N$
system (pion-two-nucleon mechanisms). The set of parameters \{$b_0,
b_1, c_0,$ and $c_1$\} originates from the $\pi N\!\rightarrow\!\pi N$
elementary amplitudes while all other parameters--excluding the
kinematic factors--have their origin in the second and higher order
corrections to the optical potential. Nuclear effects enter in the
optical potential through the nuclear density $\rho(r)$, and through
the neutron-proton density difference (isovector density)
$\delta\rho(r)$. Moreover, $A$ is the atomic number, $\lambda$ is the
Ericson-Ericson effect parameter, $k_o$ is the pion lab momentum,
$\omega$ is the pion energy in the pion-nucleus center of mass system,
and $I$ is the so-called $1/r_{correlation}$ function. The $B$ and $C$
parameters arise from true pion absorption.

\section{Coherent Process in a Relativistic Distorted Wave Impulse Approximation
Approach}

The analysis in the previous section provides us with the meson wavefunction in the presence of distortions. In this section, we will
examine the result of including this wavefunction in our formalism
for the coherent process. Recall that we have two parametrization for
the coherent process: the tensor and vector parametrizations; we must
implement the distortions in two independent fashions. We find that
distortions affect these parametrizations differently. This is yet
another manifestation of the off-shell ambiguity where two equivalent
parametrization on-shell are vastly different off their mass
shell. Since the derivations in this section are very involved, I will
give only an overview on how to implement these final-state
interactions.

\subsection{Distortions in the Tensor Parameterization}
\label{subsec:distten}

In our formalism for the coherent process (Chapter
\ref{ch:CoherentTheory}) we arrived at an integral of the following
form:
\begin{equation}
{\rho_T}^{0i}({\mathbf k},{\mathbf  k'}) = \int d^3{x} \; {\phi_{\mathbf
k'}^{*\,(-)}} \; {\rho_T}^{0i}({\mathbf x}) \; e^{i {\bf k\cdot x}}\;,
\label{tenint}
\end{equation}
In the plane-wave limit (no distortions), the meson wavefunction
(${\phi_{\mathbf k'}^{(-)}}$) takes the plane-wave form $ e^{i{\bf
k'\cdot x}}$ and so the integral in 
nothing but the Fourier transform of the tensor density with respect
to the momentum transfer ${\bf q = k - k'}$. In the presence of
distortion the integral is far more complicated. Now we have to
expand each term in the integral in terms of angular-momentum
eigenfunctions as following:
\begin{equation}
 {\phi_{\mathbf k'}^{(-)}} = \sum_{l,m} 4 \pi (i)^l \;
 {\phi_{lk'}^{(-)}}(r) \; Y_{lm}(\hat{\mathbf r}) \; {Y_{lm}^*}(\hat{\mathbf
 k'})\;,
\label{phiexpansion2}
\end{equation}
\begin{equation}
 {\rho_T}^{0i}({\mathbf x}) = \rho_T(r) \; \left( \frac{4
 \pi}{3}\right)^{\frac{1}{2}} \; Y_{1\mu}(\hat{\mathbf{r}})\;,
\end{equation}
and
\begin{equation}
  e^{i {\bf k\cdot x}} = \sum_{l,m} 4 \pi (i)^l \; {j_l}(kr) \;
 Y_{lm}(\hat{\mathbf{r}}) \; {Y_{lm}^*}(\hat{\mathbf{k'}})\;.
\label{Aexpansion}
\end{equation}
Here, ${j_l}(kr)$ is the spherical Bessel function of order $l$.

The above expansions are then substituted in the integral of Equation
\ref{tenint}. We obtain then an expression saturated with spherical
harmonics, and so we resort to the use of the theory of angular
momentum to recouple these harmonics in such a way to reduce
these sums to a tractable form for numerical calculations. We
eventually arrive at the following expression in the c.m. frame for the scattering
matrix element:
\begin{eqnarray}
   T &=& - 2 {F_T}_{0i} \; {\rho_T}^{0i}({\bf k, k'}) \\ \nonumber
     &=& i A_1(\tilde{s},t) \left[ \sum_{ll'} \frac{I_{ll'}(k,k')}{k'}
     {{P'}_{l'}} (cos\theta) \right] \left[ k {k'} {\hat{\bf
     k'}}(\hat{\bf k} \times \epsilon_\lambda({\hat{\bf k}})\right]\;,
\end{eqnarray}
where $k \equiv |{\bf k}|$, $k^\prime \equiv |{\bf k}^\prime|$, $l' = l \pm 1$, ${{P'}_{l'}}(cos\theta)$ is the derivative
of Legendre polynomial of order $l'$, $\theta$ is the scattering
angle, and
\begin{equation}
{I_{ll'}(k,k')} = 4 \pi {\int_0}^\infty r^2 dr \; {\phi_{l'k'}^{(+)}}(r)\;
\rho_T(r) \; {j_l}(kr)\;.
\end{equation}

It is notable in the above expression that since ${{P'}_0} \equiv 0$,
the $l' = 0$ component of the meson wavefunction does not contribute
in the coherent process. This fact is a consequence of parity and
Lorentz transformation properties of the scattering matrix
element. Moreover, this implies that close enough to threshold, when
the centrifugal barrier to a large extent screens the potential,
distortion effects are minimal.

\subsection{Distortions in the Vector Parameterization}

In a analogous fashion to the case for the tensor representation we
have here to evaluate an integral of the form:
\begin{equation}
T = - \int d^3{x} \; {\phi_{\mathbf k'}^{*\,(-)}} \; {F_v}^{i0}
\; {\left[{\mathbf k'}\right]}_i \; {\rho_V}({r}) \; e^{i {\bf k \cdot x}}\;.
\label{vecint}
\end{equation}
Here, ${F_v}^{\alpha \beta} = \frac {i}{2 M_N}
\varepsilon^{\mu\nu\alpha\beta} \varepsilon_\mu k_\nu$. In a similar
manner to the tensor case, we expand every term in the integral in
terms of its angular-momentum eigenfunctions. A complication arises
regarding the identity of the meson momentum ${\mathbf k'}$ in the
integral. This momentum originates from the parameterization of
the elementary process. There, the meson has a well-defined momentum
as its wavefunction is nothing but a plane wave. If we incorporate
the final-state interactions, the meson has ${\mathbf k'}$ only as its
asymptotic momentum and not as its ``local'' momentum. Since the local
momentum in the interaction region is the physically relevant
quantity, we have replaced the asymptotic momentum ${\bf k'}$ by
the meson-momentum operator ($ -i \mbox{ \boldmath $\nabla$ \unboldmath
} $). In doing so the
integral evaluation becomes exceedingly more involved. The least
painful method to compute it is through angular-momentum algebra. In
this regard, we need to expand the term $\left( -i \mbox{\boldmath
$\nabla$ \unboldmath}{\phi_{\mathbf k'}^{*\,(-)}}\right)$ with respect to angular-momentum eigenfunctions and spherical tensors. This is done by
resorting to the identity~\cite{rose57}
\begin{eqnarray}	
\mbox{\boldmath $\nabla$ \unboldmath} \Phi(r) Y_{lm}(\hat{\mathbf{x}})
&=& - \left( \frac{l+1}{2l +
1}\right)^{\frac{1}{2}} \left[ \frac{d\Phi}{dr} - \frac{l}{r}
\Phi \right] {\mathbf
T}_{l,l+1,m}\\
\nonumber
&& + \left(\frac{l}{2l + 1}\right)^{\frac{1}{2}} \left[ \frac{d\Phi}{dr} +
\frac{l+1}{r} \Phi\right] {\mathbf T}_{l,l-1,m}\;,
\end{eqnarray}
where
\begin{equation}
{\mathbf T}_{l,l \pm1 , m} = {[ Y_{l \pm 1} \otimes \mbox{\boldmath $\xi_i$
\unboldmath}]}_{l,m}\;.
\end{equation}
Here $\mbox{\boldmath $\xi_i$ \unboldmath}$ are the spherical basis
vectors.

All terms in the integral now have well-defined angular-momentum
properties. Therefore, we can use the theory of angular momentum to
reduce the integral to a manageable form for numerical
calculations. Eventually, we arrive at the following result for the
scattering matrix element:
\begin{eqnarray}	
T &=&\pm (2\pi)^{3/2}\frac{|\bf k|}{M_N} \sum_{l=1}^{\infty}
  \sqrt{\frac{l(l+1)}{2l+1}} \nonumber\\ & & {Y}_{l,\pm
  1}(\hat{\mathbf{k'}}) \int r^2 dr \rho_{\lower
  3pt\hbox{$\scriptstyle V$}}(r) R_{l}(r)\;,
\end{eqnarray}
where
\begin{equation}
    R_{l}(r) = j_{l+1}(kr)\left[\frac{d}{dr}-\frac{l}{r}\right]
    {\phi_{lk'}^{(+)}}(r) +
    j_{l-1}(kr)\left[\frac{d}{dr}+\frac{l\!+\!1}{r}\right]
    {\phi_{lk'}^{(+)}}(r)\;.
\end{equation}
Here the $\pm$ sign is for positive/negative circular polarization of
the incident photon. It is worth mentioning that adopting the ${\bf
k'} \rightarrow -i \mbox{\boldmath
$\nabla$ \unboldmath}$ prescription, has resulted, as in the
tensor case (see Subsection \ref{subsec:distten}), in no s-wave ($l\!=\!0$) contribution to
the scattering amplitude. This is also in agreement with the earlier
nonrelativistic calculation of Ref.~\cite{bofmir86}.

\chapter{Results and Discussion of the Coherent Process}
\label{ch:CoherentResults}

In this chapter, I will present the main results of our study of the
coherent photoproduction from nuclei. Specifically, I will discuss the
effects of distortions, relativity, uncertainties in the nuclear
structure, $S_{11}$-resonance suppression, nuclear dependence, isovector
component, and off-shell ambiguity. Moreover, I will present a
qualitative discussion of possible violations of the impulse
approximation.

\section{Distortion Effects}
\label{sec:distor}

In this section, I will discuss effects of 
distortions in the $\pi$ and $\eta$ coherent processes.

\subsection{Distortion Effects for the Pion}

The large effect of pionic distortions can be easily seen in Figure
\ref{fig:PionDiffDistEffe}. The left panel of the graph (plotted on a
\BFIG
\centerline{\psfig{figure=PionDiffDistEffe.pstex,height=6.0in,width=6.6in}}
\caption[Pionic distortion effects on the differential cross section
for $^{40}\rm{Ca}$] {Differential cross section for the coherent pion
photoproduction reaction from $^{40}\rm{Ca}$ at $E_{\gamma}=168$ MeV
using the vector representation for the elementary amplitude with
(solid line) and without (dashed line) the inclusion of
distortions. Results on the left(right) panel are plotted using a
linear(logarithmic) scale.}
\label{fig:PionDiffDistEffe}
\end{figure}
linear scale) shows the differential cross section for the coherent
photoproduction of neutral pions from $^{40}\rm{Ca}$ at a laboratory
energy of $E_{\gamma}\!=\!168$~MeV. The solid line displays our
results using a relativistic distorted-wave impulse approximation
(RDWIA) formalism, while the dashed line displays the corresponding
plane-wave result (RPWIA). The calculations have been done using a
vector representation for the elementary $\gamma N \rightarrow \pi^{0}
N$ amplitude. Note that this is only one of the many possible
representations of the elementary amplitude that are equivalent
on-shell. A detailed discussion of these off-shell ambiguities is
deferred to Section \ref{sec:off-shell}.  At this specific photon
energy---one not very far from threshold---the distortions have more
than doubled the value of the differential cross section at its
maximum. Yet, the shape of the angular distribution seems to be
preserved. However, upon closer examination (the right panel of the
graph shows the same calculations on a logarithmic scale) we observe
that the distortions have caused a substantial back-angle enhancement
due to a different sampling of the nuclear density, relative to the
plane-wave calculation. This has resulted in a small---but not
negligible---shift of about $10^{\circ}$ in the position of the
minimum. The back-angle enhancement, with its corresponding shift in
the position of the minimum, has been seen in our calculations also at
different incident photon energies.

The effect of distortions on the total photoproduction cross section
from $^{40}\rm{Ca}$ as a function of the photon energy is displayed in
Figure \ref{fig:PionTotCroDisEff}. The behavior of the distorted
cross section is
\BFIG
\centerline{\psfig{figure=PionTotCroDisEff.pstex,height=6.0in,width=6.0in}}
\caption[Pionic distortion effects on the total cross section
for $^{40}\rm{Ca}$] {Total cross section for the coherent pion
photoproduction reaction from $^{40}\rm{Ca}$ as a function of the
photon energy in the laboratory frame with (solid line) and without
(dashed line) including pionic distortions. A vector representation
for the elementary part of the amplitude is used.}
\label{fig:PionTotCroDisEff}
\end{figure}
explained in terms of a competition between the attractive real
(dispersive) part and the absorptive imaginary part of the optical
potential. Although the optical potential encompasses very complicated
processes, the essence of the physics can be understood in terms of
$\Delta$-resonance dominance.  Ironically, the behavior of the
dispersive and the absorptive parts are caused primarily by the same
mechanism: $\Delta$-resonance formation in the nucleus. The mechanism
behind the attractive real part is the scattering of the pion from a
single nucleon---which is dramatically increased in the
$\Delta$-resonance region. In contrast, the absorptive imaginary part
is the result of several mechanisms, such as nucleon knock-out,
excitation of nuclear states, and two-nucleon processes. At very low
energies some of the absorptive channels are not open yet, resulting
in a small imaginary part of the potential. This in turn provides a
chance for the attractive real part to enhance the coherent cross
section. As the energy increases, specifically in the
$\Delta$-resonance region, a larger number of absorptive channels
become available leading to a large dampening of the cross section.
Although the attractive part also increases around the
$\Delta$-resonance region, this increase is more than compensated by
the absorptive component, which greatly reduces the probability for the
pion to interact elastically with the nucleus.

Since understanding pionic distortions constitutes our first step
towards a comprehensive study of the coherent process, it is
instructive to examine the sensitivity of our results to various
theoretical models. To this end, we have calculated the coherent cross
section using different optical potentials, all of which fit
$\pi$-nucleus scattering data as well as the properties of pionic
atoms. We have started by calculating the coherent cross section using
the optical potential developed by Carr and collaborators~\cite{SMC}.
It should be noted that although our optical potential originates from
the work of Carr and collaborators, there are still significant
differences between the two sets of optical potentials. These
differences arise from the manner in which some parameters are
determined and from effects that were not --- at least explicitly ---
included in their model.

In addition to the above potentials, we have calculated the coherent
cross section using a simple 4-parameter Kisslinger potential of the
form~\cite{SMC}:
\begin{equation}
	2 \omega U \!=\!-4\pi \left[ b_{\rm eff} \rho(r) - c_{\rm
        eff} \mbox{\boldmath ${\nabla}$ \unboldmath}
\cdot \rho(r) \mbox{\boldmath ${\nabla}$ \unboldmath}
 + c_{\rm eff}
        \frac{\omega}{2 M_N} {\nabla}^2\rho(r) \right].
\end{equation} 
Note that we have used two different sets of parameters for this
Kisslinger potential, denoted by K1 and K2~\cite{SMC}.  Both sets were constrained by $\pi$-nucleus scattering data and by
the properties of pionic atoms.  However, while the K1 fit was
constrained to obtain $b_{\rm eff}$ and $c_{\rm eff}$ parameters that
did not deviate much from their pionic-atom values, the K2 fit allowed
them to vary freely, so as to obtain the best possible fit.

Results for the coherent photoproduction cross section from
$^{40}\rm{Ca}$ at a photon energy of $E_{\gamma}\!=\!186$~MeV
(resulting in the emission of a 50~MeV pion) for the various
optical-potential models are shown in Figure
\ref{fig:PionDistUncertainty}.  In the plot,
\BFIG
\centerline{\psfig{figure=PionDistUncertainty.pstex,height=6.0in,width=6.0in}}
\caption[Effects of uncertainties in the $\pi$-nucleus optical potential on the
vector representation for the coherent process] {Differential cross
section for the coherent pion-photoproduction reaction from
$^{40}\rm{Ca}$ at $E_{\gamma}\!=\!186$~MeV (resulting in the emission
of a 50~MeV pion) using different optical-potential models. All of
these models are equivalent insofar as they fit properties of pionic
atoms and $\pi$-nucleus scattering data. A vector representation for
the elementary part of the amplitude is used.}
\label{fig:PionDistUncertainty}
\end{figure}
our results are labeled full-distortions (solid line) while those of
Carr, Stricker-Bauer, and McManus as CSM (short dashed line); those
obtained with the 4-parameter Kisslinger potential are labeled K1
(long-dashed line) and K2 (dot-dashed line), respectively. It can be
seen from the figure that our calculation differs by at most 30\%
relative to the ones using earlier forms of the optical
potential. Note that these results are computed using the vector
parameterization of the elementary amplitude. Similar calculations
done with the tensor amplitude as can be seen in Figure
\ref{fig:PionDistenUncertainty}, 
\BFIG
\centerline{\psfig{figure=PionDistenUncertainty.pstex,height=6.0in,width=6.0in}}
\caption[Effects of uncertainties in the $\pi$-nucleus optical potential on the
tensor representation for the coherent process] {A similar plot to
Figure \ref{fig:PionDistUncertainty} but now with the tensor
representation for the elementary part of the amplitude.}
\label{fig:PionDistenUncertainty}
\end{figure}
display optical-model uncertainties far smaller (of the order of 5\%)
than the ones reported in Figure \ref{fig:PionDistUncertainty}. In
conclusion, although there seems to be a non-negligible uncertainty
arising from the optical potential, these uncertainties pale in
comparison to the large off-shell ambiguity, to be discussed later on
in this chapter.

\subsection{Distortion Effects for the Eta}

Distortion effects in the $\eta$ coherent process are also crucial in
understanding this process. However, distortions here are not as
strong and dramatic as in the case for the pion. Figure
\ref{fig:etaDistEffec} displays the
\BFIG
\centerline{\psfig{figure=etaDistEffec.pstex,height=6.0in,width=6.75in}}
\caption[Distortion effects on the differential cross section at
various incident photon energies for the $\eta$ coherent process from
$^{40}\rm{Ca}$] {The coherent $\eta$ photoproduction differential
cross section from ${}^{40}$Ca at photon laboratory energies of
$E_{\gamma} = 625, 700$, and $800$~MeV. The solid lines represent the
calculations with no distortions (RPWIA), while the dashed lines
represent the same calculations but now with distortions (RDWIA).}
\label{fig:etaDistEffec}
\end{figure}
coherent $\eta$ photoproduction cross section from ${}^{40}$Ca at
photon laboratory energies of $E_{\gamma} = 625, 700,$ and
$800$~MeV. The solid lines represent the calculations with no
distortions (RPWIA), while the dashed lines represent the same
calculations but now with distortions (RDWIA).

The effect of distortions as a function of incident photon energy is
manifest in the plot. At low $\eta$ energies (see panel one in Figure \ref{fig:etaDistEffec}) the real part of the optical potential is
attractive which creates a competition with the absorptive imaginary
part of the potential that results in a distorted-wave cross section
relatively close to its plane-wave value. At higher energies
(see panel two in Figure \ref{fig:etaDistEffec}) the real part of the optical
potential turns repulsive, and this, in addition to a relatively large
imaginary part, results in a large quenching of the distorted cross
sections relative to their plane-wave estimations. Finally at
$E_\gamma \approx 800$~MeV, the effect of distortions is reduced due
to a distortion unfavorable strong energy dependence in the $\eta N \rightarrow \eta N$
scattering matrix~\cite{bentan90}

\section{Relativistic Effects}

Having discussed the effects of distortions, I turn now to the effect
of relativity. Figure \ref{fig:CoherentRelEff} shows the differential
cross section
\BFIG
\centerline{\psfig{figure=CoherentRelEff.pstex,height=6.0in,width=6.0in}}
\caption[Relativistic effects in the coherent process]
{The Differential cross section for the coherent pion-photoproduction
reaction from $^{40}\rm{Ca}$ at $E_{\gamma}\!=\!168$~MeV calculated
using relativistic and nonrelativistic formalisms for both the tensor
and vector parameterizations.}
\label{fig:CoherentRelEff}
\end{figure}
for the coherent pion-photoproduction reaction from $^{40}\rm{Ca}$ at
$E_{\gamma}\!=\!168$~MeV calculated using relativistic and
nonrelativistic formalisms for both the tensor and vector
parameterizations. The nonrelativistic calculations were obtained by
using the free-space relation to relate the lower component to the
upper one in the bound-nucleon wavefunction, while keeping the upper
component essentially intact apart from a small normalization
correction. This method of constructing the nonrelativistic version is
our best attempt at reproducing standard nonrelativistic calculations
which employ free, on-shell spinors to affect the nonrelativistic
reduction of the elementary amplitude.

Figure \ref{fig:CoherentRelEff} draws a sharp contrast in the way
relativistic effects influence the tensor and vector representations:
the tensor parameterization is very sensitive to relativity while the
vector one shows outright apathy for it. The tensor cross section
experiences a relativistic enhancement by a factor of two. This result
is no surprise once we examine the basic definitions of these
densities. The tensor one is given by the expression:
\begin{equation}
   \rho_{\lower 3pt \hbox{$\scriptstyle T$}}(r) = \sum_{a}^{\rm occ}
   \left({2j_{a}+1 \over 4\pi r^{2}}\right) 2g_{a}(r)f_{a}(r) \,,
   \label{rhoten}
\end{equation}
while the vector one is expressed as:
\begin{equation}
   \rho_{\lower 3pt \hbox{$\scriptstyle V$}}(r) = \sum_{a}^{\rm occ}
   \left({2j_{a}+1 \over 4\pi r^{2}}\right) \Big[g_{a}^2(r) +
   f_{a}^2(r)\Big] \,.
\label{rhovec}
\end{equation}
Note that the tensor density is linear in the lower (or small)
component of the single-particle wavefunction; this is in contrast to
the vector density where the lower component enters as an $(f/g)^{2}$
correction. The mean-field approximation to Walecka model for the
nuclear structure is characterized by the existence of large Lorentz
scalar and vector potentials that are responsible for a substantial
enhancement of the lower components of the single-particle
wavefunctions. This enhancement is at the heart of the
phenomenological success enjoyed by the Walecka model and is usually
referred to as the ``$M^*$ effect'' since the effective mass of the
bound nucleon is greatly reduced due to the presence of the large
attractive scalar potential. Thus, the large relativistic enhancement
of the tensor density represents an inescapable prediction of this
model. As we will see throughout this chapter, this is only one facet
of the off-shell ambiguity in using one of these representations as
opposed to the other one.

\section{Effects of Uncertainties in the Nuclear Structure}

In order to examine the effects of uncertainties in the nuclear
structure, we evaluate the differential cross section using two
versions of Walecka model: QHD-I and QHD-II. As has been indicated in
Chapter \ref{ch:NuclearStructure}, in the QHD-I model the $NN$
interaction is mediated by the $\sigma$ and $\omega$ mesons while in
the QHD-II theory we add the photon ($\gamma$) and the $\rho$-meson
contributions as well. Figure
\ref{fig:CoherentUncNucStr} shows the results of our calculations.
\BFIG
\centerline{\psfig{figure=CoherentUncNucStr.pstex,height=6.0in,width=6.0in}}
\caption[Effects of uncertainties in the nuclear structure]
{The coherent $\eta$ photoproduction cross section from ${}^{40}$Ca at
a photon laboratory energy of $E_{\gamma}=700$~MeV. The cross section
has been calculated using Walecka QHD-I ($\sigma, \omega$) and QHD-II
($\sigma, \omega, \rho, \gamma$) models.}
\label{fig:CoherentUncNucStr}
\end{figure}
As can be easily seen, the differential cross section is rather
insensitive to which of the two models is used; the results of QHD-I
and QHD-II are within ten percent of each other.

\section{${\bf S_{11}}$ Resonance Suppression in the $\eta$ Coherent
Process}

One of the remarkable results of our study is the suppression of the
$S_{11}$ resonance in the $\eta$ coherent process. Figure
\ref{fig:S11suppresion}
\BFIG
\centerline{\psfig{figure=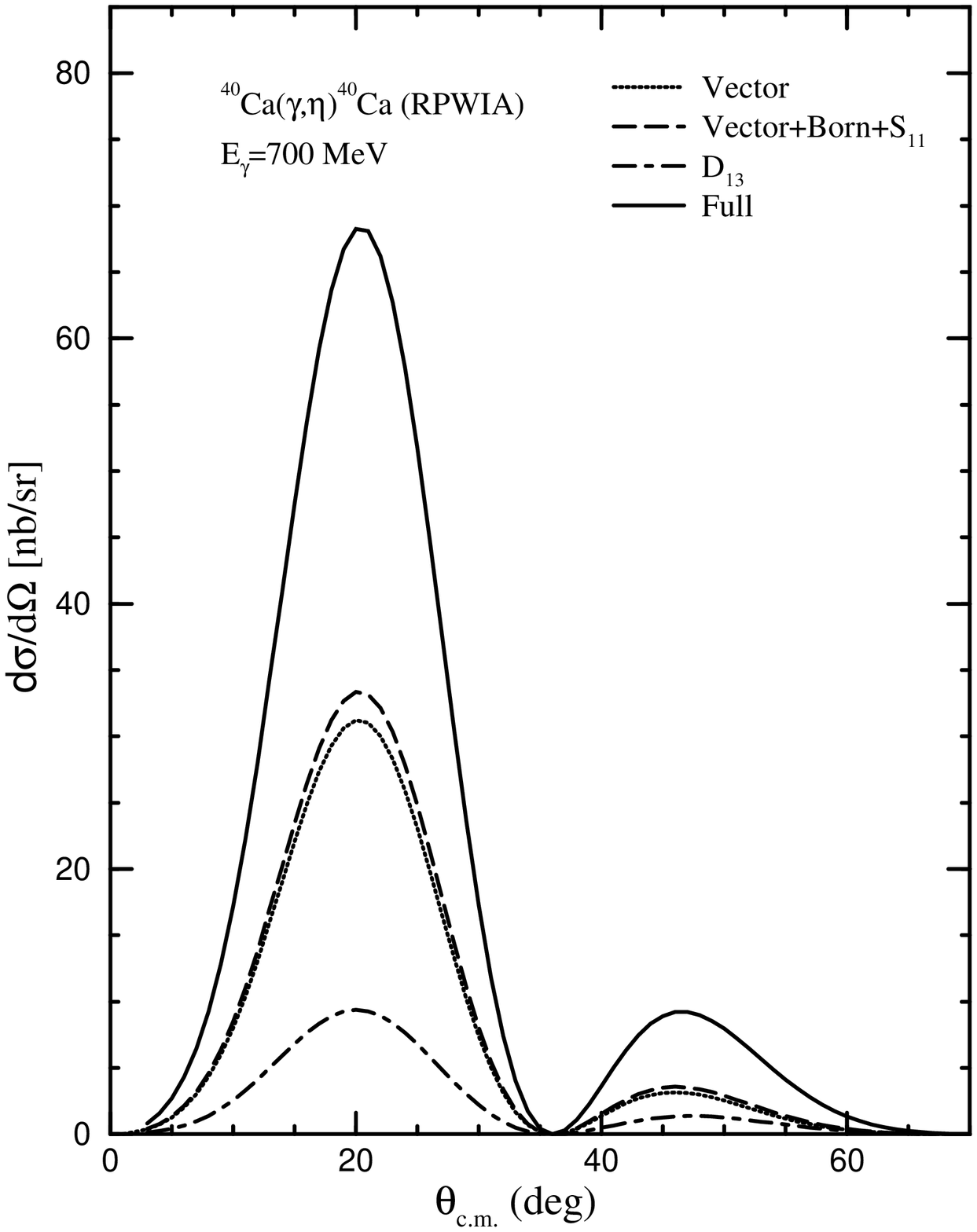,height=6.0in,width=6.0in}}
\caption[The suppression of the
$S_{11}$ resonance in the $\eta$ coherent process] {Breakdown of the
elementary contributions to the coherent $\eta$-photoproduction cross
section from ${}^{40}$Ca at a photon laboratory energy of
$E_{\gamma}=700$~MeV. All curves were generated in a relativistic
plane-wave-impulse approximation.}
\label{fig:S11suppresion}
\end{figure}
displays a breakdown of the elementary contributions to the
differential cross section for ${}^{40}$Ca using 
the RPWIA. It is clear that a
significant portion of the strength arises from the individual
contributions of the $D_{13}$(1520) excitation and the t-channel
exchange of vector mesons, while very little strength is contributed
by the $S_{11}$(1535) resonance or the Born terms. Also note that the
constructive interference between the $D_{13}$(1520) resonance and
vector-meson exchange results in a cross section substantially
stronger than their incoherent sum.

This result is remarkable because the $S_{11}$ resonance which almost
perfectly dominates the cross section for the elementary process (see
Figure \ref{fig:EtaElementaryProcessData}), is strongly suppressed in
the coherent process. Thus, the contributions that languish in the
darkness of the background in the elementary interaction, spring to
life in the coherent process. Consequently, one in principle can use
the coherent process to probe and study contributions that cannot be
disentangled in elementary processes. For example, it is not clear
whether other contributions from resonances that are not included in the
elementary process, can be significant in this process. The coherent process can provide us
with a tool to determine these background contributions.

\section{Effects of Isospin}

In most of our calculations of the coherent process, we ignored the
isovector component of the reaction amplitude (see Section
\ref{sec:isospin}). We have done so because we studied this interaction for nuclei with
equal number of protons and neutrons. One valid
criticism of our work\footnote{Most enthusiastically expressed by
D. Robson.} is that although these nuclei have small isovector
densities, there still may be a large isovector component in the
reaction amplitude because the $S_{11}$ resonance, which dominates the
elementary process but is suppressed in the isoscalar channel, may
have a large contribution in the isovector part of the amplitude. For
this purpose we have examined the effect of this isovector component
as can be seen in Figure \ref{fig:CoherentIsoSpineff}. The figure
decisively shows
\BFIG
\centerline{\psfig{figure=CoherentIsoSpineff.pstex,height=6.0in,width=6.0in}}
\caption[Effects of the isovector component in the reaction amplitude]
{The coherent $\eta$ photoproduction cross section from ${}^{40}$Ca at
a photon laboratory energy of $E_{\gamma}=700$~MeV. The cross section
has been evaluated with no isovector component (solid line) and with
isovector component (dashed line).}
\label{fig:CoherentIsoSpineff}
\end{figure}
that this component is insignificant even for ${}^{40}$Ca; the
nucleus with the largest isovector density among the nuclei that we
studied here.

\section{Nuclear Dependence of the Coherent Process}

The coherent $\eta$ photoproduction differential cross section from
${}^{4}$He, ${}^{12}$C, and ${}^{40}$Ca is shown in Figure
\ref{fig:NuclearDep} at
\BFIG
\centerline{\psfig{figure=NuclearDep.pstex,height=6.0in,width=6.6in}}
\caption[Nuclear dependence in the coherent process]
{The nuclear dependence of the coherent $\eta$-photoproduction cross
section at photon laboratory energies of $E_{\gamma}=625$, $700$ and
$800$~MeV in a relativistic distorted-wave-impulse approximation
(RDWIA). The calculations have been achieved for three nuclei:
${}^{40}$Ca, ${}^{12}$C, and ${}^{4}$He. ${}^{12}$C has the largest
cross section of
the three nuclei, while ${}^{4}$He has the smallest. This result for ${}^{12}$C is an example of the open-shell
effect (see Section \ref{sec:offshellambiguity}). All calculations have been made using the tensor
parameterization.}
\label{fig:NuclearDep}
\end{figure}
photon laboratory energies of $625$, $700$, and $800$~MeV,
respectively. All of these calculations have been achieved using the
tensor parameterization. The open-shell effect discussed in Section
\ref{sec:offshellambiguity} is manifest in this diagram where the
cross section for ${}^{12}$C is significantly larger than that for
${}^{4}$He and ${}^{40}$Ca. Note that ${}^{4}$He and ${}^{40}$Ca are
closed-shell (spin-saturated) nuclei while ${}^{12}$C is an open-shell
nucleus because its $1p^{1/2}$ orbital is empty while the $1p^{3/2}$
orbital is occupied. It is important to stress here that this
open-shell effect is a result of using the tensor representation. It
is not present if we would have used the vector one. In this aspect, this is
yet another manifestation of the off-shell ambiguity.

The relativistic results shown in this figure differ significantly
from those obtained in nonrelativistic calculations. Indeed, Bennhold
and Tanabe~\cite{bentan90} have predicted that ${}^{4}$He would have
the largest cross section of the three nuclei, due to its largest
charge form factor. This, we believe, might have been an important
reason for the selection of ${}^{4}$He for the first experimental measurement of
the coherent process~\cite{ahrens93}. However, this finding is at odds with our
relativistic results, which instead show ${}^{4}$He to have the
smallest cross section. There are two main reasons for this
difference. First, in relativistic calculations the ratio of
upper-to-lower components is determined dynamically in the Walecka
model, rather than from its free-space relation. Second, the
elementary $\eta N$ interaction used in this work~\cite{benm92,bmz95}
is different from the one used by Bennhold and Tanabe~\cite{bentan90},
in particular the non-resonant contributions were not considered in
the latter. Although both models seems to give an adequate description
of the elementary process, important differences emerge in the
calculation of the coherent reaction. This is primarily due to the
fact that the coherent process from spin-saturated nuclei becomes
insensitive to the dominant $S_{11}$(1535) intermediate-resonance
contribution, and therefore quite sensitive to the details of other
resonant and non-resonant background contributions such as the
$D_{13}(1520)$ and vector mesons. Note that our calculations for
${}^{4}$He are similar to the nonrelativistic ones reported recently
by Fix and Arenh\"ovel's~\cite{fix97}. However, this agreement seems
to be fortuitous, since neither their nuclear-structure model nor
their elementary amplitude are similar to ours; their coherent process
is dominated by $\omega$-meson exchange, while ours contains, in
addition, a significant contribution from the $D_{13}$(1520)
resonance.

Finally, there are no experimental data available for the $\eta$ coherent
process. However, the theoretical studies of this process have motivated
considerable experimental interest which have culminated in
an attempt to measure the coherent $\eta$ photoproduction cross
section from ${}^{4}$He at the Mainz Microtron
facility~\cite{ahrens93}. Possibilities for extensions to
higher energies and other nuclei exist, both at the Bonn ELSA
facility and at TJNAF~\cite{adampro99}.

\section{Effects of Off-Shell Ambiguity}
\label{sec:off-shell}

Now we come to the major obstacle in this work: the off-shell
ambiguity.  We have already in the previous sections discussed some
aspects of this ambiguity as it relates to distortion and relativistic effects and
to nuclear dependence. Here, we will discuss other angles of the
problem.

We start first by presenting in Figure \ref{fig:Offshelldiffcross} the
\BFIG
\centerline{
\psfig{figure=Offshelldiffcross.pstex,height=6.0in,width=6.0in}}
\caption[Effect of the off-shell ambiguity on the differential cross section]
{Differential cross section for the coherent pion photoproduction
reaction from $^{40}\rm{Ca}$ at $E_{\gamma}\!=\!230$~MeV with (RDWIA)
and without (RPWIA) pionic distortions. Tensor and vector
parameterizations of the elementary amplitude are used.}
\label{fig:Offshelldiffcross}
\end{figure}
differential cross section for the coherent photoproduction of neutral
pions from $^{40}\rm{Ca}$ at a photon energy of
$E_{\gamma}\!=\!230$~MeV. Both tensor and vector parameterizations of the elementary
amplitude were used, and the cross section was calculated with (RDWIA)
and without (RPWIA) pionic distortions. The off-shell ambiguity is
immense; factors of two (or more) are observed when comparing the
vector and tensor representations. It is important to stress that
these calculations were done by using the same nuclear-structure
model, the same pionic distortions, and two elementary amplitudes that
are identical on-shell. The very large discrepancy between the two
theoretical models emerges from the dynamical modification of the
Dirac spinors in the nuclear medium, and not from changes to the
elementary production amplitude (assessing the impact of medium
modifications to the elementary amplitude remains an important open
question). Moreover, the large discrepancy between the calculations
cannot be attributed to an improper treatment of gauge invariance, as
gauge invariance is strictly maintained in all of our calculations
(see Equations \ref{lorentzrep} and \ref{lorentzrepcoef}).

We have compared our theoretical results to preliminary and
unpublished data (not shown) provided to us courtesy of
B.~Krusche~\cite{krusche98}. The data follows the same shape as our
calculations but the experimental curve seems to straddle the
two calculations, although the vector calculation appears closer to
the experimental data. This behavior---a closer agreement of the
vector calculation to data---has been observed in all of the
comparisons that we have done so far.

In Figure \ref{fig:OffshelvsE} we present results for the differential
\BFIG
\centerline{
\psfig{figure=OffshelvsE.pstex,height=6.0in,width=6.7in}}
\caption[Effect of the off-shell ambiguity on the energy dependence of
the differential cross section] {Differential cross section for the
coherent pion photoproduction reaction for $^{40}\rm{Ca}$ at a variety
of photon energies using a RDWIA formalism.  Tensor (dashed line) and
vector (solid line) parameterizations of the elementary amplitude are
used.}
\label{fig:OffshelvsE}
\end{figure}
cross section from $^{40}\rm{Ca}$ at a variety of photon energies,
while in Figure \ref{fig:OffshellTotCross} we display results for the
total cross
\BFIG
\centerline{
\psfig{figure=OffshellTotCross.pstex,height=6.0in,width=6.7in}}
\caption[Effect of the off-shell ambiguity on the total cross section]
{Total cross section for the coherent pion photoproduction reaction
	 from $^{40}\rm{Ca}$ as a function of the photon energy with
	 (right panel) and without (left panel) pionic distortions.
	 Tensor (dashed line) and vector (solid line)
	 parameterizations of the elementary amplitude are used.}
\label{fig:OffshellTotCross}
\end{figure}
section. By examining these graphs one can infer that the tensor
parameterization always predicts a large enhancement of the cross
section---irrespective of the photon incident energy and the
scattering angle---relative to the vector predictions. Moreover, the
convolution of the tensor and vector densities with the pionic
distortions gives rise to similar qualitative, but quite different
quantitative, behavior on the energy dependence of the corresponding
coherent cross sections.

In Figure \ref{fig:C12OffshellCompExp} we show the differential cross
section
\BFIG
\centerline{
\psfig{figure=C12OffshellCompExp.pstex,height=6.0in,width=6.0in}}
\caption[Off-shell ambiguity and comparison with experimental data for $^{12}$C]
{Differential cross section for the coherent pion photoproduction
	 reaction from $^{12}\rm{C}$ at $E_{\gamma}\!=\!173$
	 MeV. Tensor (dashed line) and vector (solid line)
	 parameterizations of the elementary amplitude are used. The
	 experimental data are from Ref.~\protect\cite{gothe95}.}
\label{fig:C12OffshellCompExp}
\end{figure}
for the coherent process from $^{12}$C at a photon energy of
$E_{\gamma}\!=\!173$~MeV. The off-shell ambiguity for this case is
striking; at this energy the tensor result is five times larger than
the vector prediction.  The additional enhancement observed here
relative to $^{40}$Ca is easy to understand on the basis of the
open-shell effect. Figure \ref{fig:C12OffshellCompExp} also shows a
comparison of our results with the experimental data of
Ref.~\cite{gothe95}. It is clear from the figure that the vector
representation is closer to the data; note that the tensor calculation
has been divided by a factor of five.  Even so, the vector calculation
also overestimates the data by a considerable amount.

For further comparison with experimental data we have calculated the
coherent cross section from ${}^{12}$C at photon energies of
$E_{\gamma}\!=\!235$, $250$, and $291$~MeV. In Table \ref{table1} we
\begin{table}
  \caption{Maxima of the differential cross section (in $\mu$b) for
	   the coherent pion photoproduction reaction from ${}^{12}$C
	   at various energies.}
\vspace{.5cm}
\begin{center}
\thicklines 
\begin{tabular}{cccc}
\hline
\hline 
$E_{\gamma}$~(MeV) & Tensor & Vector& Experiment \\
\hline
$235$ & $694$ & $116$ & $105$ \\ 
$250$ & $731$ & $133$ & $190$ \\ 
$291$ & $786$ & $186$ & $175$ \\
\hline
\hline	   
\end{tabular} 
\end{center}
\label{table1}
\end{table}
have collated our calculations with experimental data published by Arends and collaborators~\cite{Arends83} for $E_{\gamma}\!=\! 235$
and $291$ MeV, and with data presented by Booth~\cite{Booth1978} and
Nagl, Devanathan, and \"Uberall~\cite{ndu91} for $E_{\gamma
\rm{lab}}\!=\!250$~MeV.  The experimental data exhibits similar
patterns as our calculations (not shown) but the values of the maxima
of the cross section are different. The tensor calculations continue
to predict large enhancement factors (of five and more) relative to
the vector calculations. More importantly, these enhancement factors
are in contradiction with experiment. The experimental data appears to
indicate that the maximum in the differential cross section from
$^{12}\rm{C}$ is largest at about $250$ MeV, while our calculations
predict a maximum around $295$ MeV.  It is likely that this energy
``shift'' might be the result of the formation and propagation of the
$\Delta$-resonance in the nuclear medium. Clearly, in an
impulse-approximation framework, medium modifications to the
elementary amplitude---arising from changes in resonance
properties---can not be accounted for. Yet, a binding-energy
correction of about $40$~MeV due to the $\Delta$-nucleus interaction
has been suggested before~\cite{Gaarde91}. Indeed, such a shift would also explain the
discrepancy in the position of our theoretical cross sections in
$^{40}$Ca, relative to the (unpublished) data by Krusche and
collaborators~\cite{krusche98}.  Moreover, such a shift---albeit of
only 15 MeV---was invoked by Peters, Lenske, and Mosel~\cite{Peters98b}
in their recent calculation of the coherent pion-photoproduction cross
section. Yet, a detailed study of modifications to hadronic properties
in the nuclear medium must go beyond the impulse approximation; a
topic outside the scope of the present work. However, a brief
qualitative discussion of possible violations to the impulse
approximation is given in the next section.

We conclude this section by presenting in Figures
\ref{fig:C12Offshelltheta60} and
\BFIG
\centerline{
\psfig{figure=C12Offshelltheta60.pstex,height=6.0in,width=6.0in}}
\caption[Additional figure for $^{12}$C showing the Off-shell ambiguity and comparison with experimental data]
{Differential cross section for the coherent pion photoproduction
	 reaction from $^{12}\rm{C}$ as a function of photon energy at
	 a fixed laboratory angle of $\theta_{\rm lab}=60^{\circ}$,
	 with and without pionic distortions. Tensor (dashed lines)
	 and vector (solid lines) parameterizations of the elementary
	 amplitude are used. The experimental data are from
	 Ref.~\protect\cite{schmitz96}.}
\label{fig:C12Offshelltheta60}
\end{figure}
\ref{fig:C12EnergyShift}, a comparison between our plane- and distorted-wave
\BFIG
\centerline{
\psfig{figure=C12EnergyShift.pstex,height=6.0in,width=6.0in}}
\caption[Energy shift for the differential cross section of $^{12}$C]
{Differential cross section for the coherent pion photoproduction
	 reaction from $^{12}\rm{C}$ as a function of photon energy at
	 a fixed laboratory angle of $\theta_{\rm lab}=60^{\circ}$,
	 with pionic distortions and using only a vector
	 parameterization of the elementary amplitude.  The same
	 calculation---including a shift of 25 MeV is also included
	 (dashed line). The experimental data are from
	 Ref.~\protect\cite{schmitz96}.}
\label{fig:C12EnergyShift}
\end{figure}
calculations with experimental data for the coherent cross section
from $^{12}$C as a function of photon energy for a fixed angle of
$\theta_{\rm lab}\!=\!60^{\circ}$. The experimental data from MAMI is
contained in the doctoral dissertation of M. Schmitz~\cite{schmitz96}.

Perhaps the most interesting feature in these figures is the very good
agreement between our RDWIA calculation using the vector
representation and the data---if we were to shift our results by
+25~MeV. Indeed, this effect is most clearly appreciated in Figure
\ref{fig:C12EnergyShift} where the shifted calculation is now
represented by the dashed line. In our treatment of the coherent
process, the detailed shape of the cross section as a function of
energy results from a delicate interplay between several effects
arising from: a) the elementary amplitude---which peaks at the
position of the delta resonance ($E_{\gamma}\simeq 340$~MeV from a
free nucleon and slightly lower here because of the optimal
prescription~\cite{pisabe97}), b) the nuclear form factor---which
peaks at low-momentum transfer, and c) the pionic distortions---which
strongly quench the cross sections at high energy, as more open
channels become available. We believe that the pionic distortions (see
Section \ref{sec:distor}) as well as the nuclear form factor have been
modeled accurately in the present work. The elementary amplitude,
although obtained from a recent phase-shift analysis by the VPI
group~\cite{Arnphn}, remains one of the biggest uncertainties, as no
microscopic model has been used to estimate possible medium
modifications to the on-shell amplitude.  Evidently, an important
modification might arise from the production, propagation, and decay
of the $\Delta$-resonance in the nuclear medium. Indeed, a very
general result from hadronic physics, obtained from analyses of
quasielastic $(p,n)$ and $({}^{3}He,t)$ experiments~\cite{Gaarde91},
is that the position of the $\Delta$-peak in nuclear targets is lower
relative to the one observed from a free proton target.
 
However, it is also well known that such a shift is not observed when
the $\Delta$-resonance is excited electromagnetically~\cite{Gaarde91}.
This apparent discrepancy has been attributed to the different dynamic
responses that are being probed by the two processes. In the case of
the hadronic process, it is the (pion-like) spin-longitudinal response
that is being probed, which is known to get ``softened'' (shifted to
lower excitation energies) in the nuclear medium. Instead,
quasielastic electron scattering probes the spin-transverse
response---which shows no significant energy shift.  Unfortunately,
in our present local-impulse-approximation treatment it becomes
impossible to assess the effects associated with medium modifications
to the $\Delta$-resonance. A detailed study of possible violations to
the impulse approximation and to the local assumption remains an
important open problem for the future. A qualitative discussion is
presented in the next section.

\section{Violations to the Impulse Approximation}
\label{sec:impviol}

In this section, we address an additional ambiguity in the formalism,
namely, the use of the impulse approximation. The basic assumption
behind the impulse approximation is that the interaction in the medium
is unchanged relative to its free-space value. The immense
simplification that is achieved with this assumption is that the
elementary interaction now becomes model independent, as it can be
obtained directly from a phase-shift analysis of the experimental data
(see, for example, Ref.~\cite{Arnphn}). The sole remaining question to
be answered is the value of $s$ at which the elementary amplitude
should be evaluated, as now the target nucleon is not free but rather
bound to the nucleus (see Figure \ref{fig:CoherentImpulse}). This
question is resolved by using the ``optimal'' prescription of Gurvitz,
Dedonder, and Amado~\cite{GDA79}, which suggests that the elementary
amplitude should be evaluated in the Breit frame. Then, this optimal
form of the impulse approximation leads to a factorizable and local
scattering amplitude---with the nuclear-structure information
contained in a well-determined form factor. Moreover, as the
final-state interaction between the outgoing meson and the nucleus is
well constrained from other data, a parameter-free calculation of the
coherent photoproduction process ensues.

This form of the impulse approximation has been used with great
success in hadronic processes, such as in $(p,p')$ and $(p,n)$
reactions, and in electromagnetic processes, such as in electron
scattering. Perhaps the main reason behind this success is that the
elementary nucleon-nucleon or electron-nucleon interaction is mediated
exclusively by $t$-channel exchanges---such as arising from $\gamma$-,
$\pi$-, or $\sigma$-exchange. This implies that the local
approximation (i.e., the assumption that the nuclear-structure
information appears exclusively in the form of a local nuclear form
factor) is well justified. For the coherent process this would also be
the case if the elementary amplitude would be dominated by the
exchange of mesons, as in the last Feynman diagram in Figure
\ref{fig:CoherentFeyn}. However, it is well known---at least for the
\BFIG
\centerline{
\psfig{figure=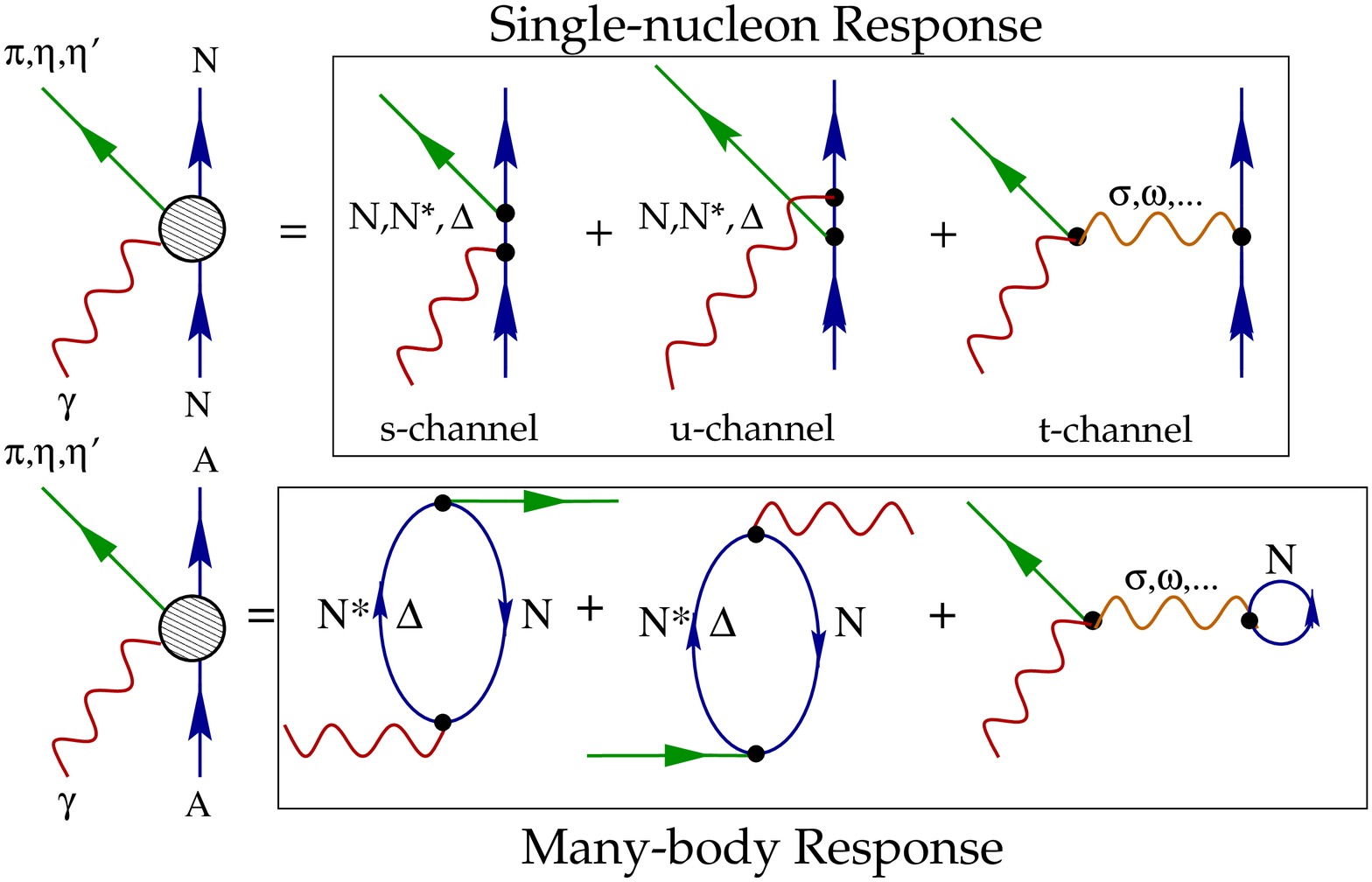,height=6.5in,width=6.5in}}
\caption[Characteristic s-, u-, and t-channel Feynman 
	 diagrams for the photoproduction of pseudoscalar mesons from
	 a single nucleon and---coherently---from the nucleus]
	 {Characteristic s-, u-, and t-channel Feynman diagrams for
	 the photoproduction of pseudoscalar mesons from a single
	 nucleon (upper panel) and---coherently---from the nucleus
	 (lower panel).}
\label{fig:CoherentFeyn}
\end{figure}
kinematical region of current interest---that the elementary
photoproduction process is dominated by resonance ($N^{\star}$ or
$\Delta$) formation, as in the $s$-channel Feynman diagram of Figure
\ref{fig:CoherentFeyn}. This suggests that the coherent reaction
probes, in addition to the nuclear density, the polarization structure
of the nucleus (depicted by the ``bubbles'' in Figure \ref{fig:CoherentFeyn}). As
the polarization structure of the nucleus is sensitive to the ground-
as well as to the excited-state properties of the nucleus, its proper
inclusion could lead to important corrections to the local
impulse-approximation treatment. Indeed, Peters, Lenske, and Mosel
have lifted the local assumption and have reported---in contrast to
all earlier local studies---that the $S_{11}(1535)$ resonance does
contribute to the coherent photoproduction of $\eta$-meson for
open-shell (non-spin-saturated) nuclei like $^{12}\rm{C}$~\cite{Peters98b}. They also reported a significant smearing of the cross
section in the case of the $\pi$ coherent process~\cite{Peters98a}. Clearly,
understanding these additional contributions to the coherent process
is an important area for future work.

\chapter{Conclusions for the Coherent Process}
\label{ch:ConclusionsCoherent}

We have studied the coherent photoproduction of pseudoscalar mesons in
a relativistic-impulse-approximation approach. We have placed special
emphasis on the ambiguities underlying most of the current theoretical
approaches. Although our conclusions are of a general nature, we have
focused our discussions on the photoproduction of neutral pions due to
the ``abundance'' of data relative to the other pseudoscalar channels.

We have employed a relativistic formalism for the elementary amplitude
as well as for the nuclear structure. We believe that, as current
relativistic models of nuclear structure rival some of the most
sophisticated nonrelativistic ones, there is no longer a need to
resort to a nonrelativistic reduction of the elementary
amplitude. Rather, the full relativistic structure of the coherent
amplitude should be maintained~\cite{pisabe97,apsm98,raddad99}.

We have also extended our treatment of the pion-nucleus interaction to
the $\Delta$-resonance region. As most of the details about the optical
potential will be reported shortly~\cite{radcar}, we summarize briefly
some of our most important findings. As expected, pionic distortions
are of paramount importance. Indeed, we have found a factor-of-two
enhancement (at low energies) and up to a factor-of-five reduction
(at high energies) in the coherent cross section relative to the
plane-wave values. Yet, ambiguities arising from the various choices
of optical-model parameters are relatively small; of at most 30\%.

We have found important discrepancies vis-a-vis nonrelativistic
results~\cite{bofmir86,cek87,bentan90,ndu91,tryfik94,fix97}. Part of
these discrepancies stem from the fact that we have used a fully
relativistic approach---with no resort to a nonrelativistic
reduction. Moreover, the elementary amplitudes used in our model are
different from those used in other theoretical calculations. We
found also that the cross section is sensitive to two
nuclear-structure quantities: {\it i)} the ground-state vector density
and {\it ii)} the ground-state tensor density. The tensor density is
as fundamental as the vector density used in the nonrelativistic
treatments, although it is not as well constrained by experiment.

By far the largest uncertainty in our results emerges from the
ambiguity in extending the many---actually infinite---equivalent
representations of the elementary amplitude off the mass shell. While
all these choices are guaranteed to give identical results for
on-shell observables, they yield vastly different predictions
off-shell.  Yet, it is worth mentioning that the off-shell ambiguity
emerges mainly from our insistence in using the impulse
approximation. With an effective microscopic model---calibrated to
reproduce two- and many-body scattering amplitudes---the off-shell
ambiguity can, to a large extent, be removed. This task, however,
remains a formidable one---forcing us, as well as most existing
theoretical approaches, to rely on the impulse approximation.

In this work we have investigated two on-shell-equivalent
representations of the elementary amplitude: a tensor and a
vector. The tensor representation employs the ``standard'' form of the
elementary amplitude~\cite{bentan90,cgln57} and generates a coherent
photoproduction amplitude that is proportional to the isoscalar tensor
density. However, this form of the elementary amplitude, although
standard, is not unique. Indeed, through a simple manipulation of
operators between on-shell Dirac spinors, the tensor representation
can be transformed into the vector one, so-labeled because the
resulting coherent amplitude becomes proportional now to the isoscalar
vector density. The tensor and vector densities were computed in a
self-consistent, mean-field approximation to the Walecka
model~\cite{serwal86}. The Walecka model is characterized by the
existence of large Lorentz scalar and vector potentials that are
responsible for a large enhancement of the lower components of the
single-particle wave functions. This so-called
``$M^{\star}$-enhancement'' generates a large increase in the tensor
density, as compared to a scheme in which the lower component is
computed from the free-space relation. No such enhancement is observed
in the vector representation, as the vector density is insensitive to
the $M^{\star}$-effect. As a result, the tensor calculation predicts
coherent photoproduction cross sections that are up to an order of magnitude larger than the vector results.  These large
enhancement factors are not consistent with existent experimental
data. Still, it is important to note that the vastly different
predictions of the two models have been obtained using the same pionic
distortions, the same nuclear-structure model, and two sets of
elementary amplitudes that are identical on-shell.

Finally, we have addressed---in a qualitative fashion---violations to the
impulse approximation. In the impulse approximation one assumes that
the elementary amplitude may be used without modification in the
nuclear medium. Moreover, by adopting the optimal prescription of
Ref.~\cite{GDA79}, one arrives at a form for the coherent amplitude
that is local and factorizable.  Indeed, such an optimal form has been
used extensively---and with considerable success---in electron and
nucleon elastic scattering from nuclei. We suggested here that the
reason behind such a success is the $t$-channel--dominance of these
processes. In contrast, the coherent-photoproduction process is
dominated by resonance formation in the $s$-channel. In the nuclear
medium a variety of processes may affect the formation, propagation,
and decay of these resonances. Thus, resonant-dominated processes may
not be amenable to treatment via the impulse-approximation. Further,
in $s$-channel--dominated processes, it is not the local nuclear
density that is probed, but rather, it is the (non-local) polarization
structure of the nucleus.  This can lead to important deviations from
the naive local picture.  Indeed, by relaxing the local assumption,
Peters and collaborators have reported a non-negligible contribution
from the $S_{11}(1535)$ resonance to the coherent 
$\eta$ process for open-shell nuclei~\cite{Peters98b}, and a significant smearing of the cross
section in the case of the $\pi$ coherent process~\cite{Peters98a}.

In summary, we have studied a variety of sources that challenge
earlier studies of the coherent photoproduction of pseudoscalar
mesons. Without a clear understanding of these issues, erroneous
conclusions are likely to be extracted from the wealth of experimental
data that will soon become available. Undoubtedly, there is still a
lot of work to be done both experimentally and theoretically. Indeed,
many challenging and interesting lessons have yet to be learned before
a deep understanding of the coherent-photoproduction process will
emerge. We hope that with the advent of new powerful and sophisticated
facilities, such as TJNAF and MAMI, the validity of the different
theoretical models can be tested.

\chapter{Theory of the Quasifree Meson Photoproduction
from Nuclei}
\label{ch:QuasifreeTheory}

This chapter and the next two will be
devoted to a study of the theory of the quasifree meson photoproduction
from nuclei. This process consists of a photon ($\gamma$-ray)
incident on a nucleus. The photon interacts with the nucleus and as a
result a pseudoscalar meson is produced (like $K^+$, $\pi$, or $\eta$)
through knocking out one of the nucleons. In the case of
the $K^+$ quasifree process that we will study here, a proton is
knocked out into one of its excited states: the $\Lambda$
hyperon. Thus, we start the interaction with a photon and some nucleus, and end up with a meson, a free nucleon or a
hyperon, and a new recoil nucleus. This process is labeled as
``quasifree'' because the interaction occurs in a similar kinematic
setting to the free process of $N(
\gamma , PS\,\,meson )N (Y)$. Specifically, the energy transfer $\omega$ is
related to the momentum transfer ${\mathbf q}$ (as in the free
process) according to the equation
\begin{equation}
\omega\!=\!\sqrt{{\mathbf q}^2+M_{\Lambda}^{2}}-M_{N}\;.
\end{equation}
This equation defines what is usually called the ``quasifree condition''.

Furthermore, this process is perceived to take place from only one of
the nucleons in the nucleus. In this aspect, it is identical to the
elementary process except in the fact that the target nucleon is bound
as opposed to being free. Due to this similarity, it
is no surprise that this interaction is our best attempt to obtain
insights into the nature of the elementary process in slightly
different circumstances.

\section{Basic Ingredients}

As in the case for the coherent process, we employ the relativistic
impulse approximation. However, we do not incorporate any distortions to
the emitted meson or to the outgoing nucleon (hyperon). In other words, we study this
interaction in the framework of the relativistic plane-wave impulse
approximation (RPWIA). Figure \ref{fig:QuasifreeImpulse} provides a
schematic diagram for the kaon quasifree process within this approximation.
\BFIG
\centerline{\psfig{figure=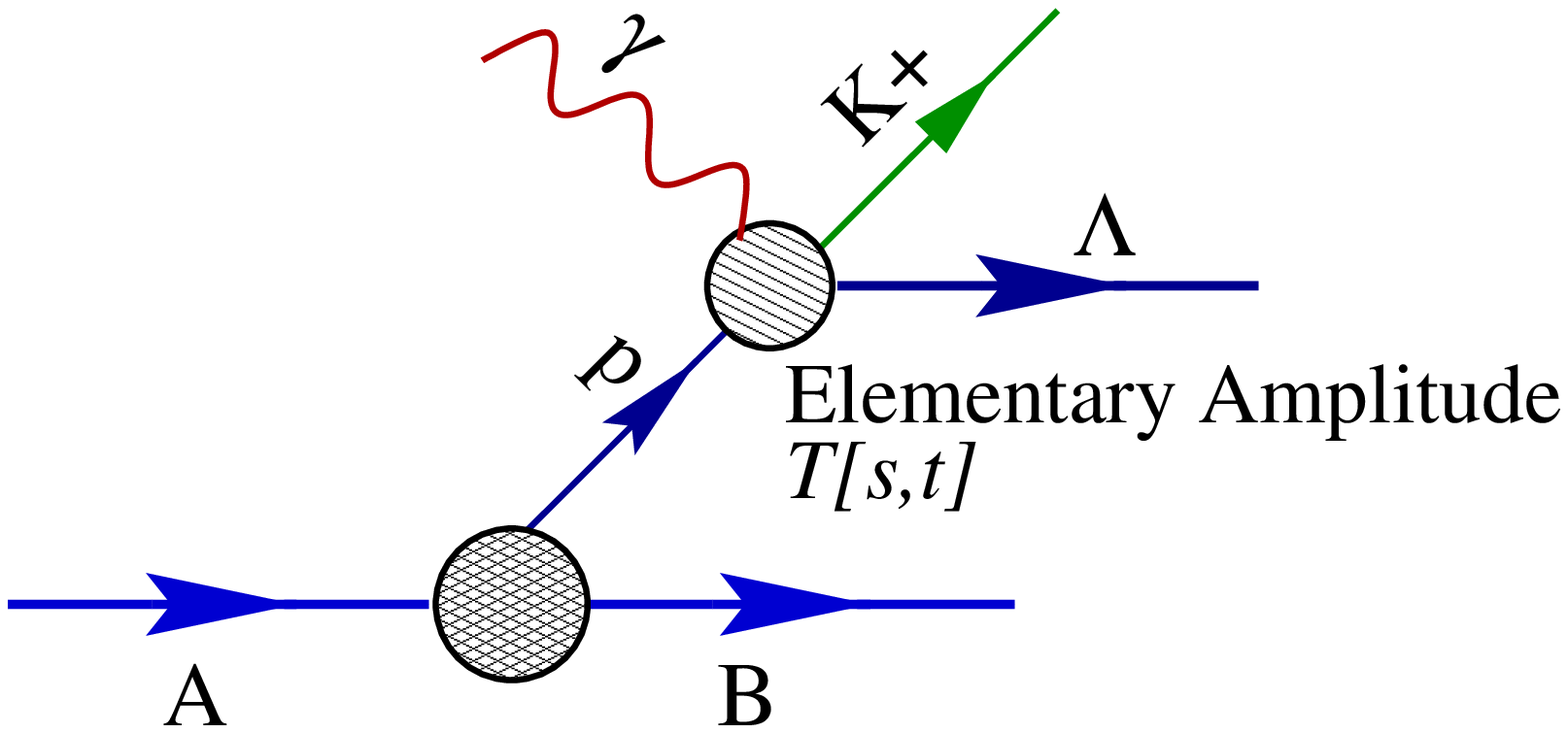,height=9.0cm,width=14.5cm}}
\caption[Schematic diagram for the quasifree process $A(\gamma,K^+
\Lambda)B$]{A representation of the quasifree photoproduction 
	  of a $\Lambda$-hyperon in a plane-wave impulse-approximation
	  approach.}
\label{fig:QuasifreeImpulse}
\EFIG
The rationale for not including distortions is due to our interest in
the polarization observables which are insensitive to
distortion effects. Indeed, earlier nonrelativistic
calculations~\cite{lwb93,lwbt96,blmw98} have demonstrated that two
important polarization observables --- the 
recoil polarization of the ejected nucleon (hyperon) and the photon asymmetry ---
are largely insensitive to distortion effects. Moreover, they seem to
be also independent of the mass of the target nucleus. Polarization observables carry the richest information about the fundamental
physics in this process, and they are far more effective
discriminators of subtle dynamics than the unpolarized cross
section. Finally, and in practical terms, ignoring distortions
results in an enormous simplification in the formalism of this
process.

We maintain the full relativistic structure whether in the elementary photoproduction or in the nuclear
structure. This approach forms a significant departure from the
traditional nonrelativistic studies~\cite{lwb93,lwbt96,blmw98}. Indeed, RPWIA calculations have been successful in
identifying physics not present at the nonrelativistic level. For
example, relativistic effects have been shown to contaminate any
attempt to infer color transparency from a measurement of the
asymmetry in the $(e,e'p)$ reaction~\cite{gp94}. Further, the
well-known factorization limit of nonrelativistic plane-wave
calculations has been shown to break down due to the presence of
negative-energy components in the bound-nucleon
wavefunction\cite{cdmu98}.

\section{Observables}

The differential cross section is derived using well-established
procedures~\cite{ManSha}. Note however that we have here three
particles in the final state as opposed to two in the coherent
process. As a result, the differential cross section (in the lab
system) is expressed as:
\begin{equation}
{\left[ d^5 \sigma (s^{\prime},\epsilon)\right]}_{\rm lab}=
   \frac{1}{2 E_\gamma} \; (2\pi)^4 \; \delta^4 (k + p_A - k^{\prime} -
   p^{\prime} - p_B) \; {\big|{\cal M}\big|}^2 \;
\frac{d^3{\bf k}^{\prime}}{(2\pi)^3 2E_{{\bf k}^{\prime}}} 
\frac{M_{N (\Lambda)}d^3 {\bf p}^{\prime}}{(2\pi)^3 2E_{{\bf p}^{\prime}}} 
\frac{d^3 {\bf p}_B}{(2\pi)^3}\;, 
\label{d5sigma1}
\end{equation}
where $k$ is the four-momentum of the incident photon, while
$k^{\prime}$ and $p^{\prime}$ are the momenta of the produced meson and
nucleon (hyperon), respectively. Here, $p_A$($p_B$) represents the
momentum of the target(residual) nucleus. Finally, $s^{\prime}$ is the
spin of the emitted nucleon (hyperon), $\epsilon$ is the polarization of
the incident photon, $M_{N (Y)}$ is the nucleon (hyperon) mass, and ${\cal M}$ is the
transition matrix element. 

By integrating over the delta function, we
obtain the following form for the differential cross section:
\begin{equation}
\left(\frac{d^5\sigma (s^{\prime},\epsilon)}
  {d{k}^{\prime}d\Omega_{{\bf k}^{\prime}} d\Omega_{{\bf
   p}^{\prime}}}\right)_{\rm lab} = \frac{2 \pi}{2 E_\gamma} \mbox{ }
   \frac{|{\bf k}^{\prime}|^2}{(2\pi)^3\,2E_{{\bf k}^{\prime}}}\mbox{
   } \frac{M_{N (Y)}|{\bf p}^{\prime}|}{(2\pi)^3}\mbox{ } {\big|{\cal
   M}\big|}^2 \;. \label{d5sigma}
\end{equation}

The unpolarized differential expression can be obtained by summing
over the two possible components of the spin of the nucleon (hyperon) and
averaging over the transverse photon polarization.  That is,
\begin{equation}
  \left(\frac{d^5\sigma} {d{k}^{\prime}d\Omega_{{\bf k}^{\prime}}
  d\Omega_{{\bf p}^{\prime}}}\right)_{\rm lab} =
  \frac{1}{2}\sum_{s^{\prime},\epsilon} \left(\frac{d^5\sigma
  (s^{\prime},\epsilon)} {d{k}^{\prime}d\Omega_{{\bf
  k}^{\prime}} d\Omega_{{\bf p}^{\prime}}}\right)_{\rm lab} \;.
\end{equation}
Yet, our prime interest in this work is the calculation of
polarization observables: the recoil $N(Y)$-polarization (${\cal
P}$) and the photon asymmetry ($\Sigma$). The former is defined
as~\cite{ndu91,wcc90,dfls96}
\begin{equation}
{\cal P} = \sum_{\epsilon} \left( \frac{{d^5\sigma}(\uparrow) -
 {d^5\sigma}(\downarrow)} {{d^5\sigma}(\uparrow) +
 {d^5\sigma}(\downarrow)} \right)_{\rm lab} \;,
\label{p}
\end{equation}
while the latter by~\cite{lwbt96,blmw98,dfls96}
\begin{equation}
 {\Sigma} = \sum_{s^{\prime}} \left( \frac{{d^5\sigma}(\perp) -
  {d^5\sigma}(\parallel)} {{d^5\sigma}(\perp) +
  {d^5\sigma}(\parallel)} \right)_{\rm lab} \;.
\label{sigma}
\end{equation}
In these expressions $\uparrow$ and $\downarrow$ represent the
projection of the spin of the $N(Y)$-hyperon with respect to the
normal to the scattering plane (${\bf k}\times{\bf k}^{\prime}$),
while $\perp$($\parallel$) represents the out-of-plane(in-plane)
polarization of the photon.

It can be shown that the differential element $d^5\sigma$ can be written as $ Z ^\mu Z_\nu
\epsilon_\mu \epsilon^\nu$, where $Z_\mu$ is some coefficient . This fact allows us to derive a more
useful expression for $\Sigma$ as
\begin{equation}
 {\Sigma} = \left( \frac{ 2 {d^5\sigma}(\perp)/ d{
  k}^{\prime}d\Omega_{{\bf k}^{\prime}} d\Omega_{{\bf p}^{\prime}}}
  {{d^5\sigma}/ d{k}^{\prime}d\Omega_{{\bf k}^{\prime}}
  d\Omega_{{\bf p}^{\prime}}} \right)_{\rm lab} - 1\;.
\label{photnsigmanew}
\end{equation}
Therefore, this observable is related to the ratio of twice the cross
section for out-of-plane polarization divided by the unpolarized cross
section.

\section{Elementary ($\gamma p \rightarrow K^{+}\Lambda$) Amplitude}
\label{sec:kaoneleamp}

As I have indicated above, we will study in this work only the kaon
quasifree process: $A(\gamma,K^+ \Lambda)B$. For the elementary photoproduction amplitude we have used the standard
model-independent parameterization (see Chapter
\ref{ch:ElementaryProcess}) as following:
\begin{equation}
 T(\gamma p \rightarrow K^+ \Lambda) = F^{\alpha \beta}_{T}
 \sigma_{\alpha \beta} + iF_P\gamma_5 + F^{\alpha}_{A} \gamma _\alpha
 \gamma_5 \;,
\label{elemampl}
\end{equation}
where we have tensor, pseudoscalar, and axialvector components.  Note
that in the presently-adopted parameterization, no scalar nor vector
invariants appear.

For our calculations we use various different models for the
elementary process. These include the hadronic model developed by
Williams, Ji, and Cotanch\cite{wcc90}. These authors impose crossing
symmetry in their model to develop phenomenologically consistent
strong-coupling parameterizations which simultaneously describe the
kaon-photoproduction and radiative-capture reactions. Although these are
theoretically sound, other choices for the elementary amplitude ---
more sophisticated and up to date --- have also been adopted. In
particular, we use the ``Saclay-Lyon-Collaboration'' model developed
by David, Fayard, Lamot, and Saghai\cite{dfls96}. This model is based
on an isobaric treatment using low-order Feynman amplitudes that
include nucleonic (spin $\leq$~5/2), hyperonic (spin 1/2), and kaonic
resonances. Recently, this model has been extended by T. Mizutani,
C. Fayard, G.H. Lamot, and B. Saghai to incorporate off-shell effects
implied in any treatment of fermions with spin $\geq
3/2$~\cite{mfls98}. In their approach two different models were
obtained. The first one (labeled Model B) is based on a simplified
version of the Saclay-Lyon-Collaboration model --- the $N(1440)$ and
$N(1675)$ resonances have been removed --- but it includes an
off-shell treatment for the only retained spin-3/2 resonance
[$N(1720)$] in the reaction mechanism. The second one (Model C) is
identical to Model B, except for the addition of an extra spin-3/2
hyperonic resonance [$\Lambda(1890)$] and its off-shell behavior.  In
referring to the various models we have adopted the following
conventions: the model of the Saclay-Lyon Collaboration is labeled by
SL, while Model B and Model C are labeled as SLB and SLC,
respectively. Finally, WJC labels the model by Williams, Ji, and
Cotanch. Note that the SL model will be used in all of our
calculations, unless stated otherwise.

\section{Scattering Matrix Element}

In a similar fashion to the coherent process, the most general
expression for the scattering matrix element in the framework of the
RPWIA can be written as a multiple integral in the following form:
\begin{equation}
 \int d^4x_1 \ldots d^4x_N \; \overline{{\psi}} \; A^\mu J_\mu
(x_1,\ldots,x_N)\; {\cal U}_{\alpha}\; \phi\;,
\label{impulemateleqf}
\end{equation}
where ${\cal U}_{\alpha}$ is a single-particle Dirac spinor for the
bound nucleon, $\psi$ is the Dirac spinor for the outgoing nucleon (hyperon), $A^\mu$ is the photon wavefunction, and $\phi$
is the pseudoscalar meson wavefunction. The number $N$ of the
independent variables to be integrated over depends on the nature of
the effective field theory employed. It can be shown that this
expression can be reduced to the following form:
\begin{equation}
 \big|{\cal M}\big|^2 = \delta (p^0 + k^0 - p'^0 -k'^0) \sum_m \Big|
 \overline{\cal U}({\bf p}^{\prime},s^{\prime})\mbox{ } T(s,t)\mbox{
 }{\cal U}_{\alpha,m}({\bf p}) \Big|^2 \;.
\label{Msquare}
\end{equation}     
Here ${\cal U}({\bf p}^{\prime},s^{\prime})$ is the free Dirac spinor
for the emitted $\Lambda$-hyperon and ${\cal U}_{\alpha,m}({\bf p})$
is the Fourier transform of the relativistic spinor for the bound
nucleon [$\alpha$ denotes the collection of all quantum numbers
necessary (besides $m$) to specify the single-particle orbital]. Note that since we
assume that the impulse approximation is valid, we employ the on-shell
photoproduction operator $T(s,t)$ as given by Equation \ref{elemampl}.

The nucleon bound-state wavefunction can be expressed in a two
component representation as following (see Chapter
\ref{ch:NuclearStructure}),
\begin{equation}
 {\cal U}_{E \kappa m}({\bf x}) = \frac {1}{x} \left[ \begin{array}{c}
 g_{E \kappa}(x) {\cal Y}_{+\kappa \mbox{} m}(\hat{\bf x}) \\ if_{E
 \kappa}(x) {\cal Y}_{-\kappa \mbox{} m}(\hat{\bf x}) \end{array}
 \right]\;,
\end{equation}
where the spin-angular functions are defined as:
\begin{equation}
 {\cal Y}_{\kappa\mbox{} m}(\hat{\bf x}) \equiv \langle{\hat{\bf
 x}}|l{\scriptstyle\frac{1}{2}}jm>\;; \quad j = |\kappa| - \frac
 {1}{2} \;; \quad l = \cases{ \kappa\;, & if $\kappa>0\;$; \cr
 -1-\kappa\;, & if $\kappa<0\;$. \cr} \label{curlyy}
\end{equation}
The Fourier transform of the relativistic bound-state wavefunction can
now be evaluated . We obtain,
\begin{equation}
  {\cal U}_{E\kappa m}({\bf p}) \equiv \int d{\bf x} \; e^{-i{\bf
      p}\cdot{\bf x}} \; {\cal U}_{E\kappa m}({\bf x}) = {4\pi \over
      p} (-i)^{l} \left[ \begin{array}{c} g_{E\kappa}(p) \\
      f_{E\kappa}(p) ({\bf \sigma}\cdot{\hat{\bf p}}) \end{array}
      \right] {\cal{Y}}_{+\kappa m}(\hat{\bf{p}}) \;, \label{uofp}
\end{equation}       
where we have written the Fourier transforms of the radial wavefunctions as
\begin{eqnarray}
   g_{E\kappa}(p) &=& \int_{0}^{\infty} dx \,g_{E\kappa}(x)
    \hat{\jmath}{\hbox{\lower 3pt\hbox{$_l$}}}(px) \;, \\
    f_{E\kappa}(p) &=& ({\rm sgn}\kappa) \int_{0}^{\infty} dx
    \,f_{E\kappa}(x) \hat{\jmath}{\hbox{\lower 3pt\hbox{$_{l'}$}}}(px)
    \;.
\label{gfp}
\end{eqnarray}
Note that in the above expressions we have introduced the
Riccati-Bessel function in terms of the spherical Bessel
function~\cite{Taylor72}: $\hat{\jmath}{\hbox{\lower
3pt\hbox{$_{l}$}}}(z)= zj{\hbox{\lower 3pt\hbox{$_{l}$}}}(z)$, and that
$l'$ is the orbital angular momentum corresponding to $-\kappa$ (see
Equation \ref{curlyy}).

Since the scattering matrix element is proportional to the bound-nucleon
wavefunction in momentum space, it is instructive to examine the
momentum content of the wavefunction.
\BFIG
\centerline{\psfig{figure=PsiFourier.pstex,height=6.0in,width=6.0in}}
\caption[Momentum content in the bound-nucleon wavefunction]
{$g_{E\kappa}(p)$ and $f_{E\kappa}(p)$: the radial components of the
bound-nucleon wavefunction in momentum space for the
$1p^{3/2}$ orbital of ${}^{12}$C.}
\label{fig:psi(p)}
\end{figure}
Figure \ref{fig:psi(p)} shows $g_{E\kappa}(p)$ and $f_{E\kappa}(p)$ as
a function of momentum for the $1p^{3/2}$ orbital of ${}^{12}$C. It is
evident here that the wavefunction has its maximum around $100$~MeV
and that it is appreciable only for $p \leq 300$~MeV.

\section{Closed-form Expression for the Photoproduction Amplitude}

Having introduced all relevant quantities, we are now in a position to
evaluate the (square of the) photoproduction amplitude (Equation
\ref{Msquare}). Without distortions, the evaluation of the $\Lambda$
propagator is now standard due to an algebraic ``trick'' that appears
to be used for the first time by Casimir~\cite{Pais86,g87}:
\begin{equation}
 S(p^{\prime}) \equiv \sum_{s^{\prime}} {\cal U } ({\bf
            p}^{\prime},s^{\prime}) \, \overline{{\cal U }}({\bf
            p}^{\prime},s^{\prime}) =
            \frac{\rlap/p^{\prime}+M_{\Lambda}}{2M_{\Lambda}} \;;
            \quad \left(p^{\prime\,0}\equiv E_{\Lambda}({\bf
            p}^{\prime})= \sqrt{{\bf
            p}^{\prime\,2}+M_{\Lambda}^2}\right) \;.
\label{sfree}
\end{equation}
Subsequently others --- Feynman being apparently the first one ---
used this trick to reduce the ``complex'' computation of covariant
matrix elements to a simple and elegant evaluation of traces of Dirac
$\gamma$-matrices. These trace-techniques have been used here to
compute free polarization observables (note that free polarization
observables will serve as the baseline for comparison against
bound-nucleon calculations). In principle, one does not expect that
these useful trace-techniques will generalize once the nucleon goes
off its mass shell. Yet, simple algebraic manipulations --- first
performed to our knowledge by Gardner and Piekarewicz~\cite{gp94} ---
show that a trick similar to that of Casimir holds even for bound
spinors. Indeed, the validity of their result rests on the following
simple identity:
\begin{equation}
  \sum_{m} {\cal{Y}}_{+\kappa m}(\hat{\bf{p}})
   {\cal{Y}}^{*}_{\pm\kappa m}(\hat{\bf{p}}) =
   \pm\frac{2j+1}{8\pi} \cases{1 \cr {\bf \sigma}\cdot\hat{\bf p} \cr}
   \;,
\end{equation}
which enables one to introduce the notion of a ``bound-state
propagator''. That is,
\begin{eqnarray}
  S_{\alpha}({\bf p}) &\equiv& {1 \over 2j+1} \sum_{m} {\cal
  U}_{\alpha,m}({\bf p}) \, \overline{\cal U}_{\alpha,m}({\bf p})
  \nonumber \\ &=& \left({2\pi \over p^{2}}\right) \left(
  \begin{array}{cc} g^{2}_{\alpha}(p) &
  -g_{\alpha}(p)f_{\alpha}(p){\bf \sigma}\cdot{\hat{\bf p}} \\
  +g_{\alpha}(p)f_{\alpha}(p){\bf \sigma}\cdot{\hat{\bf p}} &
  -f^{2}_{\alpha}(p) \end{array} \right) \nonumber \\ &=&
  ({\rlap/{p}}_{\alpha} + M_{\alpha}) \;, \quad
  \Big(\alpha=\{E,\kappa\}\Big) \;.  \label{salpha}
\end{eqnarray}
Note that we have defined the above mass-, energy-, and momentum-like
quantities as
\begin{eqnarray}
  M_{\alpha} &=& \left({\pi \over p^{2}}\right) \Big[g_{\alpha}^{2}(p)
                  - f_{\alpha}^{2}(p)\Big] \;, \\ E_{\alpha} &=&
                  \left({\pi \over p^{2}}\right)
                  \Big[g_{\alpha}^{2}(p) + f_{\alpha}^{2}(p)\Big] \;,
                  \label{epm} \\ {\bf p}_{\alpha} &=& \left({\pi \over
                  p^{2}}\right) \Big[2 g_{\alpha}(p)
                  f_{\alpha}(p)\hat{\bf p} \Big] \;,
\end{eqnarray}
which satisfy the ``on-shell relation''
\begin{equation}
  p_{\alpha}^{2}=E_{\alpha}^{2}-{\bf p}_{\alpha}^{2} =M_{\alpha}^{2}
                \;.  \label{onshell}
\end{equation}
The evident similarity in structure between the free and bound
propagators (Equations \ref{sfree} and \ref{salpha}) results in an
enormous simplification; we can now employ the powerful trace
techniques developed by Feynman to evaluate all polarization
observables --- irrespective of whether the nucleon is free or bound to a
nucleus.  It is important to note, however, that this enormous
simplification would have been lost if distortion effects would have
been incorporated in the propagation of the emitted $\Lambda$-hyperon.

It is informative to examine the behavior and significance of the
mass-, energy-, and momentum-like quantities: $M_{\alpha}$,
$E_{\alpha}$, and ${\bf p}_{\alpha}$. Figure \ref{fig:EpMtilda}
exhibits these variables as a function of momentum.
\BFIG
\centerline{\psfig{figure=EPMtilda.pstex,height=6.0in,width=6.0in}}
\caption[Effective mass-, energy-, and momentum-like quantities: $M_{\alpha}$,
$E_{\alpha}$, and $p_{\alpha}$] {The effective mass-, energy-, and
momentum-like quantities: $M_{\alpha}$, $E_{\alpha}$, and $|{\bf p}_{\alpha}|$
as a function of the momentum ($p$).}
\label{fig:EpMtilda}
\end{figure}
Note that $|{\bf p}_{\alpha}| \ll E_{\alpha}$ and $M_{\alpha} \simeq
E_{\alpha}$. This is a consequence of the fact that $f_{\alpha}(p)$,
although enhanced in the nuclear medium, is still much smaller than
$g_{\alpha}(p)$. Since the cross section is proportional to the term
(${\rlap/{p}}_{\alpha} + M_{\alpha}$), and since $|{\bf p}_{\alpha}|
\ll E_{\alpha}$ and $M_{\alpha} \simeq E_{\alpha}$, the cross section
becomes directly proportional to $E_{\alpha}$. This is a pleasant
outcome as $E_{\alpha}$ has a simple interpretation as the
bound-nucleon density in momentum space (see Equation \ref{epm}). Therefore, we deduce that the
quasifree process provides us with a direct probe of the momentum
distribution in the bound-nucleon wavefunction.

To provide a feeling for the enormous simplification entailed by the
above trick, we derive now an example assuming, for mere simplicity,
that the photoproduction amplitude contains only the tensor term
($\sigma_{\alpha \beta}$ term in Equation \ref{elemampl}). For this case the square of the unpolarized photoproduction matrix
element (Equation \ref{Msquare}) becomes proportional to:
\begin{eqnarray}
 \big|{\cal M}\big|^2 &\rightarrow& |A_1|^2 \,
 \left(-\frac{1}{2}g_{\mu\nu}\right)\, T{\hbox{\lower 2pt\hbox{$r$}}}
 \Big[\gamma^{5}\gamma^{\mu}\rlap/k
 \left(\rlap/p_{\alpha}+M_{\alpha}\right)
 \gamma^{5}\gamma^{\nu}\rlap/k
 \left(\rlap/p^{\prime}+M_{\Lambda}\right) \Big] \nonumber \\ &=&
 \frac{1}{2}\,|A_1|^2 \left[ T{\hbox{\lower 2pt\hbox{$r$}}}
 \Big(\gamma^{\mu}\rlap/k\rlap/p_{\alpha}
 \gamma_{\mu}\rlap/k\rlap/p^{\prime} \Big) - M_{\alpha}M_{\Lambda}
 T{\hbox{\lower 2pt\hbox{$r$}}} \Big(\gamma^{\mu}\rlap/k
 \gamma_{\mu}\rlap/k \Big) \right]\nonumber \\ &=& 8\,|A_1|^2 (k \cdot
 p_{\alpha})(k\cdot p^{\prime}) \;.
\label{example}
\end{eqnarray}    
This result is, indeed, simple and illuminating. Although including
the full complexity of the elementary amplitude requires the
evaluation of thirty-two such terms (not all of them independent) the
evaluation of any one of those terms is not much more complicated than
the one presented above. Yet, to automate this straightforward but
lengthy procedure, we rely on the {\it FeynCalc 1.0}\cite{mh92}
package with {\it Mathematica 2.0} to calculate all traces involving
$\gamma$-matrices. The output from these symbolic manipulations was
then fed into a FORTRAN code to obtain the final numerical values for
all different polarization observables. In the appendix, I
include full calculations of the traces of $\gamma$-matrices for the
generalized case involving two fermions, one of which is polarized
while the other one is not.

\section{Kinematics in the Quasifree Process}

The kinematics for the quasifree production of a $\Lambda$-hyperon
through the photoproduction reaction $A(\gamma,K^+ \Lambda)B$ is
constrained by two conditions. First, there is an overall energy-momentum
conservation:
\begin{equation}
k + p_A = k^{\prime} + p^{\prime} + p_B\;.
\end{equation}
Moreover, since we are studying the photoproduction process within the
framework of the impulse approximation (see Figure
\ref{fig:QuasifreeImpulse}) there is a second kinematical constraint
arising from energy-momentum conservation at the $\gamma N
\!\rightarrow\! K^{+} \Lambda$ vertex:
\begin{equation}
k + p = k^{\prime} + p^{\,\prime} \;,
\end{equation}
where $p$ is the four-momentum of the bound nucleon, whose space part
is known as the missing momentum:
\begin{equation}
{\bf p}_{m}\equiv {\bf p}^{\,\prime}-{\bf q} \;; \quad ({\bf
                 q}\equiv{\bf k}-{\bf k}^{\prime}) \;.
\end{equation}
Thus, as in most semi-inclusive processes --- such as in the $(e,e'p)$
reaction --- the quasifree production process becomes sensitive to the
nucleon momentum distribution.

The kinematic structure of the quasifree process is much richer than
that of the free process. The reason is that the target nucleon is
bound and thus has a distribution of momentum states as opposed to
only one specific state. This adds more degrees of freedom for the
outgoing particles which can now be in states that are not
permitted in the free process. Moreover, as opposed to being
constrained to one plane, the quasifree process allows out-of-plane
scattering events due to the three-dimensional nature of the momentum
distribution of the bound nucleon. These events however, have smaller
cross section~\cite{lwbt96} and will not be investigated in this work.

Let me remind the reader once more that we study this process in the
impulse approximation, and so the interaction is assumed to proceed
from only one of the bound nucleons. The rest of the nucleons act
merely as
spectators. For a clearer picture of the quasifree kinematics, Table
\ref{compkinm} provides a comparison between the kinematics in the free and
\begin{table}
\caption{Comparison between the free and quasifree kinematics.}
\vspace{.5cm}
\begin{center}
\thicklines
\begin {tabular} {lcc}
\hline
\hline
\thinlines
& Free Process & Quasifree Process\\
\hline
\thicklines
Kinematic variables & $k, p, k^{\prime}, p^{\prime}$ & $k, p,
k^{\prime}, p^{\prime}$ \\ &&\\ Number of degrees of freedom & 16 &
16\\ On-mass-shell condition & 4 & 3 \\ Energy-momentum conservation &
4 & 4 \\ Observables fixed by experiment & 8 & 9 \\
\hline
\hline
\end{tabular}
\end{center}
\label{compkinm}
\end{table}
quasifree processes. At the interaction vertex, we have four
kinematic variables. These are identical to those in the
free process; the four-momenta of the photon, target nucleon, emitted
meson, and outgoing nucleon. Thus, we have a total of sixteen degrees of
freedom. This number is then reduced by four in the free process and
by three in the quasifree due to the on-mass-shell conditions. Note
that the quasifree case has only three on-mass-shell conditions since the
bound nucleon is off its mass shell. The number of degrees of freedom
is further reduced by four for each of these cases because of
energy-momentum conservation. Therefore, in the free case we have eight
degrees of freedom to be fixed by experiment. These are \{${\mathbf
p},{\mathbf k}, \hat{\mathbf k^{\prime}}$\}. Note that as far as
outgoing particles are concerned, it is only the direction of the
outgoing-meson momentum that can be probed by experiment.

In the quasifree setting, we have nine available degrees of freedom to
be fixed by experiment. Since we cannot fix the momentum state of the
target nucleon, the fixed quantities are \{$E_{\rm
bound}$, $\hat{{\mathbf
p}^\prime},{\mathbf k}, {\mathbf k^{\prime}}$\}. Here $E_{\rm bound}$
is the binding energy of the bound nucleon and is fixed by setting the
kinematics in such a way to knock a proton from a specific orbital in
the nucleus. It
is evident then that the quasifree interaction offers a richer
experimental output for the outgoing particles. By tuning these
kinematics, we can accordingly probe a specific momentum state of the
bound nucleon.

As a consequence of the kinematic richness, we can study the quasifree process in
more than one kinematic setting; unlike in the free
process. We have used two of these kinematical
settings in this work. In the first one, we tune $\hat{{\mathbf p}^\prime}$ in
such a way to fix the the missing momentum (probed bound-nucleon
momentum) at the maximum of the momentum distribution of the bound
nucleon, and then vary the scattering angle between ${\mathbf k}$ and
${\mathbf k^{\prime}}$.  Knowing that the cross section is
proportional to the momentum ditribution of the bound nucleon, this
allows us to maximize the measured cross section.

In the second kinematic setting, we vary $\hat{{\mathbf p}^\prime}$
and so effectively modify the probed momentum state of the nucleon. In
this condition, we are scanning the strength of the different momentum
components in the nucleon wavefunction. These two kinematic settings
will be further discussed in the context of the results in the next
chapter.

As has been indicated earlier, the calculation of the scattring
matrix element reduces to an evaluation of traces of $\gamma$-matrices
similar to the example of Equation \ref{example}. In deriving the
expression for the 
unpolarized cross section, we find that the cross section most
generally depends on the amplitudes $A_1, A_2, A_3,\; {\rm and}\; A_4$ and the
following set of scalar products: $\{ k \cdot p$, $k \cdot
k^{\prime}$, $ k \cdot p^{\prime}$, $k \cdot p_\alpha$, $k \cdot k = 0
$, $p \cdot k^{\prime}$, $p \cdot p^{\prime}$,  $p \cdot
p_\alpha$, $p \cdot p$, $k^{\prime} \cdot p^{\prime}$, $k^{\prime}
\cdot p_\alpha$, $k^{\prime} \cdot k^{\prime} = {m^2}_{\rm meson}$,
$p^{\prime} \cdot p_\alpha$, $p^{\prime} \cdot p^{\prime} =
{M^2}_{\Lambda}$, $p_\alpha \cdot p_\alpha \}\;$.

Analogously, when we derive the expression for the the recoil
$\Lambda$-polarization (${\cal P}$), we find that most generally this observable
depends also on the following set of scalar products: $\{
eps[s^{\prime},k,p,k^{\prime}]$, $eps[s^{\prime},k,p,p^{\prime}]$,
$eps[s^{\prime},k,p,p_\alpha]$,
$eps[s^{\prime},k,k^{\prime},p^{\prime}]$, 
$eps[s^{\prime},k,k^{\prime},p_\alpha]$,
$eps[s^{\prime},k,p^{\prime},p_\alpha]$,
$eps[s^{\prime},p,k^{\prime},p^{\prime}]$,
$eps[s^{\prime},p,k^{\prime},p_\alpha]$,
$eps[s^{\prime},p,p^{\prime},p_\alpha]$, \newline $eps[s^{\prime},k^{\prime},p^{\prime},p_\alpha]\}\;$. Here,
$eps[R_1,R_2,R_3,R_4] \equiv \varepsilon^{\mu \nu \gamma \delta} 
{R_1}_\mu {R_2}_\nu {R_3}_\gamma {R_4}_\delta\;$.

In finding the expression for the photon asymmetry ($\Sigma$), no
additional scalar products appear apart from $\epsilon \cdot
\epsilon^{*} = -1 $. This is a consequence of using Equation
\ref{photnsigmanew} for this observable, and as a result of taking
${\bf k}$,
${\bf p}$, ${\bf k}^{\prime}$, and ${\bf p}^{\prime}$ to be in the
same scattering plane.

By using the formalism depicted in this chapter, we have arrived at
closed-form expressions for the different observables as functions of
various scalar products. Now by deriving expressions for these scalar
products in terms of the experimentally given quantities, we can
numerically evaluate these observables. The results of these
calculations will be presented in the next chapter.

\chapter{Results and Discussion of the Quasifree Process}
\label{ch:ResultQuasifree}

In this chapter I will present the main results of our study of the
quasifree kaon photoproduction from nuclei. Particularly, the
sensitivity of the quasifree process to relativistic effects, nuclear
target effects, and to the elementary amplitude will be
investigated. Additionally, the process will be studied in two
kinematic regimes where the relation of the interaction to the momentum
distribution in the bound-nucleon wavefunction will be
explored. Please note the following conventions in this chapter: $q$
stands for $|{\bf q}|$, and $p_m$ stands for $|{\bf p}_m|$.

\section{Relativistic Effects}

We start the discussion of our results by examining the role of the
relativistic dynamics on the polarization observables. On Figure
\ref{fig:KaonRelEff} we display the recoil polarization (${\cal P}$)
of the
\BFIG
\centerline{\psfig{figure=KaonRelEff.pstex,height=6.0in,width=7.0in}}
\caption[Relativistic effects on the polarization observables in the
quasifree process] {Comparison between relativistic and
nonrelativistic calculations of the recoil polarization of the
$\Lambda$-hyperon (${\cal P}$) and the photon asymmetry ($\Sigma$) as
a function of the kaon scattering angle for the knockout of a proton
from the $p^{3/2}$ orbital in ${}^{12}$C. The SL model for the
elementary amplitude is used here.}
\label{fig:KaonRelEff}
\end{figure}
$\Lambda-$hyperon and the photon asymmetry ($\Sigma$) as a function of
the kaon scattering angle for the knockout of a proton from the
$p^{3/2}$ orbital in ${}^{12}$C using the SL model for the elementary
amplitude. The polarization observables were evaluated at a photon
energy of $E_{\gamma}\!=\!1400$~MeV and at a missing momentum of
$p_m\!=\!120$~MeV (this value is close to the maximum in the momentum
distribution of the $p^{3/2}$ orbital; see Figure
\ref{fig:EpMtilda}). Note that in this figure --- and all throughout
this chapter --- we compute all observables in the laboratory system
using the quasifree condition:
$\omega\!=\!\sqrt{q^2+M_{\Lambda}^{2}}-M_{N}$. The insensitivity of
our results to the relativistic dynamics is evident. Indeed, the
relativistic and nonrelativistic curves can barely be resolved in the
figure. We have also examined in Figure \ref{KaonCrossRelEff} these
effects on the unpolarized cross
\BFIG
\centerline{\psfig{figure=KaonCrossRelEff.pstex,height=6.0in,width=6.0in}}
\caption[Relativistic effects on the differential cross section in the
quasifree process] {Comparison between relativistic and
nonrelativistic calculations of the differential cross section as a
function of the kaon scattering angle for the knockout of a proton
from the $p^{3/2}$ orbital in ${}^{12}$C. The SL model for the
elementary amplitude is used here.}
\label{KaonCrossRelEff}
\end{figure}
section and found them insignificant as well. The main reason behind this
insensitivity is that in the quasifree process all CGLN amplitudes
including the tensor, pseudoscalar, and axialvector contributions
participate in the process as opposed to the tensor one only (which is
very sensitive to the relativistic enhancement) in the
coherent process. Note that our
``nonrelativistic'' results were obtained by adopting the free-space
relation in the determination of the lower-component of the
bound-state wavefunction. This represents our best attempt at
reproducing nonrelativistic calculations, which employ free, on-shell
spinors to effect the nonrelativistic reduction of the elementary
amplitude.

\section{Nuclear Target Effects}

Next we examine the nuclear dependence of the polarization
observables. Figure \ref{KaonNucTarEff} displays the recoil
polarization and the
\BFIG
\centerline{\psfig{figure=KaonNucTarEff.pstex,height=6.0in,width=7.0in}}
\caption[Nuclear target effects on the polarization observables in the
quasifree process] {The recoil polarization of the $\Lambda$-hyperon
(${\cal P}$) and the photon asymmetry ($\Sigma$) as a function of the
kaon scattering angle for the knockout of a valence proton from a
variety of nuclei.  The photoproduction from a free proton is depicted
with the filled circles. The SL model for the elementary amplitude is
used here.}
\label{KaonNucTarEff}
\end{figure}
photon asymmetry for the knockout of a valence proton for a variety of
nuclei, ranging from ${}^{4}$He all the way to ${}^{208}$Pb. That is,
we have computed the knockout from the $1s^{1/2}$ orbital of
${}^{4}$He, the $1p^{3/2}$ orbital of ${}^{12}$C, the $1p^{1/2}$
orbital of ${}^{16}$O, the $1d^{3/2}$ orbital of ${}^{40}$Ca, and the
$3s^{1/2}$ orbital of ${}^{208}$Pb. We have included also polarization
observables from a single free proton to establish a baseline for
comparison against our bound--nucleon calculations. The sensitivity of
the polarization observables to the nuclear target is rather small.
Indeed, it seems that as soon as the quasifree process takes place
from a proton bound to a ``lump'' of nuclear matter, the polarization
observables become largely insensitive to the fine details of the
lump. In other words, the polarization observables are not sensitive
to the fine details of the bound-nucleon wavefunction.  Moreover, the
deviations from the free value (shown with the filled circles) are
significant. This indicates important modifications to the elementary
process in the nuclear medium.

\section{Sensitivity to the Elementary Amplitude}

Having established the independence of polarization observables to
relativistic effects and to a large extent to the nuclear target, we
are now in a good position to discuss the sensitivity of these
observables to the elementary amplitude (note that an insensitivity of
polarization observables to final-state interactions has been shown
specifically for the kaon quasifree process in Ref.~\cite{blmw98}). We
display in Figure \ref{KaonCrossp} the differential
\BFIG
\centerline{\psfig{figure=KaonCrossp.pstex,height=6.0in,width=6.0in}}
\caption[Elementary amplitude effects on the differential cross
section in the quasifree process] {The differential cross section as a
function of the kaon scattering angle for the knockout of a proton
from the $p^{3/2}$ orbital in ${}^{12}$C using various models for the
elementary amplitude.}
\label{KaonCrossp}
\end{figure}
cross section as a function of the kaon scattering angle for the
knockout of a proton from the $p^{3/2}$ orbital in ${}^{12}$C using
four different models for the elementary amplitude (see Section
\ref{sec:kaoneleamp}). Again, the
photon incident energy and the missing momentum have been fixed at
$1400$~MeV and $120$~MeV, respectively. Although there are noticeable
differences between the models, primarily at small angles, these
differences are relatively small. This behavior has been confirmed by
a recent calculation that suggests that the kaon-photoproduction cross
section --- as a function of the energy of the photon beam --- is
slightly model dependent~\cite{LMST98}. Much more significant,
however, are the differences between the various sets for the case of
the polarization observables displayed in Figure \ref{KaonPolp}. The
added
\BFIG
\centerline{\psfig{figure=KaonPolp.pstex,height=6.0in,width=7.0in}}
\caption[Elementary amplitude effects on the polarization observables in the
quasifree process] {The recoil polarization of the $\Lambda$-hyperon
(${\cal P}$) and the photon asymmetry ($\Sigma$) as functions of the
kaon scattering angle for the knockout of a proton from the $p^{3/2}$
orbital in ${}^{12}$C using various models for the
elementary amplitude.}
\label{KaonPolp}
\end{figure}
sensitivity to the choice of amplitude exhibited by the polarization
observables should not come as a surprise; unraveling subtle details
about the dynamics is the hallmark of polarization observables. In
particular, polarization observables show a strong sensitivity to the
inclusion of the off-shell treatment for the various high-spin
resonances, as suggested in Ref.~\cite{mfls98}

\section{Observables and Momentum Distribution in the Bound
Nucleon Wavefunction}

We display in Figure \ref{KaonCrossq} the cross section as a
function
\BFIG
\centerline{\psfig{figure=KaonCrossq.pstex,height=6.0in,width=6.0in}}
\caption[Differential cross section as a function of the missing
momentum in the quasifree process] {The differential cross section as
a function of the missing momentum for the knockout of a proton from
the $p^{3/2}$ orbital of ${}^{12}$C using two parameterizations of the
elementary amplitude. The figure includes also the $E_\alpha$
parameter (up to an arbitrary scale) which is proportional to the
momentum distribution of the bound-proton wavefunction.}
\label{KaonCrossq}
\end{figure}
of the missing momentum for the $p^{3/2}$ orbital in ${}^{12}$C using
a different kinematical setting. Here we have kept the photon incident
energy fixed at $1400$~MeV but have set the momentum transfer $q$ at
$400$~MeV. To a large extent the cross sections represents --- up to
an overall normalization factor --- the momentum distribution of the
$p^{3/2}$ orbital. Indeed, the peak in the cross section is located at
$p_{m}\!\approx\!110$~MeV, which is also the position of the maximum
in the momentum distribution. To further appreciate the similarities
between the two we have included the energy-like parameter $E_\alpha$,
up to an arbitrary scale. As seen from Equation \ref{epm}, $E_\alpha$
is directly proportional to the momentum distribution of the
bound-proton wavefunction. The similarities between the cross section
and the momentum distribution are indisputable. Note that the cross
section dies out for $p_{\rm m}\!>\!250$~MeV. This region of
high-momentum components is sensitive to short-range correlations,
which are beyond the scope of our simple mean-field description. Thus,
the tail of the photoproduction cross section can be used to test more
sophisticated models of nuclear structure.

The sensitivity of the cross section to the momentum content in the
bound-nucleon wavefunction suggests an innovative use for quasifree
processes (although not using kaon photoproduction): probing neutron
densities in halo nuclei. Halo nuclei are neutron-saturated nuclei
where the valence neutrons (called also halo neutrons) are barely
bound in the nucleus. Nuclear densities are usually studied using electron scattering which
is a very clean tool, but one that discriminates against the neutrally
charged neutron. Generally speaking, in meson photoproduction
processes, the photon couples with comparable strengths to the protons
and neutrons. Thus, the quasifree process is a direct tool for
probing the neutron wavefunction in halo nuclei. This aspect is
particularly appealing because the quasifree cross section is
typically large enough to be measured experimentally, and there is a wealth
of observables (such as polarization observables) to
study. Furthermore, the process is only sensitive to the wavefunction
of the knocked-out neutron, that is the process is minimally polluted by
the other constituents in the nucleus. However, we cannot be ambitious
to probe the fine details of the neutron wavefunction using this
process. Studying halo nuclei using the quasifree process may encounter an
experimental challenge as most of these nuclei have two halo neutrons and
probably it is difficult to knock out only one of these neutrons
without affecting the other; considering the fragility of the
binding.

For completeness, Figure \ref{KaonPolObsqkin} displays the polarization
observables
\BFIG
\centerline{\psfig{figure=KaonPolObsqkin.pstex,height=6.0in,width=7.0in}}
\caption[Polarization observables as functions of the missing
momentum in the quasifree process] {The recoil polarization of the
$\Lambda$-hyperon (${\cal P}$) and the photon asymmetry ($\Sigma$) as
functions of the missing momentum for the knockout of a proton from
the $p^{3/2}$ orbital of ${}^{12}$C using two parameterizations of the
elementary amplitude.}
\label{KaonPolObsqkin}
\end{figure}
as functions of the missing momentum. The sensitivity of these
observables to the elementary amplitude is manifest in the
figure. It is notable however that these observables have a small
magnitude and are rather insensitive to changes in the missing momentum. The
evident sensitivity at larger values of the missing momentum is not to
be taken seriously; the cross section at these values is too small for
the ratios defining the polarization observables to be meaningful (see
Equations \ref{p} and \ref{sigma}).

\chapter{Conclusions of the Quasifree Process}
\label{ch:ConclusionQuasifree}

We have computed polarization observables --- the recoil polarization
of the $\Lambda$-hyperon and the photon asymmetry --- for the
quasifree $K^{+}$ photoproduction reaction from nuclei~\cite{abpi2000}.  Motivated by
the large insensitivity of polarization observables to distortion
effects, a relativistic plane-wave impulse approximation was
developed. For the elementary amplitude we used a variety of models
while for the nuclear structure we employed a relativistic mean-field 
approximation to the Walecka model~\cite{serwal86}. In this manner the 
quasifree amplitude was evaluated without recourse to a
nonrelativistic reduction, as the full relativistic structure of the 
amplitude was maintained.

By assuming the validity of the relativistic plane-wave impulse
approximation an enormous simplification ensued: by introducing the 
notion of a bound-state propagator --- as was done for the first
time by Gardner and Piekarewicz in Ref.~\cite{gp94} --- the
mathematical structure of all quasifree observables was cast in a 
manner analogous to that of the elementary process. Thus, we brought 
the full power of Feynman's trace techniques to bear into the problem.
We stress that the relativistic formalism presented here can be
applied with minor modifications to most quasifree knockout studies,
at least in the plane-wave limit. In particular, the application of
this formalism carries a prominent promise in the study of the quasifree processes of other
pseudoscalar mesons like $\pi$ and $\eta$ mesons, as well as its
use in quasifree electron scattering. Furthermore, the appropriateness
of this process in probing neutron densities in halo nuclei is appealing. 

In addition of being largely insensitive to distortions effects, we
found polarization observables insensitive to relativistic effects
and mostly independent of the target nucleus. Polarization observables
appear to only be sensitive to the elementary 
amplitude.  As free polarization observables provide a baseline
against which possible medium effects may be inferred, we conclude 
that quasifree polarization observables might be one of the
cleanest tools for probing modifications to the elementary amplitude
in the nuclear medium. Deviations from their free values are likely to
stem from a modification of the elementary interaction inside the
nuclear medium due, for example, to a change in resonance parameters.
Indeed, for the kinematics adopted in this work
($E_{\gamma}\!=\!1.4$~GeV or $\sqrt{s}\!\approx\!1.9$~GeV) one should
be very sensitive to the formation, propagation, and decay of the
$P_{13}(1900)$ and $F_{17}(1990)$
$N^{\star}-$resonances~\cite{PDG98}. The meson photoproduction (and
electroproduction) programs at various experimental facilities ---
such as TJNAF, NIKHEF, and MAMI --- should shed light on the physics
of this interesting and fundamental problem.

Shortly after submitting our work on the quasifree photoproduction for
publication~\cite{abpi2000}, a comprehensive study of
kaon-photoproduction observables was reported by Lee, Mart, Bennhold,
and Wright~\cite{lmbw99}. The authors have presented a very detailed
analysis of the effect of distortions on various photoproduction
observables. One of the central results from their study is that
polarization observables are not as insensitive to distortion effects
as once believed. Yet they showed categorically that for certain
kinematical situations --- such as those adopted in our present work
--- the effect of distortions on the polarization observables is
insignificant indeed.

\chapter{General Conclusions}
\label{ch:GeneralConclusion}

This manuscript presents the core of our studies in the field of
pseudoscalar meson photoproduction from nuclei. We studied two
processes: the coherent and the quasifree reactions. We have found
that the current treatments of this process suffer from various
sources of ambiguities and uncertainties. Among these problems are
final-state interactions, relativistic effects, off-shell ambiguities,
and violations to the impulse approximation. By far the largest
uncertainty emerges from the ambiguity in extending the many
on-shell-equivalent representations of the elementary amplitude off
the mass shell. Thus one must be very cautious in interpreting the
wealth of experimental data that will be available soon.

The coherent process can be a very useful tool in investigating
nucleon resonances and their modifications in nuclear medium. In order
to do so, the difficulties in this process must be addressed. Much
work remains to be done in investigating the off-shell ambiguities and
the violations to the impulse approximation. Developing a formalism
that addresses all of these intricacies not only will help us
understand these processes, but may have significant impact on our
understanding of quantum chromodynamics (QCD) and its implicit
symmetries. An example in this line of thought is our recent attempt
to use the coherent eta photoproduction process as a probe of the
chiral symmetry mirror assignment~\cite{sumins99}.

As for the quasifree process, we have developed a powerful formalism
for studying these processes, at least in the plane-wave limit. We
found that the most useful tools are the the
polarization observables which are mostly insensitive to distortion,
nuclear target, and relativistic effects. However, these observables
are very sensitive to the fundamental physics behind the elementary
process and to any modification in the nuclear medium. Thus, we have
powerful tools at our disposal to study such aspects of
photoproduction processes.

The quasifree formalism that we developed can be easily extended to
studies of the quasifree processes of other mesons like the eta or the
pion. In fact, we plan to embark on such a study. Furthermore, the
formalism can be easily extended to electron-scattering quasifree
processes and we plan also to study such processes. Finally, the most
exciting is the study of the possibility of kaon condensation. This is
done by investigating the inclusive kaon photoproduction from
nuclei. This can be done using the single-particle response or the
random-phase-approximation (RPA) response. Comparison of the
theoretical results with expected experimental data may lead us to
find a signature for this condensation.

\appendix           
\chapter{Generalized Calculations of Traces of $\gamma$-Matrices}
\label{ap:Traces}

As has been indicated in Chapter \ref{ch:QuasifreeTheory}, the
calculation of the square of the scattering matrix element reduces to a
calculation of traces of $\gamma$-matrices in the free and quasifree
processes. In this appendix, I include full calculations of the
traces of $\gamma$-matrices for the generalized case involving two
fermions, one of which is unpolarized with the propagator:
\begin{equation}
 S_{\rm unpolarized}(p) \equiv \sum_{s} {\cal U } ({\bf p},s) \,
            \overline{{\cal U }}({\bf p},s) = \frac{\rlap/p+M}{2M} \;;
            \quad \left(p^{0}\equiv E({\bf p})= \sqrt{{\bf
            p}^{2}+M^2}\right),
\label{prounpol}
\end{equation}
while the other is polarized leading to the propagator:
\begin{equation}
  S_{\rm polarized} \equiv {\cal U } ({\bf p},s) \, \overline{{\cal U
            }}({\bf p},s) = \frac{\rlap/p+M}{2M} \; \frac {1}{2} \; (1 +
            \gamma ^5 \rlap/s) \;; \quad \left(p^{0}\equiv E({\bf p})=
            \sqrt{{\bf p}^{2}+M^2}\right).
\label{propol}
\end{equation}
Hence, we can write the generalized structure of the traces as following:
\begin{eqnarray}
Tr\left[ \;\{{\cal L}\} \; (\rlap/p_1 + m_1) \; \{{\cal R}\} \;
(\rlap/p_2 + m_2) \; \frac
{1}{2} \; (1 + \rlap/s_2 \gamma ^5) \; \right],
\end{eqnarray}
where ${p_1}$ is the four-momentum of the unpolarized fermion,
${p_2}$ is the four-momentum of the polarized one, whereas ${s_2}$ is
the {\it negative} of the four-spin of the polarized fermion. The author
apologizes for this awkward definition of the spin, but this has to do
with the ``history'' of writing this appendix.

In this algebraic structure, ${\cal L}$ can be any item of the set
\begin{eqnarray}
\{{\cal L}\} \equiv \{{1, \gamma ^\mu, \gamma ^\mu \gamma ^5, i \gamma ^5,
\sigma ^{\mu \nu}}\}\;, 
\end{eqnarray}
while $ {\cal R}$ can be any item of the set
\begin{eqnarray}
\{ {\cal R}\} \equiv \{ {1, \gamma ^\alpha, \gamma ^\alpha \gamma ^5,
i \gamma ^5, \sigma ^{\alpha \beta}}\}\;.
\end{eqnarray}
The item $\sigma ^{\delta \rho}$ in the above sets is defined as
\begin{eqnarray}
\sigma ^{\delta \rho} \equiv \frac {i}{2} [\gamma ^\delta, \gamma
^\rho]\;.
\end{eqnarray}
The symbolic calculations of these traces have been achieved using
{\it FeynCalc 1.0}\cite{mh92} package of {\it Mathematica 2.0}. The package has
been written by Mertig and Hubland specifically for high-energy physics
calculations. The results of the analysis are shown in the
following five tables. Each table corresponds to one item of the set
$\{ {\cal L}\} $ with each item of the set $\{ {\cal R}\}$.
 
\begin{table}
\caption[Generalized traces of $\gamma$ matrices: ${1}$
traces]{${1}$} 
\vspace{.5cm}
\begin{center}
\begin {tabular} { |l||c| }
\hline
& $1$\\
\hline 
$1$
& $2 m_1 m_2 + 2 p_1.p_2$\\

&\\

$\gamma ^\alpha$ 
& $- 2 i \varepsilon ^{\alpha \gamma \delta
\rho} {p_1}_\gamma {p_2}_\delta {s_2}_\rho + 2 m_2 {p_1}^\alpha + 2 m_1
{p_2}^\alpha$\\

&\\

$\gamma ^\alpha \gamma ^5$ 
& $ -2 p_2.s_2 {p_1}^{\alpha} + 2 p_1.s_2 {p_2}^{\alpha} - 2 m_1 m_2
{s_2}^{\alpha} - 2 p_1.p_2 {s_2}^{\alpha}$\\ 

&\\

$i\gamma ^5$
 & $2 i m_2 p_1.s_2
+2 i m_1 p_2.s_2 $\\

&\\

$\sigma ^{\alpha \beta}$
 & $2 m_2 \varepsilon ^{\alpha \beta \delta \rho}
{p_1}_\delta {s_2}_\rho
+ 2 m_1 \varepsilon ^{\alpha \beta \delta \rho}
{p_2}_\delta {s_2}_\rho 
+ 2 i {p_1}^\beta {p_2}^\alpha
- 2 i {p_1}^\alpha {p_2}^\beta$\\
\hline
\end{tabular}
\end{center}
\end{table}
\begin{table}
\caption[Generalized traces of $\gamma$ matrices:
 $\gamma^\mu$  
traces]{$\gamma^\mu$}
\vspace{.5cm}
\begin{center}
\begin {tabular} { |l||c| }
\hline
& $\gamma ^\mu$\\
\hline 
$1$
& $2 i \varepsilon ^{\mu \gamma \delta \rho} {p_1}_\gamma {p_2}_\delta
{s_2}_\rho 
+ 2 m_2 {p_1}^\mu + 2 m_1 {p_2}^\mu$\\

&\\

$\gamma ^\alpha$ 
& $2 i m_2 \varepsilon ^{\alpha \mu \delta \rho} {p_1}_\delta
{s_2}_\rho 
- 2 i m_1 \varepsilon ^{\alpha \mu \delta \rho} {p_2}_\delta
{s_2}_\rho 
+ 2 m_1 m_2 g^{\alpha \mu}$\\
& $- 2 g^{\alpha \mu} p_1.p_2 
+ 2 {p_1}^\mu {p_2}^\alpha 
+ 2 {p_1}^\alpha {p_2}^\mu$\\

&\\

$\gamma ^\alpha \gamma ^5$ 
& $-2 i \varepsilon ^{\alpha \mu \delta \rho} {p_1}_\delta {p_2}_\rho
+ 2 m_2 g^{\alpha \mu} p_1.s_2 
- 2 m_1 g^{\alpha \mu} p_2.s_2 
+2 m_2 {p_1}^\mu {s_2}^\alpha$\\
& $- 2 m_1 {p_2}^\mu {s_2}^\alpha 
- 2 m_2 {p_1}^\alpha {s_2}^\mu 
+ 2 m_1 {p_2}^\alpha {s_2}^\mu$\\ 

&\\

$i\gamma ^5$
&  $2 i p_2.s_2 {p_1}^\mu 
+ 2 i p_1.s_2 {p_2}^\mu 
+ 2 i m_2 m_1  {s_2}^\mu
- 2 i p_1.p_2 {s_2}^\mu$\\

&\\

$\sigma ^{\alpha \beta}$
& $2 m_1 m_2 \varepsilon ^{\alpha \beta \mu \rho} {s_2}_\rho 
+ 2 \varepsilon ^{\alpha \beta \mu \rho} {p_1}_\rho p_2.s_2 
- 2 \varepsilon ^{\beta \mu \delta \rho} {p_2}_\delta {s_2}_\rho
{p_1}^\alpha$\\
& $ - 2 i m_2 g^{\beta \mu} {p_1}^\alpha
+ 2 \varepsilon ^{\alpha \mu \delta \rho} {p_2}_\delta {s_2}_\rho
{p_1}^\beta 
+ 2 i m_2 g^{\alpha \mu} {p_1}^\beta$\\
& $ + 2 i m_1 g^{\beta \mu} {p_2}^\alpha 
- 2 i m_1 g^{\alpha \mu} {p_2}^\beta 
+ 2 \varepsilon ^{\alpha \beta \delta \rho} {p_1}_\delta {s_2}_\rho
{p_2}^\mu$\\
\hline
\end{tabular}
\end{center}
\end{table}
\begin{table}
\caption[Generalized traces of $\gamma$ matrices: $\gamma^\mu \gamma ^5$ 
traces]{$\gamma^\mu \gamma ^5$}
\vspace{.5cm}
\begin{center}
\begin {tabular} { |l||c| }
\hline
& $\gamma ^\mu \gamma ^5$\\
\hline 
$1$
& $2 p_2.s_2 {p_1}^\mu  
+ 2 p_1.s_2 {p_2}^\mu 
- 2 m_1 m_2{s_2}^\mu 
- 2 p_1.p_2 {s_2}^\mu$\\

&\\

$\gamma ^\alpha$ 
& $ -2 i \varepsilon ^{\alpha \mu \delta \rho} {p_1}_\delta {p_2}_\rho
+ 2 m_2 g^{\alpha \mu} p_1.s_2 
+ 2 m_1 g^{\alpha \mu} p_2.s_2 
- 2  m_2 {p_1}^\mu {s_2}^\alpha$\\
& $ + 2  m_1 {p_2}^\mu {s_2}^\alpha 
- 2  m_2{p_1}^\alpha {s_2}^\mu 
- 2  m_1 {p_2}^\alpha {s_2}^\mu$\\

&\\

$\gamma ^\alpha \gamma ^5$ 
& $2 i m_2 \varepsilon ^{\alpha \mu \delta \rho} {p_1}_\delta {s_2}_\rho 
+ 2 i m_1 \varepsilon ^{\alpha \mu \delta \rho} {p_2}_\delta {s_2}_\rho 
- 2 m_1 m_2 g^{\alpha \mu}$\\
& $ - 2  g^{\alpha \mu} p_1.p_2  
+  2{p_1}^\mu {p_2}^\alpha 
+  2 {p_1}^\alpha {p_2}^\mu$\\ 

&\\

$i\gamma ^5$
& $ -2 \varepsilon ^{\mu \gamma \delta \rho} {p_1}_\gamma {p_2}_\delta
{s_2}_\rho  
+ 2 i m_2 {p_1}^\mu 
- 2 i m_1 {p_2}^\mu$\\

&\\

$\sigma ^{\alpha \beta}$
& $ -2 m_2 \varepsilon ^{\alpha \beta \mu \rho}
{p_1}_\rho 
- 2 m_1 \varepsilon ^{\alpha \beta \mu \rho}
{p_2}_\rho 
- 2 i g^{\beta \mu} p_2.s_2 {p_1}^\alpha 
+ 2 i g^{\alpha \mu} p_2.s_2 {p_1}^\beta$\\
& $ + 2 i g^{\beta \mu} p_1.s_2 {p_2}^\alpha 
- 2 i g^{\alpha \mu} p_1.s_2 {p_2}^\beta 
- 2 i m_1 m_2  g^{\beta \mu}  {s_2}^\alpha 
- 2 i g^{\beta \mu} p_1.p_2 {s_2}^\alpha$\\
& $ + 2 i{p_1}^\mu {p_2}^\beta {s_2}^\alpha 
+ 2 i{p_1}^\beta {p_2}^\mu {s_2}^\alpha 
+ 2 i m_1 m_2  g^{\alpha \mu}  {s_2}^\beta 
+ 2 i g^{\alpha \mu} p_1.p_2 {s_2}^\beta $\\
& $- 2 i{p_1}^\mu {p_2}^\alpha {s_2}^\beta 
- 2 i{p_1}^\alpha {p_2}^\mu {s_2}^\beta 
- 2 i{p_1}^\beta {p_2}^\alpha {s_2}^\mu 
+ 2 i{p_1}^\alpha {p_2}^\beta {s_2}^\mu$\\
\hline
\end{tabular}
\end{center}
\end{table}
\begin{table}
\caption[Generalized traces of $\gamma$ matrices: $i\gamma ^5$
traces]{$i\gamma ^5$}
\vspace{.5cm}
\begin{center}
\begin {tabular} { |l||c| }
\hline
& $i\gamma ^5$\\
\hline 
$1$
& $- 2 i m_2 p_1.s_2 
+ 2 i m_1p_2.s_2$\\

&\\

$\gamma ^\alpha$ 
& $2 i p_2.s_2 {p_1}^\alpha 
- 2 i p_1.s_2 {p_2}^\alpha 
+ 2i{s_2}^\alpha (p_1.p_2 
- m_1 m_2)$\\

&\\

$\gamma ^\alpha \gamma ^5$ 
& $ -2 \varepsilon ^{\alpha \gamma \delta \rho} {p_1}_\gamma
{p_2}_\delta {s_2}_\rho 
- 2 i m_2 {p_1}^\alpha 
+ 2 i m_1 {p_2}^\alpha$\\ 

&\\

$i\gamma ^5$
&  $-2 m_1 m_2 
+ 2 p_1.p_2$\\

&\\

$\sigma ^{\alpha \beta}$
& $- 2 i \varepsilon ^{\alpha \beta \delta \rho}
{p_1}_\delta {p_2}_\rho 
+ 2 m_2 {p_1}^\beta {s_2}^\alpha 
- 2 m_1 {p_2}^\beta {s_2}^\alpha 
- 2 m_2 {p_1}^\alpha {s_2}^\beta 
+ 2 m_1 {p_2}^\alpha {s_2}^\beta $\\
\hline
\end{tabular}
\end{center}
\end{table}
\begin{table}
\caption[Generalized traces of $\gamma$ matrices: $\sigma^{\mu\nu}$  
traces]{$\sigma^{\mu\nu}$}
\vspace{.5cm}
\begin{center}
\begin {tabular} { |l||c| }
\hline
& $\sigma ^{\mu \nu}$\\
\hline 
$1$
& $2 m_2 \varepsilon ^{\mu \nu \delta \rho} {p_1}_\delta
{s_2}_\rho 
+ 2 m_1 \varepsilon ^{\mu \nu \delta \rho} {p_2}_\delta
{s_2}_\rho 
- 2 i {p_1}^\nu {p_2}^\mu 
+ 2 i {p_1}^\mu {p_2}^\nu$\\

&\\

$\gamma ^\alpha$ 
& $2 m_1 m_2 \varepsilon ^{\alpha \mu \nu \rho} {s_2}_\rho
+\varepsilon ^{\nu \gamma \delta \rho} {p_1}_\gamma {p_2}_\delta 
{s_2}_\rho g^{\alpha \mu} 
- \varepsilon ^{\mu \gamma \delta \rho} {p_1}_\gamma {p_2}_\delta
{s_2}_\rho g^{\alpha \nu}  
- 2 \varepsilon ^{\alpha \mu \nu \rho} {p_1}_\rho p_2.s_2$\\
& $ + 2 \varepsilon ^{\mu \nu \delta \rho} {p_2}_\delta {s_2}_\rho
{p_1}^\alpha 
+ \varepsilon ^{\alpha \nu \delta \rho} {p_2}_\delta {s_2}_\rho
{p_1}^\mu 
+ 2 i m_2 g^{\alpha \nu} {p_1}^\mu$\\
& $ - \varepsilon ^{\alpha \mu \delta \rho} {p_2}_\delta {s_2}_\rho
{p_1}^\nu 
-  2 i m_2 g^{\alpha \mu} {p_1}^\nu 
+ \varepsilon ^{\alpha \nu \delta \rho} {p_1}_\delta {s_2}_\rho
{p_2}^\mu$\\
& $ - 2 i m_1 g^{\alpha \nu} {p_2}^\mu 
- \varepsilon ^{\alpha \mu \delta \rho} {p_1}_\delta {s_2}_\rho
{p_2}^\nu 
+ 2 i m_1 g^{\alpha \mu} {p_2}^\nu$\\

&\\

$\gamma ^\alpha \gamma ^5$ 
& $ - 2 m_2 \varepsilon ^{\alpha \mu \nu \rho} {p_1}_\rho 
- 2 m_1\varepsilon ^{\alpha \mu \nu \rho} {p_2}_\rho 
- 2 i g^{\alpha \nu} p_2.s_2 {p_1}^\mu 
+ 2 i g^{\alpha \mu} p_2.s_2 {p_1}^\nu$\\
& $ - 2 i g^{\alpha \nu} p_1.s_2 {p_2}^\mu 
+ 2 i g^{\alpha \mu} p_1.s_2 {p_2}^\nu 
+ 2 i {p_1}^\nu {p_2}^\mu {s_2}^\alpha 
- 2 i {p_1}^\mu {p_2}^\nu {s_2}^\alpha$\\
& $ + 2 i m_1 m_2 g^{\alpha \nu} {s_2}^\mu 
+ 2 i g^{\alpha \nu} p_1.p_2 {s_2}^\mu 
- 2 i {p_1}^\nu {p_2}^\alpha {s_2}^\mu 
- 2 i {p_1}^\alpha {p_2}^\nu {s_2}^\mu$\\
& $ - 2 i m_1 m_2 g^{\alpha \mu} {s_2}^\nu 
- 2 i g^{\alpha \mu} p_1.p_2 {s_2}^\nu 
+ 2 i {p_1}^\mu {p_2}^\alpha {s_2}^\nu 
+ 2 i {p_1}^\alpha {p_2}^\mu {s_2}^\nu$\\ 

&\\

$i\gamma ^5$
& $2 i \varepsilon ^{\mu \nu \delta \rho} {p_1}_\delta {p_2}_\rho 
+ 2 m_2 {p_1}^\nu {s_2}^\mu 
- 2 m_1 {p_2}^\nu {s_2}^\mu 
- 2 m_2 {p_1}^\mu {s_2}^\nu 
+ 2 m_1 {p_2}^\mu {s_2}^\nu$\\

&\\

$\sigma ^{\alpha \beta}$
& $- i m_1  \varepsilon ^{\beta \nu \delta \rho} {p_2}_\delta
{s_2}_\rho g^{\alpha \mu}
+ i m_1  \varepsilon ^{\beta \mu \delta \rho} {p_2}_\delta
{s_2}_\rho g^{\alpha \nu} 
+ i m_1  \varepsilon ^{\alpha \nu \delta \rho} {p_2}_\delta
{s_2}_\rho g^{\beta \mu}
- 2 m_1 m_2 g^{\alpha \nu} g^{\beta \mu}$\\
& $ - i m_1  \varepsilon ^{\alpha \mu \delta \rho} {p_2}_\delta
{s_2}_\rho g^{\beta \nu} 
+ 2 m_1 m_2 g^{\alpha \mu} g^{\beta \nu} 
- 2 g^{\alpha \nu} g^{\beta \mu} p_1.p_2 
+ 2 g^{\alpha \mu} g^{\beta \nu} p_1.p_2$\\
& $ - 2 i m_1  \varepsilon ^{\alpha \beta \mu \nu} p_2.s_2 
- 2 i m_2  \varepsilon ^{\beta \mu \nu \rho} {s_2}_\rho {p_1}^\alpha 
+ 2 i m_2  \varepsilon ^{\alpha \mu \nu \rho} {s_2}_\rho {p_1}^\beta 
- 2 g^{\beta \nu} {p_1}^\mu {p_2}^\alpha$\\
& $ + 2 g^{\beta \mu} {p_1}^\nu {p_2}^\alpha 
+ 2 g^{\alpha \nu} {p_1}^\mu {p_2}^\beta 
- 2 g^{\alpha \mu} {p_1}^\nu {p_2}^\beta 
- i m_1 \varepsilon ^{\alpha \beta \nu \rho} {s_2}_\rho {p_2}^\mu$\\
& $ - 2 g^{\beta \nu} {p_1}^\alpha {p_2}^\mu 
+ 2 g^{\alpha \nu} {p_1}^\beta {p_2}^\mu 
+ i m_1 \varepsilon ^{\alpha \beta \mu \rho} {s_2}_\rho {p_2}^\nu 
+ 2 g^{\beta \mu} {p_1}^\alpha {p_2}^\nu$\\
& $- 2 g^{\alpha \mu} {p_1}^\beta {p_2}^\nu 
- i m_2 \varepsilon ^{\alpha \beta \nu \rho} {p_1}_\rho {s_2}^\mu 
+ i m_2 \varepsilon ^{\alpha \beta \mu \rho} {p_1}_\rho {s_2}^\nu $\\
\hline
\end{tabular}
\end{center}
\end{table}


\bibliography{thesis}
\biosketch{

\renewcommand{\baselinestretch}{1} \small
\begin{center}
{\scriptsize {\it Department of Physics and School of Computational
 Science and Information Technology\\ 
Florida State University, Tallahassee, FL  32306 \\
Phone: (850) 644-8347\\
Fax: (850) 644-0098\\
e-mail: raddad@csit.fsu.edu\\
URL: www.csit.fsu.edu/\~{}raddad/}} \\*[2.0cm]
\end{center}

{\small
\noindent
{\large {\underline{\bf Personal}}}\\*[.3cm]
\begin{tabular}{p{2.7cm}p{.01cm}p{11.2cm}}
Date of Birth: & &{\bf June 3, 1970.} \\
Place of Birth: & &{\bf Amman, Jordan.} \\
Marital Status: & &{\bf Married, with one son.} \\
Languages: & &{\bf Arabic and English; limited Spanish.} \\
Nationality: & &{\bf Jordanian.} \\
Immigration Status: & & {\bf Permanent resident of the U.S.A.}\\
\end{tabular}\\*[1.5cm]

\noindent
{\large {\underline{\bf Education}}}\\*[.3cm]
\begin{tabular}{p{2.7cm}p{.01cm}p{11.2cm}}
2000 & &{\bf Ph.D. in physics}, specialization in
computational theoretical \linebreak nuclear physics, Florida State
University. G.P.A. 4.0. \newline {\bf Dissertation:} ``Photoproduction of
Pseudoscalar Mesons from Nuclei''. \\
1994 & &{\bf M.Sc. in physics}, specialization in experimental
condensed \linebreak matter physics, Miami University. G.P.A. 4.0. \newline
{\bf Thesis:} ``Perturbed Angular Correlation Spectroscopy Study of the
Recrystallization of Natural Zircon''.\\
1992& &{\bf B.Sc. in physics}, University of Jordan. G.P.A. 90.4 \%.  \\
\end{tabular}\\*[1.7cm]

\newpage
\vspace*{1.0cm}
\noindent
{\large {\underline{\bf Experience}}}\\*[.3cm]
\begin{tabular}{p{2.7cm}p{.01cm}p{11.2cm}}
2000 to present & &{\bf Postdoctoral Research Associate}, School of
Computational Science and Information Technology\footnotemark[1] (CSIT), Florida State University. \\
1995 to 2000 & &{\bf Research Assistant}, School of Computational
Science and Information Technology\footnotemark[1] (CSIT), Florida State University. \\
July 97 & &{\bf Research Assistant}, Thomas Jefferson National
Accelerator Facility (TJNAF). \\
1994 to 1995 & &{\bf Teaching Assistant}, Department of Physics,
Florida State University. \\
1992 to 1994 & &{\bf Teaching Assistant}, Department of Physics, Miami
University. \\
\end{tabular}\\*[3.0cm]

\footnotetext[1]{Formerly known as the Supercomputer
Computations Research Institute (SCRI).}

\noindent
{\large {\underline{\bf Honors and Awards}}}\\*[.3cm]
\begin{tabular}{p{2.7cm}p{.01cm}p{11.2cm}}
April 2000 & & {\bf Dirac-Hellmann Award in Theoretical Physics},
Florida State University. \\
June to August 1999 & & {\bf Summer Institute in Japan Fellowship},
National Science Foundation (NSF) and Science and Technology Agency of
Japan (STA).\\
1996 to 1998 & &{\bf SURA/Jefferson Laboratory Fellowship}, Thomas
Jefferson National Accelerator Facility (TJNAF). \\
August 1997 & &{\bf Ninth Annual Summer School in Nuclear Physics \linebreak
Scholarship}, Yale University. \\
June 1996 & &{\bf HUGS at Jefferson Laboratory Fellowship}, Thomas
Jefferson National Accelerator Facility (TJNAF). \\
July 1992 & &{\bf Ranked first in the B.Sc. degree}, University of
Jordan. \\
 & &{\bf Member of $\Sigma \Pi \Sigma$ and $\Phi K \Phi$ honor societies.} \\
\end{tabular}\\*[1.7cm]

\newpage
\vspace*{1.0cm}
\noindent
{\large {\underline{\bf Analytical and Computational Skills}}}
\begin{description}
\item[Analytical Skills:] Scientific research using
various aspects of applied mathematics such as mathematical simulation and modeling, numerical analysis,
differential and integral equations, Green's functions and boundary
value problems, group theory, calculus of variation, variational
methods, and perturbation theory.

\item[Computational Skills]: Scientific computing involving programming, symbolic and logical
analysis, Monte Carlo simulations, algorithm development, data analysis, and software
development.\\ 
	\begin{itemize}
	\item {\bf Programming Languages:} FORTRAN and C$^{++}$.
	\item {\bf Symbolic Manipulators:} Mathematica and Maple.	
	\item {\bf Operating Systems:} UNIX.\\*[1.3cm]
	\end{itemize}
\end{description}

\noindent 
{\large {\underline{\bf General Research Interests}}}\\*[.3cm]
Theoretical nuclear physics, mathematical physics, and computational
physics.\\*[.8cm]   

\vspace*{1.0cm}
\noindent 
{\large {\underline{\bf Ph.D. Dissertation Advisor}}}\\*[.3cm]
{\bf Jorge Piekarewicz} at Florida State University.\\*[1.3cm]  

\noindent
{\large {\underline{\bf Additional Research Advisors}}}\\*[.3cm]
{\bf Makoto Oka} at Tokyo Institute of Technology during the NSF Summer
Institute in Japan \linebreak Fellowship.\\
{\bf Jos\'e Goity} at Thomas Jefferson National Accelerator Facility
(TJNAF) and Hampton University
during the research assistantship at Jefferson Laboratory.

\newpage
\noindent
{\large {\underline{\bf Publications}}}
\begin{enumerate}
\item {\bf L.J. Abu-Raddad and J. Piekarewicz}, ``Quasifree kaon
photoproduction from nuclei in a relativistic approach'',
Phys. Rev. C {\bf (61)}, 014604, (2000).

\item {\bf L.J. Abu-Raddad, J. Piekarewicz, A.J. Sarty, and
R.A. Rego}, ``Lessons to be learned from the coherent photoproduction
of pseudoscalar mesons'', Phys. Rev. C {\bf (60)}, 054606 (1999).

\item {\bf L.J. Abu-Raddad, J. Piekarewicz, A.J. Sarty, and
M. Benmerrouche}, ``Nuclear dependence of the coherent $\eta$
photoproduction reaction in a relativistic approach'', Phys. Rev. C
{\bf (57)}, 2053 (1998).

\item {\bf H. Jaeger, L.J. Aburaddad, and D. Wick}, ``TDPAC study of
structural disorder in metamict zircon'', App. Radiat. Isotopes {\bf
48 (8)}, 1083 (1997).

\item {\bf L. Abu-Raddad, R. Klindworth, and L. Zhang}, ``Two theories
of baryons'', Proceedings of the 11th HUGS at CEBAF Workshop, 22 (1996).

\item {\bf H. Jaeger, L.J. Aburaddad}, ``Two-detector coincidence
routing circuit for personal computer-based multichannel analyzer'',
Rev. Sci. Intrum. {\bf 66 (4)}, 3069 (1995).
\end{enumerate}

\noindent
{\large {\underline{\bf Presentations and Conferences}}}
\begin{enumerate}

 \item {\bf L.J. Abu-Raddad}, ``Photoproduction of pseudoscalar mesons
from nuclei'', TRIUMF, February 2000, Vancouver, Canada.

 \item {\bf L.J. Abu-Raddad and J. Piekarewicz}, ``Quasifree kaon
photoproduction from nuclei in a relativistic approach'', Fall Meeting
of the Division of Nuclear Physics of the American Physical Society,
October 1999, Asilomar, California.

 \item {\bf L.J. Abu-Raddad}, ``Photoproduction of pseudoscalar mesons
from nuclei'', presented at the July Mini Workshop, Tokyo Institute of
Technology, July 1999, Tokyo, Japan. It has been also presented at the
Research Center for Nuclear Physics (RCNP), August 1999, Osaka
University, Osaka, Japan, and at the Department of Physics, Osaka City
University, August 1999, Osaka, Japan.

\item {\bf L.J. Abu-Raddad, J. Piekarewicz, and A.J. Sarty},
``Coherent photoproduction of neutral pions'', Centennial Meeting of
the American Physical Society, March 1999, Atlanta, Georgia.

\item {\bf L.J. Abu-Raddad, J. Piekarewicz, A.J. Sarty, and
M. Benmerrouche}, ``Nuclear dependence of the coherent $\eta$
photoproduction of from nuclei'', Fall Meeting of the Division of
Nuclear Physics of the American Physical Society, October 1997,
Whistler, Canada.

\item {\bf L.J. Abu-Raddad and H. Jaeger}, ``TDPAC study of natural
zircon'', Spring Meeting of the Ohio Section of the American Physical
Society, May 1994, Cleveland, Ohio.

\end{enumerate}
}

\end{document}